\documentclass[preprint,12pt]{elsarticle}

\usepackage{graphicx}

\usepackage{lineno}

\usepackage{amsmath,amssymb}
\usepackage{bm}

\usepackage{float}

\usepackage{geometry}
\usepackage{threeparttable}
\biboptions{numbers,sort&compress}
\usepackage{subfigure}

\usepackage[colorlinks, citecolor=red]{hyperref}
\usepackage{xcolor}  

\journal{Elsevier}
\usepackage{CJK} 
\begin{document}

\begin{frontmatter}
	
	\title{Further acceleration of multiscale simulation of rarefied gas flow via a generalized boundary treatment}
	
	\author[sustech,hkust]{Wei Liu}
	\author[sustech]{Yanbing Zhang}
	\author[sustech]{Jianan Zeng}
	\author[sustech]{Lei Wu\corref{cor1}}
	\ead{wul@sustech.edu.cn}
	\cortext[cor1]{Corresponding author}

	\address[sustech]{Department of Mechanics and Aerospace Engineering, Southern University of Science and Technology, Shenzhen, 518055, China}
	\address[hkust]{Division of Emerging Interdisciplinary Areas, The Hong Kong University of Science and Technology, Clear Water Bay, Hong Kong, China}
	
\begin{abstract}
		
The recently-developed general synthetic iterative scheme (GSIS) is efficient in simulating multiscale rarefied gas flows due to the coupling of mesoscopic kinetic equation and macroscopic synthetic equation: for linearized Poiseuille flow where the boundary flux is fixed at each iterative step, the steady-state solutions are found within dozens of iterations in solving the gas kinetic equations, while for general nonlinear flows the iteration number is increased by about one order of magnitude, caused by the incompatible treatment of the boundary flux for the macroscopic synthetic equation. In this paper, we propose a generalized boundary treatment (GBT) to further accelerate the convergence of GSIS. The main idea is, the truncated velocity distribution function at the boundary, similar to that used in the Grad 13-moment equation, is reconstructed by the macroscopic conserved quantities from the synthetic equation, and the high-order correction of non-equilibrium stress and heat flux from the kinetic equation; therefore, in each inner iteration solving the synthetic equation, the explicit constitutive relations facilitate real-time updates of the macroscopic boundary flux, driving faster information exchange in the flow field, and consequently achieving quicker convergence. Moreover, the high-order correction derived from the kinetic equation can compensate the approximation by the truncation and ensure the boundary accuracy. The one-dimensional Fourier flow, two-dimensional hypersonic flow around a cylinder, three-dimensional pressure-driven pipe flow and the flow around the hypersonic technology vehicle are simulated. The accuracy of GSIS-GBT is validated by the direct simulation Monte Carlo method, the previous versions of GSIS, and the unified gas-kinetic wave-particle method. For the efficiency, in the near-continuum flow regime and slip regime, GSIS-GBT can be faster than the conventional iteration scheme in the discrete velocity method and the previous versions of GSIS by two- and one-order of magnitude, respectively. 
\end{abstract}
	\begin{keyword}
		Rarefied gas flow  \sep Boltzmann kinetic equation \sep Finite volume method \sep general synthetic iterative scheme \sep hypersonic flow \sep low-speed internal flow	\end{keyword}	
\end{frontmatter}


\newpage 

\section{Introduction}\label{S:1}

Multiscale rarefied gas flows are encountered in many engineering problems, such as the high-altitude aerothermodynamics of space vehicles~\cite{ivanov_computational_1998, blanchard_aerodynamic_2012}, micro-electro-mechanical systems~\cite{gad-el-hak_flow_2001}, and gas transportation within ultra-tight shale rock~\cite{klinkenberg_permeability_1941, darabi_gas_2012}. 
For instance, supersonic flows around the X38-like vehicle encompass the free-molecular to continuum flow regimes, where the Knudsen number (Kn, the ratio of mean free path of gas molecules to characteristic flow length) traverses a range of five orders of magnitude~\cite{jiang_implicit_2019}.

While the classic Navier-Stokes (NS) equations are only applicable up to the slip flow regime where $Kn\lessapprox 0.1$, the Boltzmann equation transcends the limitations of continuum assumption and describes the gas flows from the continuum, slip, transition and free-molecular regimes. The discrete velocity method (DVM) and the direct simulation Monte Carlo (DSMC) method are the two representative numerical methods to solve the Boltzmann equation~\cite{goldstein_investigations_1989, bird_molecular_1994}, which are efficient when the Knudsen number is large. However, a significant challenge arises when the rarefaction effects gradually diminish and the gas become denser. Hence, the spatial cell size and the time step must be smaller than the limits imposed by the molecular mean free path and mean collision time, respectively~\cite{neumann_hybrid_2014}, which not only requires huge amount of computer memory, but also converges slowly to the steady-state solution.

The common feature of the DVM and DSMC is that the streaming and collision operators in the Boltzmann equation are handled separately, which results in large numerical dissipation when the Knudsen number is small, so that the cell size and time step have to be small. 
Based on the integral solution of the Boltzmann-BGK equation, the unified gas-kinetic scheme (UGKS) has been proposed to simultaneously handle the streaming and collision, via the multiscale flux reconstruction in the finite-volume framework~\cite{xu_unified_2010}. As a result, it has the asymptotic preserving property to overcome the constraints in cell size and time step.  This concept and technique have been further explored, and many multiscale schemes are developed~\cite{guo_discrete_2013, zhu_unified_2019, fei_time-relaxed_2023}.

Even though the numerical dissipation in near-continuum flows is solved, the slow convergence remains a persistent problem. In time-dependent solvers, steady-state solutions are usually found after $10^5$ to $10^6$ steps~\cite{wang_comparative_2018}. In the implicit solver such as the conventional iterative scheme (CIS, which evolves the temporal derivatives, the streaming term, and the loss part of the collision operator at the current time step, while the gain part of the collision operator is computed based on macroscopic quantities from the previous iteration step), steady-state solutions can be found within tens to hundreds iterations for flows in the free-molecular and transition regimes; however, the iteration number in simulating near-continuum flows experiences a substantial surge, e.g., to about $10^6$ steps when $\text{Kn}\approx0.001$~\cite{wang_comparative_2018}. This poses a huge computational burden in the DVM since the velocity distribution function in the Boltzmann equation is defined in a six-dimensional phase space. For efficient simulation, the iteration number should be controlled within $10^3$ or even $10^2$, at all Knudsen numbers.

Building on Larsen's work in accelerating the convergence in neutron transport~\cite{larsen_numerical_1983}, the concept of mesoscopic-macroscopic coupling for iteration acceleration has been extended to rarefied gas dynamics, including the synthetic iterative scheme~\cite{Valougeorgis2023Siam,szalmas_fast_2010}, the moment-guided Monte Carlo method~\cite{degond_moment-guided_2011}, the high-order/low-order method~\cite{chacon_multiscale_2017}, and the general synthetic iterative scheme (GSIS)~\cite{su_can_2020}. Particularly, the GSIS been successfully applied to gas transportation in porous media~\cite{su_gsis_2020}, polyatomic gas flows~\cite{su_multiscale_2021,zeng_general_2023-1}, and unsteady multiscale flows~\cite{zeng_general_2023}. Moreover, it has been extended to accelerate the convergence of low-variance DSMC~\cite{luo_boosting_2023}. Rigorous mathematical analysis has shown that the GSIS is able to  asymptotically preserve the Navier-Stokes limit when the cell size is much larger than the molecular mean free path, and facilitate rapid convergence to the steady state~\cite{su_fast_2020}. 
The key to the success of GSIS lies in its incorporation of the mesoscopic kinetic equation alongside the macroscopic synthetic equation. The stress and heat flux in the macroscopic synthetic equations are expressed as combinations of the Newton law of viscosity and the Fourier law of heat conduction, as well as the high-order constitutive relations extracted from the kinetic equation. Providing higher-order corrections to the synthetic equations ensures the iterations to converge to accurate results, particularly in strong rarefied flows. The explicit inclusion of the Newton and Fourier constitutive relations enable the macroscopic equations to efficiently exchange the flow information in the near-continuum regime, thereby accelerating convergence, see the Fourier flow in Ref.~\cite{su_can_2020}. 

However, the coupling acceleration between mesoscopic and macroscopic treatments is applied only in the interior computational domain, e.g., it has been pointed out that ``the zeroth-order moment is not accelerated at the boundaries and its estimate is based on the values of the distribution function at the boundaries obtained at the first stage of the iteration"~\cite{szalmas_fast_2010}. This does not affect the speed of convergence in the Poiseuille flow since the distribution function reflected from the solid wall is always zero, so that the numerical flux is fixed. For general nonlinear flows, this is not the case, and since the synthetic equation plays an important role in guiding the gas system towards the steady state, its boundary condition (flux) is critical in accelerating the convergence. In the first work to extend the GSIS to nonlinear flows, the numerical fluxes for macroscopic equations at the boundaries are directly calculated from the DVM, which remains unchanged when solve the macroscopic synthetic equations~\cite{zhu_general_2021}. In the second work of nonlinear GSIS, an incremental feedback of macroscopic variables from the cell centers to wall is used to update the boundary flux~\cite{zeng_general_2023}. Although both methods are faster than the CIS in near-continuum flow regimes, converged solutions for hypersonic flows are found after hundreds of iterations, which is about 10 times slower than the GSIS for linear flows where the steady-state solutions can be found within dozens of iterations~\cite{su_can_2020}. Thus, for general nonlinear flows, the boundary flux needs to be properly treated, to further reduce the iteration steps.

The boundary conditions are of paramount importance in numerical simulations, and this importance is significant in rarefied gas flows. Many critical phenomena in rarefied flows, such as the velocity slip, temperature jump, and thermal creep~\cite{maxwell_vii_1997, sharipov_data_2011,smoluchowski_von_smolan_ueber_1898, ohwada_numerical_1989,sone_flow_1972,sone_asymptotic_2000}, appear within a few mean free path away from the wall. The conventional no-slip wall boundary conditions of macroscopic equations are physically inconsistent with these rarefied phenomena, and their direct application can lead to unstable and inaccurate results in macroscopic equations~\cite{zhao_application_2023}. 
Fortunately, the macroscopic boundary conditions are solved in the moment equations for slightly-rarefied gas. For instance, Grad proposed the wall boundary condition for the 13-moment equations by integrating the truncated distribution function~\cite{grad_kinetic_1949}, which is further extended to the regularized 13- and 26-moment equations by Gu \textit{et al.}~\cite{gu_computational_2007, gu_high-order_2009}. 
Studies demonstrated that the velocity slip can be automatically obtained by these boundary conditions in the moment equations. 

Inspired by the boundary condition in the moment methods~\cite{grad_kinetic_1949,gu_computational_2007, gu_high-order_2009}, we shall propose in the present work a generalized boundary treatment (GBT) to further accelerate the convergence of GSIS. The rest of the paper is organized as follows. In section~\ref{S:2}, the mesoscopic kinetic equation and macroscopic synthetic equation are introduced, together with their numerical schemes in the finite-volume framework. In section~\ref{S:3}, the GBT is proposed in great details. In section~\ref{S:4}, the accuracy and efficiency of GSIS-GBT is assessed in planar Fourier flow, hypersonic flow past a cylinder, pressure-driven flow through three circular pipes, and the multiscale flow around a hypersonic technology vehicle. Finally, conclusions are given in section~\ref{S:5}.


\section{Methodology}
\label{S:2}

In this section, the mesoscopic kinetic equation for the polyatomic gas with translational and rotational degrees of freedom is firstly introduced, as well as the finite-volume scheme; then, the macroscopic synthetic equation is derived from the kinetic equation, which is also solved by the finite-volume method.  Finally, the flowchart of GSIS is outlined. 

\subsection{The kinetic equation}
\label{SS:2-1}

To simultaneously account for real gas effects while avoiding complex kinetic models, we consider the kinetic model of polyatomic gas with rotational degrees of freedom $d_r$, excluding vibrational and electronic degrees of freedom. The gas system can be described by the distribution function $f^{T}(t, \boldsymbol{x}, \boldsymbol{\xi}, I_r) $, where the variable $\mathbf{x}$ denotes spatial position, $t$ represents time, $\bm{\xi}$ represents the molecular velocity, and $I_r$ is the rotational energy. Two velocity distribution functions (VDFs), $f_0(\mathbf{x}, \bm{\xi}, t)=\int f^{T}dI_r$ and $f_1(\mathbf{x}, \bm{\xi}, t)=\int I_rf^{T}dI_r$, can be introduced to reduce the rotational energy variable, and the final dimensionless kinetic model in the absence of external force can be expressed as~\cite{rykov_model_1976,li_uncertainty_2021}
\begin{equation}
	\begin{aligned}
		& \frac{\partial f_0}{\partial t}+\bm{\xi} \cdot \frac{\partial f_0}{\partial \mathbf{x}}=\frac{g_{0 t}-f_0}{\tau}+\frac{g_{0 r}-g_{0 t}}{Z_r \tau}, \\
		& \frac{\partial f_1}{\partial t}+\bm{\xi} \cdot \frac{\partial f_1}{\partial \mathbf{x}}=\frac{g_{1 t}-f_1}{\tau}+\frac{g_{1 r}-g_{1 t}}{Z_r \tau},
	\end{aligned}
	\label{eq2.1.1}
\end{equation}

\noindent where $\tau$ and $Z_r \tau$ represent the elastic (where the kinetic energy is conserved) and inelastic (energy exchange between translational and rotational motions) collision times of gas molecules, with $Z_r$ being the rotational collision number and the mean collision time is defined as 
\begin{equation} 
	{ \tau=\frac{\mu}{p} \sqrt{\frac{\pi}{2}} \text{Kn},}
	\label{eq2.1.2}
\end{equation} 

\noindent where $\mu$ is the normalized shear viscosity: if the power-law potential is used, we have 
\begin{equation}
	\mu=T_t^\omega,
	\label{eq2.1.3}
\end{equation}

\noindent with $\omega$ being the viscosity index. 

The function $g$ denotes the reference distribution function, where the subscripts $t$ and $r$ stand for the relaxation processes associated with translational and rotational motions, respectively:
\begin{equation}
	\begin{gathered}
		g_{0 t}=f_0^{e q}\left[1+\frac{2 \mathbf{q}_t \cdot \mathbf{c}}{15 T_t p_t}\left(\frac{c^2}{2 T_t}-\frac{5}{2}\right)\right], 
		\quad 
		g_{0 r}=f_1^{e q}\left[1+\frac{2 \mathbf{q}_0 \cdot \mathbf{c}}{15 {T p}}\left(\frac{c^2}{2 T}-\frac{5}{2}\right)\right], \\
		g_{1 t}=\frac{d_r}{2} T_r g_{0 t}+f_0^{e q} \frac{\mathbf{q}_r \cdot \mathbf{c}}{\rho T_t}, \quad 
		g_{1 r}=\frac{d_r}{2} T g_{0 r}+f_1^{e q} \frac{\mathbf{q}_1 \cdot \mathbf{c}}{\rho T},
	\end{gathered}
	\label{eq2.1.4}
\end{equation}
with the equilibrium distribution functions
\begin{equation}
	\begin{aligned}
		& f_0^{e q}=\rho\left(\frac{1}{2 \pi T_t}\right)^{3 / 2} \exp \left(-\frac{c^2}{2 T_t}\right), \quad 
		f_1^{e q}=\rho\left(\frac{1}{2 \pi T}\right)^{3 / 2} \exp \left(-\frac{c^2}{2 T}\right).
	\end{aligned}
	\label{eq2.1.5}
\end{equation}
It should be noted that, from Eq.~\eqref{eq2.1.1} we known that the VDFs $f_0$ and $f_1$ evolve towards the reference VDFs in Eq.~\eqref{eq2.1.4}. Meanwhile, during the evolution of $f_0$ and $f_1$, the heat fluxes in the reference VDFs gradually go to zero in isolated system. Therefore, in the final steady state, the VDFs $f_0$ and $f_1$ will be described by the corresponding equilibrium distribution functions in Eq.~\eqref{eq2.1.5}, which is consistent with the Boltzmann H-theorem.

Macroscopic quantities including the density $\rho$, velocity $\mathbf{u}$, translational temperature $T_t$, rotational temperature $T_r$, as well as the stress $\boldsymbol{\sigma}$, translational heat flux $\mathbf{q}_t$ and rotational heat flux $\mathbf{q}_r$ are expressed through the moment integration of the distribution functions as

\begin{equation}
	\begin{gathered}
		\rho=\int f_0 \mathrm{~d} \boldsymbol{\xi}, \quad 
		\rho \mathbf{u}=\int \boldsymbol{\xi} f_0 \mathrm{~d} \boldsymbol{\xi}, \\
		\frac{3}{2} \rho T_t=\int \frac{c^2}{2} f_0 \mathrm{~d} \boldsymbol{\xi}, \quad \boldsymbol{\sigma}=\int\left(\mathbf{c}\mathbf{c}-\frac{c^2}{3} \mathrm{I}\right) f_0 \mathrm{~d} \boldsymbol{\xi}, 
		\quad \frac{d_r}{2} \rho T_r=\int f_1 \mathrm{~d} \boldsymbol{\xi}, \\
		\mathbf{q}_t=\int \mathbf{c} \frac{c^2}{2} f_0 \mathrm{~d} \boldsymbol{\xi}, 
		\quad \mathbf{q}_{\mathbf{r}}=\int \mathbf{c} f_1 \mathrm{~d} \boldsymbol{\xi}.
	\end{gathered}
	\label{eq2.1.6}
\end{equation}
Furthermore, relationships pertaining to the total temperature $T$, translational pressure $p_t$, and total pressure $p$ are delineated as follows:
\begin{equation}
	\begin{aligned}
		& T=\frac{3 T_t+d_r T_r}{3+d_r}, \quad 
		p_t=\rho T_t, 
		\quad p= \rho T, 
	\end{aligned}
	\label{eq2.1.7}
\end{equation}

\noindent where $\mathbf{c}=\boldsymbol{\xi}-\mathbf{u}$ denotes the peculiar velocity. Here the relaxation rates of the heat flux $\boldsymbol{A}=\left[A_{t t}, A_{t r}, A_{r t}, A_{r r}\right]$ are derived from the DSMC simulation to determine $\mathbf{q}_0$ and $\mathbf{q}_1$ as~\cite{li_uncertainty_2021}: 
\begin{equation}
	\left[\begin{array}{l}
		\mathbf{q}_0 \\
		\mathbf{q}_1
	\end{array}\right]=\left[\begin{array}{cc}
		\left(2-3 A_{t t}\right) Z_r+1 & -3 A_{t r} Z_r \\
		-A_{r t} Z_r & -A_{r r} Z_r+1
	\end{array}\right]\left[\begin{array}{l}
		\mathbf{q}_t \\
		\mathbf{q}_r
	\end{array}\right].
	\label{eq2.1.8}
\end{equation}

It is noted that the derivations mentioned here are based on dimensionless kinetic equations. The dimensionless treatment of macroscopic variables and distribution functions is presented in \ref{A:1} for reference.

\subsection{The conventional iterative scheme}
\label{SS:2-2}

We briefly introduce the CIS to solve the kinetic equation by the implicit finite-volume method. Considering that the evolution equations for translational and rotational distribution functions in Eq.~\eqref{eq2.1.1} exhibit identical structural formulations, here we employ the notations $f$ and $g$ to represent a unified form of translational/rotational distribution functions. The discrete form of the mesoscopic kinetic equation within the control volume cell and a time step $\Delta t_f=t^{n+1}-t^n$
from the $n-$th step to the $(n+1)-$th step is shown as follows
\begin{equation}
	\frac{f_i^{n+1}-f_i^n}{\Delta t_f}+\frac{1}{\Omega_i} \sum_{j \in N(i)} {\xi}_n f_{i j}^{n+1} S_{i j}=\frac{g_{t, i}^n-f_i^{n+1}}{\tau_i^n}+\frac{g_{r, i}^n-g_{t, i}^n}{Z_r \tau_i^n},
	\label{eq2.2.1}
\end{equation}
where the subscript $i$ denotes the average variables at the cell center, while the subscript $ij$ represents the cell interface located between cells $i$ and $j$. $S_{i j}$ and $\Omega_i$ respectively denote the area of the cell interface and the volume of the cell. ${\xi}_n = \bm{{\xi}}_n \cdot \mathbf{n}_{i j}$ represents the molecular velocity component along the interface normal vector $\mathbf{n}_{i j}$. 

To facilitate matrix-free implicit iterations, an incremental form of the distribution function is introduced:
\begin{equation}\label{delta_f_dvm}
    \Delta f_i^n=f_i^{n+1}-f_i^n
\end{equation} 
which allows the rewriting of Eq.~\eqref{eq2.2.1} as 
\begin{equation}
	\left(\frac{1}{\Delta t_f}+\frac{1}{\tau_i^n}\right) \Delta f_i^n+\frac{1}{\Omega_i} \sum_{j \in N(i)} \xi_n \Delta f_{i j}^n S_{i j}
	=\underbrace{\frac{g_i^n-f_i^n}{\tau_i^n}-\frac{1}{\Omega_{i}} \sum_{j \in N(i)} \xi_n f_{i j}^n S_{i j}}_{\mathrm{RHS}_i}.
	\label{eq2.2.2}
\end{equation}

\noindent where $\mathrm{RHS}_i$ is the mesoscopic residual. The mesoscopic time step $\Delta t_f$ can be determined based on the Courant–Friedrichs–Lewy (CFL) condition as \cite{blazek_computational_2015}:
\begin{equation}
	\Delta t_f=\vartheta_f \min _i \frac{\Omega_i}{\left(\left|u_{i, x}\right|+3 \sqrt{T_i}\right) S_{i, x}+\left(\left|u_{i, y}\right|+3 \sqrt{T_i}\right) S_{i, y}+\left(\left|u_z\right|+3 \sqrt{T_i}\right) S_{i, z}},
	\label{eq2.2.3}
\end{equation}
in which $S_{i}$ is the projections of the control volume on the corresponding direction and $\vartheta_f$ is the mesoscopic CFL number. 

The CIS is found to be efficient when the Knudsen number is large, but is highly dissipative and inefficient when the Knudsen number is small~\cite{wang_comparative_2018}. To expedite the convergence of the kinetic equations, the VDF is firstly evaluated at the intermediate step, i.e., $n+1$ in Eqs.~\eqref{eq2.2.1} and~\eqref{delta_f_dvm} is replaced by $n+1/ 2$. Subsequently, macroscopic synthetic equations are introduced in GSIS which will guide the evolution of the VDF towards steady-state solutions.

\subsection{The macroscopic synthetic equation in GSIS}
\label{SS:2-3}

The proper design of  macroscopic synthetic equation is crucial in GSIS, which the guidelines are: (i) the synthetic equation should be derived exactly from the kinetic equation, (ii) the synthetic equation should facilitate efficient exchange of flow information in the whole computational domain, and (iii) the synthetic equation should be as simple as possible, so that it is compatible with the numerical techniques in the computational fluid dynamics. In this case, the second version of GSIS (which contains the evolution equations for the density, velocity, and temperature only) is used~\cite{zhu_general_2021,zeng_general_2023}, rather than the first version (which not only contains the evolution equations for the density, velocity, and temperature, but also the evolution equations for the stress and heat flux)~\cite{su_can_2020}. 

Exploiting the mesoscopic-macroscopic relationships in Eq.~\eqref{eq2.1.6} and applying them to Eq.~\eqref{eq2.1.1}, the governing equations for macroscopic quantities can be derived as
\begin{equation}
	\begin{aligned}
		& \frac{\partial \rho}{\partial t}+\nabla \cdot(\rho \mathbf{u})=0, \\
		& \frac{\partial}{\partial t}(\rho \mathbf{u})+\nabla \cdot(\rho \mathbf{u} \mathbf{u})+\nabla \cdot\left(\rho T_t \mathrm{I}+\boldsymbol{\sigma}\right)=0, \\
		& \frac{\partial}{\partial t}(\rho E)+\nabla \cdot(\rho E \mathbf{u})+\nabla \cdot\left(\rho T_t \mathbf{u}+\boldsymbol{\sigma} \cdot \mathbf{u}+\mathbf{q}_t+\mathbf{q}_r\right)=0, \\
		& \frac{\partial}{\partial t}\left(\rho E_r\right)+\nabla \cdot\left(\rho E_r \mathbf{u}\right)+\nabla \cdot \mathbf{q}_r=\frac{d_r \rho}{2} \frac{T-T_r}{Z \tau},
	\end{aligned}
	\label{eq2.1.9}
\end{equation}

\noindent where $E=\left(3 T_t+d_r T_r\right) / 2+u^2 / 2$ and $E_r=d_r T_r / 2$ are the total and rotational energies, respectively. 

When the Knudsen number is small, the stress and heat flux in Eq.~\eqref{eq2.1.9} can be approximated  via the first-order Chapman-Enskog expansion of the kinetic equation as~\cite{chapman_mathematical_1962}:
\begin{equation}
	\begin{aligned}
\boldsymbol{\sigma}\approx &\boldsymbol{\sigma}^{\mathrm{M}}=-\mu\left(\nabla \mathbf{u}+\nabla \mathbf{u}^{\mathrm{T}}-\frac{2}{3} \nabla \cdot \mathbf{u I}\right), \\
	\mathbf{q}_t\approx	&\mathbf{q}_t^{\mathrm{M}}=-\kappa_t \nabla T_t, \quad \mathbf{q}_r\approx \mathbf{q}_r^{\mathrm{M}}=-\kappa_r \nabla T_r,
	\end{aligned}
	\label{eq2.1.10}
\end{equation}
where $\kappa_t$ and $\kappa_r$ are the translational and rotational thermal conductivities:
\begin{equation}
	\left[\begin{array}{l}
		\kappa_t \\
		\kappa_r
	\end{array}\right]=\frac{p_t \tau}{2}\left[\begin{array}{ll}
		A_{t t} & A_{t r} \\
		A_{r t} & A_{r r}
	\end{array}\right]^{-1}\left[\begin{array}{c}
		5 \\
		d_r
	\end{array}\right] .
	\label{eq2.1.11}
\end{equation}

The macroscopic synthetic equation is constructed as follows:
\begin{equation}
	\begin{aligned}
		& \frac{\partial \rho}{\partial t}+\nabla \cdot(\rho \mathbf{u})=0, \\
		& \frac{\partial}{\partial t}(\rho \mathbf{u})+\nabla \cdot(\rho \mathbf{u} \mathbf{u})+\nabla \cdot\left(\rho T_t \mathrm{I}+\boldsymbol{\sigma}^{\mathrm{M}}\right)=-\nabla \cdot(\Delta \boldsymbol{\sigma}), \\
		& \frac{\partial}{\partial t}(\rho E)+\nabla \cdot(\rho E \mathbf{u})+\nabla \cdot\left(\rho T_t \mathbf{u}+\boldsymbol{\sigma}^{\mathrm{M}} \cdot \mathbf{u}+\mathbf{q}^{\mathrm{M}}_t+\mathbf{q}^{\mathrm{M}}_r\right)=-\nabla \cdot(\Delta \boldsymbol{\sigma} \cdot \mathbf{u}+\Delta \mathbf{q}_t+\Delta \mathbf{q}_r), \\
		& \frac{\partial}{\partial t}\left(\rho E_r\right)+\nabla \cdot\left(\rho E_r \mathbf{u}\right)+\nabla \cdot \mathbf{q}^{\mathrm{M}}_r-\frac{d_r \rho}{2} \frac{T-T_r}{Z \tau}=-\nabla \cdot \Delta \mathbf{q}_r,
	\end{aligned}
	\label{eq2.2.4}
\end{equation}
where
\begin{equation}
	\begin{gathered}
		\Delta \boldsymbol{\sigma}=\boldsymbol{\sigma}^{n+1 /2}-\boldsymbol{\sigma}^{\mathrm{M}, n+1 /2}=\int\left(\mathbf{c}-\frac{c^2}{3} \mathrm{I}\right) f_0^{n+1 /2} \mathrm{~d} \xi
  -\boldsymbol{\sigma}^{\mathrm{M}, n+1 /2}, \\
\Delta \mathbf{q}_t=\mathbf{q}_t^{n+1 /2}-\mathbf{q}_t^{\mathrm{M}, n+1 /2}=\int \mathbf{c} \frac{c^2}{2} f_0^{n+1 /2} \mathrm{~d} \xi-\mathbf{q}_t^{\mathrm{M}, n+1 /2}, \\
\Delta \mathbf{q}_r=\mathbf{q}_r^{n+1 /2}-\mathbf{q}_r^{\mathrm{M}, n+1 /2}=\int \mathbf{c} f_1^{n+1 /2} \mathrm{~d} \xi-\mathbf{q}_r^{\mathrm{M}, n+1 /2}.
	\end{gathered}
	\label{eq2.2.9}
\end{equation}

It should be emphasized that, in the numerical simulation, the variables at the left-hand-side of Eq.~\eqref{eq2.2.4} are evaluated at the $(n+1)$-th time step, while the one at the right-hand-side are evaluated at the $(n+1/2)$-th time step. Only when the solution finally converges (the solution at the $(n+1/2)$-th time step is the same as that at the $(n+1)$-th time step) can the synthetic equation turns back to Eq.~\eqref{eq2.1.9} which is exactly derived from the kinetic equation. In the intermediate iteration step, Eq.~\eqref{eq2.2.4} is not the same as the exact one~\eqref{eq2.1.9}, but such treatment plays an important role in boosting the convergence to the steady state.

Compared to the original macroscopic equation~\eqref{eq2.1.9}, the synthetic equation introduces high-order corrections in the stress and heat flux $\Delta \boldsymbol{\sigma}, \Delta \mathbf{q}_t, \Delta \mathbf{q}_r$ on the right-hand-side of equations, ensuring the accuracy of macroscopic equation iterations in the presence of strong rarefaction effects. On the other hand, the explicit inclusion of the Newton law of viscosity and the Fourier law of thermal conductivity in the left-hand-side of Eq.~\eqref{eq2.2.4} could lead to fast exchange of the flow field, since in the steady state the diffusion-type equations for the velocity and temperature can be obtained, where the information propagation speed is infinite.

\subsection{Finite-volume method for the synthetic equations}
\label{SS:2-4}

Under the finite-volume discretization, the macroscopic synthetic equation within the control volume cell $i$ and a inner time step $\Delta t_W=t^{m+1}-t^m$ can be formulated as follows: 
\begin{equation}
	\frac{\mathbf{W}_i^{m+1}-\mathbf{W}_i^m}{\Delta t_W}+\frac{S_{i j}}{\Omega_i} \sum_{j \in N(i)}\left[\mathbf{F}_{c, i j}^{m+1}+\mathbf{F}_{v, i j}^{m+1}\left(\boldsymbol{\sigma}^{\mathrm{M}}, \mathbf{q}^{\mathrm{M}}\right)\right]
 =\mathbf{Q}_i^{m+1}-\frac{S_{i j}}{\Omega_i} \sum_{j \in N(i)}\mathbf{F}_{v, i j}^{\mathrm{HoT}}(\Delta \boldsymbol{\sigma}, \Delta \mathbf{q}),
	\label{eq2.2.5}
\end{equation}

\noindent Here, the superscript $m$ corresponds to the iteration step of the macroscopic equations, distinguishing it from the iteration step of the mesoscopic kinetic equation; $N(i)$ denotes the set of neighbouring cells. The macroscopic time step $\Delta t_W$ for the inner iteration can be determined based on the newly macroscopic variables $\mathbf{W}_i^{m}$ in conjunction with Eq.~\eqref{eq2.2.3} by replacing the mesoscopic CFL number $\vartheta_{f}$ with macroscopic CFL number $\vartheta_{W}$. Vectors related to the macroscopic quantities $\mathbf{W}$, the convective numerical flux $\mathbf{F}_c$, and the source term $\mathbf{Q}$ are defined as 
\begin{equation}
	\mathbf{W}=\left[\begin{array}{c}
		\rho \\
		\rho u_x \\
		\rho u_y \\
		\rho u_z \\
		\rho E \\
		\rho E_r
	\end{array}\right], 
 \quad 
 \mathbf{F}_c=\left[\begin{array}{c}
		\rho u_n \\
		\rho u_x u_n+n_x p_t \\
		\rho u_y u_n+n_y p_t \\
		\rho u_z u_n+n_z p_t \\
		u_n\left(\rho E+p_t\right) \\
		u_n \rho E_r
	\end{array}\right], 
 \quad
 \mathbf{Q}=\left[\begin{array}{c}
		0 \\
		0 \\
		0 \\
		0 \\
		0 \\
		\frac{d_r}{2} \rho \frac{T-T_r}{Z_r \tau}
	\end{array}\right], 
	\label{eq2.2.6}
\end{equation}
where $u_n=\mathbf{u} \cdot \mathbf{n}_{i j}$ is the velocity along the normal direction of the interface. The viscous numerical flux $\mathbf{F}_v(\boldsymbol{\sigma}, \mathbf{q})$ is given by
\begin{equation}
	\mathbf{F}_v(\boldsymbol{\sigma}, \mathbf{q})=\left[\begin{array}{c}
		0 \\
		n_x \sigma_{x x}+n_y \sigma_{x y}+n_z \sigma_{x z} \\
		n_x \sigma_{y x}+n_y \sigma_{y y}+n_z \sigma_{y z} \\
		n_x \sigma_{z x}+n_y \sigma_{z y}+n_z \sigma_{z z} \\
		n_x \Theta_x+n_y \Theta_y+n_z \Theta_z \\
		n_x q_{x, r}+n_y q_{y, r}+n_z q_{z, r}
	\end{array}\right], 
	\label{eq2.2.7}
\end{equation}

\noindent where $\Theta_x(\boldsymbol{\sigma}, \mathbf{q})=u_x \sigma_{x x}+u_y \sigma_{x y}+u_z \sigma_{x z}+q_{x, t}+q_{r, t}$, $\Theta_y(\boldsymbol{\sigma}, \mathbf{q})=u_x \sigma_{y x}+u_y \sigma_{y y}+u_z \sigma_{y z}+q_{y, t}+q_{y, r}$ and $\Theta_z(\boldsymbol{\sigma}, \mathbf{q})=u_x \sigma_{z x}+u_y \sigma_{z y}+u_z \sigma_{z z}+q_{z, t}+q_{z, r}$. 

Similarly, an incremental form of macroscopic quantities is introduced as $\Delta \boldsymbol{W}_i^m=\boldsymbol{W}_i^{m+1}-\boldsymbol{W}_i^m$, and Eq. (\ref{eq2.2.5}) can be reformulated as follows:
\begin{equation}
	\begin{gathered}
		{\left[\frac{1}{\Delta t_W}-\left(\frac{\partial \mathbf{Q}}{\partial \mathbf{W}}\right)^m\right] \Delta \mathbf{W}_i^m+\frac{S_{i j}}{\Omega_i} \sum_{j \in N(i)} \Delta \mathbf{F}_{i j}^{m+1}=\mathbf{R}_i^m+\mathbf{R}_i^{\mathrm{HoT}},} \\
		\mathbf{R}_i^m=-\frac{S_{i j}}{\Omega_i} \sum_{j \in N(i)}\left[\mathbf{F}_{c, i j}^{m+1}+\mathbf{F}_{v, i j}^{m+1}(\boldsymbol{\sigma}^m, \mathbf{q}_t^m, \mathbf{q}_r^m)\right]+\mathbf{Q}_i^m, \\ \mathbf{R}_i^{\mathrm{HoT}}=-\frac{S_{i j}}{\Omega_i} \sum_{j \in N(i)}\mathbf{F}_{v, i j}^{\mathrm{HoT}}(\Delta \boldsymbol{\sigma}^{{n+1/2}}, \Delta \mathbf{q}_t^{{n+1/2}}, \Delta \mathbf{q}_r^{{n+1/2}}).
	\end{gathered}
	\label{eq2.2.8}
\end{equation}

\noindent Here, $\mathbf{R}_i^m$ and $\mathbf{R}_i^{\mathrm{HoT}}$ represent the macroscopic residual and high-order residual, respectively. In order to incorporate the high-order corrections including $\Delta \boldsymbol{\sigma}$, $\Delta \mathbf{q}_t$, $\Delta \mathbf{q}_r$ into the macroscopic iterations, the corresponding viscous flux $\mathbf{F}_{v, i j}^{\mathrm{HoT}}$ is computed following the same form as $\mathbf{F}_{v}$ in Eq.~\eqref{eq2.2.7}, but the stress and heat flux therein need to be replaced by higher-order corrections $\Delta \boldsymbol{\sigma}, \Delta \mathbf{q}_t, \Delta \mathbf{q}_r$.

It is worth noting that the high-order corrections in Eq.~\eqref{eq2.2.9} are computed based on the VDF and macroscopic quantities at the time step $n+1/2$, and they remain constant during inner iterations. Therefore, the high-order residual terms $\mathbf{R}_i^{\mathrm{HoT}}$ in Eq.~\eqref{eq2.2.8} should be treated as source terms in the computation.

The Rusanov scheme and the central-difference scheme are employed to compute the convective flux $\mathbf{F}_{c, i j}$ and viscous flux $\mathbf{F}_{v, i j}$ in the residual $\mathbf{R}_i^m$ of Eq. (\ref{eq2.2.8}), respectively. Meanwhile, the incremental form of the macroscopic flux $\Delta \mathbf{F}_{i j}$ can be approximated by the Eulerian form of the convective flux. Equations (\ref{eq2.2.2}) and (\ref{eq2.2.8}) can be iteratively solved using the classical Lower Upper Symmetric Gauss–Seidel (LU-SGS) technology. Given the widespread application of LU-SGS, specific details are omitted here for the sake of brevity. Relevant literature specifying the settings for implicitly solving these equations using LU-SGS are provided in the reference~\cite{zhu_general_2021, zeng_general_2023}.

\subsection{The flowchart of GSIS}
\label{SS:2-5}

The framework of the GSIS algorithm, aimed at accelerating convergence by alternately solving mesoscopic kinetic equation and macroscopic synthetic equation within one iteration step, is summarized below:
\begin{itemize}
	\setlength{\itemsep}{0pt}
	\setlength{\parsep}{0pt}
	\setlength{\parskip}{0pt}
	\item[1)]
	Based on the macroscopic quantities $\mathbf{W}_i^{n}$ and mesoscopic CFL number $\vartheta_{f}$, the time step $\Delta t_f$ for the mesoscopic kinetic equations can be set according to Eq. (\ref{eq2.2.3}).
	
	\item[2)]
	The implicit solution of the mesoscopic kinetic equations evolves the VDF to $f_{i}^{n+1/2}$ according to Eq.~\eqref{eq2.2.2}. The kinetic gas-surface boundary condition is used. 
	
	\item[3)]
	The corresponding macroscopic quantities $\mathbf{W}_i^{n+1/2}$, as well as the stress $\boldsymbol{\sigma}^{n+1/2}$ and the heat flux $\mathbf{q}_t^{n+1/2},\mathbf{q}_r^{n+1/2}$, are obtained through moment integration of the distribution function as presented in Eq. (\ref{eq2.1.6}).
	
	\item[4)]
	Calculate the derivatives of the macroscopic quantities by the least square method. Then, the high-order correction including $\Delta \boldsymbol{\sigma}^{n+1 /2}, \Delta \mathbf{q}^{n+1 /2}, \Delta \mathbf{q}^{n+1 /2}$ can be obtained from Eq.~\eqref{eq2.2.9}.
	
	\item[5)] The LU-SGS technology is employed for the implicit iteration of the macroscopic synthetic equations~\eqref{eq2.2.8}. In contrast to the kinetic equations, which are computed only once in each step, the synthetic equations involve hundreds of inner iterations to expedite the flow information exchange. Note that the traditional no-velocity-slip and no-temperature boundary conditions do not work any more in rarefied gas flows~\cite{sharipov_data_2011}, therefore the boundary condition should be carefully designed; this will be elaborated in section~\ref{S:3}.
	
	\item[6)] Based on the macroscopic quantities $\mathbf{W}_i^{k}$ from the inner iterations ($k$ is the total steps of inner iterations), the next-step VDF $f_{i}^{n+1}$ are computed as
\begin{equation}
	\begin{aligned}
		& f_{0, i}^{n+1}=f_{0, i}^{n+1 / 2}+\left[f_0^{\mathrm{eq}}\left(\mathbf{W}_i^{k}\right)-f_0^{\mathrm{eq}}\left(\mathbf{W}_i^{n+1 / 2}\right)\right], \\
		& f_{1, i}^{n+1}=f_{1, i}^{n+1 / 2}+\left[f_1^{\mathrm{eq}}\left(\mathbf{W}_i^{k}\right)-f_1^{\mathrm{eq}}\left(\mathbf{W}_i^{n+1 / 2}\right)\right].
	\end{aligned}
	\label{eq2.2.10}
\end{equation}
Then the macroscopic quantities $\mathbf{W}_i^{n+1}$ are consistent with those obtained through moment integration of $f_{i}^{n+1}$ in Eq.~(\ref{eq2.1.6}).
	
	\item[7)] 
	Repeat the steps (1-7) until the following convergence criterion is satisfied:
	
	\begin{equation}
		\left.\frac{\sqrt{\sum_i\left(\psi_i^{n+1}-\psi_i^{n}\right)^2 \Omega_i}}{\sqrt{\sum_i\left(\psi_i^{n}\right)^2 \Omega_i}}\right|_{\max }<10^{-6}, \quad \text { for } \quad \psi \in\left\{\rho, \mathbf{u}, T, T_r\right\}.
		\label{eq2.2.11}
	\end{equation}
\end{itemize}

It is noteworthy that, in the simulation of hypersonic and complex three-dimensional flows, it is a common practice to perform thousands steps of macroscopic iterations by first-order Euler equations to initialize the flow field~\cite{chen_three-dimensional_2020, fan_implementation_2023} and 10 steps of CIS iterations to initialize the VDF~\cite{zhang_efficient_2023} prior to GSIS iterations.

\section{Generalized boundary treatment}
\label{S:3}

\subsection{General considerations}

The fast convergence of GSIS lies in the use of macroscopic synthetic equation~\eqref{eq2.2.4} to guide the evolution of gas systems towards the final steady state~\cite{su_can_2020,su_fast_2020,su_multiscale_2021}. Since the deterministic mesoscopic solver is much more time-consuming to solve than the macroscopic synthetic equation, in the outer iteration, Eq.~\eqref{eq2.2.2} is only solved once at each iteration, while in the inner iteration, Eq.~\eqref{eq2.2.4} is solved multiple times or till convergence. Since the macroscopic synthetic equation can be solved quickly and for many steps, the flow field has the time to exchange information globally, and hence faster convergence to the steady state is achieved in GSIS than solving the kinetic equation solely. 

However, the macroscopic synthetic equation needs boundary conditions, but in general rarefied gas flows the no-velocity-slip and no-temperature-jump boundary conditions are not valid anymore. But, since the kinetic solver is already slow to converge when the Knudsen number is small, directly feeding its boundary information into macroscopic synthetic equation is not very efficient~\cite{zhu_general_2021}. In order to further accelerate the GSIS, we need to figure out a way to provide the boundary condition to the macroscopic synthetic equation at each inner iteration~\cite{zeng_general_2023}, and this boundary condition should be compatible with that in the kinetic equation. 

In the moment methods~\cite{grad_kinetic_1949,gu_computational_2007, gu_high-order_2009}, the boundary conditions are introduced in three steps. First, the VDF is re-constructed from conserved macroscopic quantities and high-order moments in the moment equation. Second, the reflected VDF from the wall is calculated by applying the kinetic gas-surface boundary condition. Third, the macroscopic boundary conditions are explicitly derived for the wall VDF. The similar idea can be borrowed to GSIS. To be specific, the truncated distribution function applicable to polyatomic gases with rotational degrees of freedom is derived to approximate the unknown distribution function at the wall for an explicit integration. Then, the numerical flux can be directly reconstructed recursively based on the conserved macroscopic quantities, as well as the stress and heat flux at the interface. The stress and heat flux embedded in the truncated distribution function can be closed based on explicit constitutive relations of the Newton and Fourier. Additionally, the concept of high-order correction to the constitutive equations, which is employed in GSIS, is also incorporated into this macroscopic boundary condition, ensuring stability and accuracy in hypersonic flows and strong rarefied flows.

%



\subsection{Overview of boundary conditions in previous versions of GSIS}

Within the finite-volume framework, the boundary condition can be treated as an special cell interface. Therefore, the implementation of boundary conditions requires the VDF and macroscopic numerical flux at the interface based on the gas-surface kinetic boundary condition. Taking the classical Maxwell boundary condition as an example, assuming the left side of the interface $ij$ is the wall and the right side is the interior cell $j$, the distribution function at the interface $ij$ can be determined as follows 
\begin{equation}
	\begin{gathered}
		f_{k, i j}=\left[\mathrm{H}_n^{+}\left(\alpha_M f_{k, i j}^{W}+\left(1-\alpha_M\right) f_{k, i j}^R(-\bm{{\xi}})\right)+\left(1-\mathrm{H}_n^{+}\right) f_{k, i j}^R(\bm{{\xi}})\right], \quad 
  k=0,1,
	\end{gathered}
	\label{eq23}
\end{equation}

\noindent where $\alpha_M$ is the accommodation coefficient representing the proportion of gas molecule would follow the diffuse reflection, while the remaining proportion follows the specular reflection; $\mathrm{H}_n^{+}=\left[1+\operatorname{sign}\left(\bm{{\xi}} \cdot \mathbf{n}_{i j}\right)\right]$ is the Heaviside function and $f_{k, i j}^R(\bm{{\xi}})$ represents the incident distribution function; the latter can be obtained by interpolating the VDFs at the cell center to the position of cell interface as
\begin{equation}
	\begin{gathered}
		f_{k, i j}^R(\bm{{\xi}})=f_{k, j}+\nabla f_{k, j} \cdot\left(\mathbf{x}_{i j}-\mathbf{x}_i\right), \quad 
  k=0,1.
	\end{gathered}
	\label{eq24}
\end{equation}
The $\alpha_M$ proportion of distribution function at the left side obeys diffuse reflection and thus becomes an equilibrium distribution function as
\begin{equation}
	\begin{gathered}
		f_{0, i j}^{W}=\rho^W\left(\frac{1}{2 \pi T_t^W}\right)^{3 / 2} \exp \left(-\frac{\left|\boldsymbol{\xi}-\mathbf{u}^W\right|^2}{2 T_t^W}\right), \\
		f_{1, i j}^{W}=\frac{d_r}{2} T_r^W \rho^W\left(\frac{1}{2 \pi T_t^W}\right)^{3 / 2} \exp \left(-\frac{\left|\boldsymbol{\xi}-\mathbf{u}^W\right|^2}{2 T_t^W}\right),
	\end{gathered}
	\label{eq25}
\end{equation}
while the rest of gas molecules undergo specular reflection and the corresponding VDFs follow $f_{k, i j}^R(-\bm{{\xi}})$. The superscript $W$ in Eq.~\eqref{eq25} denotes that the corresponding macroscopic quantities need to be determined based on the wall boundary conditions.

In the first GSIS for nonlinear flows~\cite{zhu_general_2021}, the numerical fluxes of the macroscopic equations at the interface $\mathbf{F}_{i j}$ are calculated by the VDFs at the interface $ij$ as
\begin{equation}
	\mathbf{F}_{i j} = \mathbf{F}_{c, i j}+\mathbf{F}_{v, i j}=\int \xi_n
 \left[\begin{array}{c}
		f_{0, i j} \\
		\xi_x f_{0, i j} \\
		\xi_y f_{0, i j} \\
		\xi_z f_{0, i j} \\
		\frac{c^2}{2} f_{0, i j}+f_{1, i j} \\
		f_{1, i j}
	\end{array}\right] 
 d \boldsymbol{\xi}.
	\label{eq26}
\end{equation}
While this boundary treatment ensures that the macroscopic equations attain consistent boundary conditions with the mesoscopic kinetic equations, it fails to maintain updates during macroscopic iterations, which does not realize the best possible performance of GSIS. In the second GSIS for nonlinear flows~\cite{zeng_general_2023}, we introduce a correction for macroscopic boundaries based on the increment of macroscopic quantities at the cell center. That is, 
in the inner iteration, after a single LU-SGS iteration, we obtain the global macro quantity increment and use a method similar to the Roe scheme to modify the interface flux using the physical quantity increment of the boundary element:
\begin{equation}\label{eq:updateMacroFlux}
    \mathbf{F}_{i j}^\star = \mathbf{F}_{i j} + \frac{F_c(\Delta \mathbf{W}_i)}{2}.
\end{equation}
This approach involves updating the boundary flux based on the updates in the interior flow, ensuring that the macroscopic wall flux evolves in tandem with the interior field to boost convergence. Nevertheless, the macroscopic boundary, unable to evolve independently, would result in a delay in the evolution of the macroscopic equations at the boundary during the inner iterations. This delay hinders the acceleration of convergence for the kinetic equations at the boundary.

\subsection{New treatment} 

Hence, with the hinder-sight, we propose a GBT for the macroscopic synthetic equation. The fundamental concept lies in approximating the incident distribution function at the interface using the truncated distribution function, thereby achieving an explicit reconstruction of the macroscopic numerical flux at the interface. In order to approximate the distribution function and close the higher-order terms in the macroscopic equations, Grad proposed a series of distribution functions based on Hermite orthogonal polynomials. Following this expansion, we present the truncated distribution function $f_{ij}^{T}(t, \boldsymbol{x}_{ij}, \boldsymbol{\xi}, I_r) $ applicable to polyatomic gas with rotational energy $I_r$ as
\begin{equation}
	f_{ij}^T\left(t, \mathbf{x}_{ij}, \xi, I_r\right)=f_{0, ij}^{e q} f_{I_r, ij}^{e q}\left[1+\frac{\boldsymbol{\sigma} \cdot \mathbf{c c}}{2 \rho\left(T_t\right)^2}+\frac{\mathbf{q}_t \cdot \mathbf{c}}{\rho\left(T_t\right)^2}\left(\frac{c^2}{5 T_t}-1\right)+\frac{\mathbf{q}_r \cdot \mathbf{c}}{\frac{d_r}{2} \rho T_t T_r}\left(\frac{I_r}{T_r}-\frac{d_r}{2}\right)\right],
	\label{eq27}
\end{equation}

\noindent where $f_{I_r, ij}^{e q}$ is defined as
\begin{equation}
	f_{I_r, ij}^{e q}=\frac{I_r^{d_r / 2-1}}{\Gamma\left(d_r / 2\right)\left(T_r\right)^{d_r / 2}} \exp \left(-\frac{I_r}{T_r}\right).
	\label{eq28}
\end{equation}

\noindent To reduce the rotational energy space for the distribution function, we have 
\begin{equation}
	\begin{aligned}
		 f_{0, ij}^T(t, \mathbf{x}_{ij}, \boldsymbol{\xi})&=\left\langle f_{ij}^T\left(t, \mathbf{x}_{ij}, \boldsymbol{\xi}, I_r\right)\right\rangle_{I_r} \\
		& =\left\langle f_{0, ij}^{e q} f_{I_r, ij}^{e q}\left[
			1+\frac{\boldsymbol{\sigma} \cdot \mathbf{c c}}{2 \rho\left(T_t\right)^2}+\frac{\mathbf{q}_t \cdot \mathbf{c}}{\rho\left(T_t\right)^2}\left(\frac{c^2}{5 T_t}-1\right) 
			+\frac{\mathbf{q}_r \cdot \mathbf{c}}{\frac{d_r}{2} \rho T_t T_r}\left(\frac{I_r}{T_r}-\frac{d_r}{2}\right)
  \right]\right\rangle_{I_r} \\
		& =f_{0, ij}^{e q}\left[1+\frac{\boldsymbol{\sigma} \cdot \mathbf{c c}}{2 \rho T_t^2}+\frac{\mathbf{q}_t \cdot \mathbf{c}}{\rho T_t^2}\left(\frac{c^2}{5 T_t}-1\right)\right], 
	\end{aligned}
	\label{eq29}
\end{equation} 
and
\begin{equation}
	\begin{aligned}
		 f_{1, ij}^T(t, \mathbf{x}_{ij}, \boldsymbol{\xi})&=\left\langle I_r f_{ij}^T\left(t, \mathbf{x}_{ij}, \boldsymbol{\xi}, I_r\right)\right\rangle_{I_r} \\
		& =\left\langle I_r f_{0,ij}^{e q} f_{I_r, ij}^{e q}\left[
			1+\frac{\boldsymbol{\sigma} \cdot \mathbf{c c}}{2 \rho\left(T_t\right)^2}+\frac{\mathbf{q}_t \cdot \mathbf{c}}{\rho\left(T_t\right)^2}\left(\frac{c^2}{5 T_t}-1\right) 
			+\frac{\mathbf{q}_r \cdot \mathbf{c}}{\frac{d_r}{2} \rho T_t T_r}\left(\frac{I_r}{T_r}-\frac{d_r}{2}\right)
		\right]\right\rangle_{I_r} \\
		& =\left(\frac{d_r}{2} T_r\right) f_{0,ij}^{e q}\left[1+\frac{\boldsymbol{\sigma} \cdot \mathbf{c c}}{2 \rho T_t^2}+\frac{\mathbf{q}_t \cdot \mathbf{c}}{\rho T_t^2}\left(\frac{c^2}{5 T_t}-1\right)\right]+f_{0,ij}^{e q} \frac{\mathbf{q}_r  \cdot \mathbf{c}}{\rho T_t},
	\end{aligned}
	\label{eq30}
\end{equation}

\noindent where $\langle\cdots\rangle_{I_r}=\int_0^{+\infty}(\cdots) d I_r$ is the integration over rotational energy space. A detailed derivation of the truncated distribution function is provided in \ref{B:1}. It is observed that the truncated distribution functions $f_{i j}^T$ in Eqs.~\eqref{eq29} and (\ref{eq30}) are expressed as polynomials in terms of macroscopic conserved variables, stress and heat flux and $f_{i j}^T$ would degenerate to $f_{i j}^{e q}$ when the stress and heat flux are zero.

With the approximation of the incident distribution function $f_{k, i j}^R$ by the truncated distribution function $f_{i j}^T$, the macroscopic numerical flux at the interface can be explicitly reconstructed. Taking the case of complete diffuse reflection ($\alpha_M$ = 1) in Eq.~\eqref{eq23} as an example, the macroscopic numerical flux at the interface can be expressed as follows
\begin{equation}
	\mathbf{F}_{i j}=\mathbf{F}_{c, i j}+\mathbf{F}_{v, i j}=\rho^L\left[\begin{array}{c}
		F_1^L \\
		n_x F_2^L+m_x F_3^L+o_x F_3^L \\
		n_y F_2^L+m_y F_3^L+o_y F_3^L \\
		n_z F_2^L+m_z F_3^L+o_z F_3^L \\
		F_5^L \\
		F_6^L
	\end{array}\right]+\rho^R\left[\begin{array}{c}
		F_1^R \\
		n_x F_2^R+m_x F_3^R+o_x F_3^R \\
		n_y F_2^R+m_y F_3^R+o_y F_3^R \\
		n_z F_2^R+m_z F_3^R+o_z F_3^R \\
		F_5^R \\
		F_6^R
	\end{array}\right].
	\label{eq31}
\end{equation}

\noindent Here, $\mathbf{n}_{i j}=\left[n_x, n_y, n_z\right]$, $\mathbf{m}_{i j}=\left[m_x, m_y, m_z\right]$, and $\mathbf{o}_{i j}=\left[o_x, o_y, o_z\right]$ constitute a set of orthogonal vectors determining the normal and two tangential directions of the interface. The definitions of the integral coefficients $F_1$ to $F_6$ at the interface are defined as follows
\begin{equation}
	\begin{aligned}
		& F_1^L=\left\langle\xi_n^1 \xi_\tau^0 \xi_s^0 f_{0, i j}^T\right\rangle_{>0}, 
  \quad
  F_1^R=\left\langle\xi_n^1 \xi_\tau^0 \xi_s^0 f_{0, i j}^T\right\rangle_{<0}, \\
		& F_2^L=\left\langle\xi_n^2 \xi_\tau^0 \xi_s^0 f_{0, i j}^T\right\rangle_{>0}, 
  \quad
  F_2^R=\left\langle\xi_n^2 \xi_\tau^0 \xi_s^0 f_{0, i j}^T\right\rangle_{<0}, \\
		& F_3^L=\left\langle\xi_n^1 \xi_\tau^1 \xi_s^0 f_{0, i j}^T\right\rangle_{>0}, 
  \quad
  F_3^R=\left\langle\xi_n^1 \xi_\tau^1 \xi_s^0 f_{0, i j}^T\right\rangle_{<0}, \\
		& F_4^L=\left\langle\xi_n^1 \xi_\tau^0 \xi_s^1 f_{0, i j}^T\right\rangle_{>0},
  \quad 
  F_4^R=\left\langle\xi_n^1 \xi_\tau^0 \xi_s^1 f_{0, i j}^T\right\rangle_{<0},
	\end{aligned}
	\label{eq32}
\end{equation}

\noindent and 
\begin{equation}
	\begin{gathered}
		F_5^L=\frac{1}{2}\left(\left\langle\xi_n^3 \xi_\tau^0 \xi_s^0 f_{0, i j}^T\right\rangle_{>0}+\left\langle\xi_n^1 \xi_\tau^2 \xi_s^0 f_{0, i j}^T\right\rangle_{>0}+\left\langle\xi_n^1 \xi_\tau^0 \xi_s^2 f_{0, i j}^T\right\rangle_{>0}\right)+F_6^L, \\
		F_5^R=\frac{1}{2}\left(\left\langle\xi_n^3 \xi_\tau^0 \xi_s^0 f_{0, i j}^T\right\rangle_{<0}+\left\langle\xi_n^1 \xi_\tau^2 \xi_s^0 f_{0, i j}^T\right\rangle_{<0}+\left\langle\xi_n^1 \xi_\tau^0 \xi_s^2 f_{0, i j}^T\right\rangle_{<0}\right)+F_6^R, \\
		F_6^L=\frac{d_r}{2} T_r\left\langle\xi_n^1 \xi_\tau^0 \xi_s^0 f_{0, i j}^T\right\rangle_{>0}+\left\langle\xi_n^1 \xi_\tau^0 \xi_s^0 f_{1, i j}^T\right\rangle_{>0}, \\
		F_6^R=\frac{d_r}{2} T_r\left\langle\xi_n^1 \xi_\tau^0 \xi_s^0 f_{0, i j}^T\right\rangle_{<0}+\left\langle\xi_n^1 \xi_\tau^0 \xi_s^0 f_{1, i j}^T\right\rangle_{<0},
	\end{gathered}
	\label{eq33}
\end{equation}

\noindent where the subscripts $n$, $\tau$, and $s$ correspond to the directions of molecular velocity after the transformation into the local coordinate system. The symbol $\langle\cdot\rangle_{<0}$ represents the integral with the interval from negative infinity to zero and the $\langle\cdot\rangle_{>0}$ represents the integral with the interval from zero to infinity:
\begin{equation}
	\begin{aligned}
		& \langle\cdots f^T\rangle_{<0}=\frac{1}{\rho}\int_{-\infty}^0 \int_{-\infty}^{+\infty} \int_{-\infty}^{+\infty}(\cdots) f^T d \xi_n d \xi_\tau d \xi_s \\
		& \langle\cdots f^T\rangle_{>0}=\frac{1}{\rho}\int_0^{+\infty} \int_{-\infty}^{+\infty} \int_{-\infty}^{+\infty}(\cdots) f^T d \xi_n d \xi_\tau d \xi_s.
	\end{aligned}
	\label{eq34}
\end{equation}

The next task is computing the moment coefficients including $\left\langle\xi_{n}^{o} \xi_{\tau}^{p} \xi_{s}^{q} f_{i j}^T\right\rangle_{>0}$ and $\left\langle\xi_{n}^{o} \xi_{\tau}^{p} \xi_{s}^{q} f_{i j}^T\right\rangle_{<0}$. Based on the binomial theorem, the coefficients associated with $f_{i j}^T$ on the right side of interface can be expressed through a two-step recursion as
\begin{equation}
	\begin{aligned}
		& \left\langle\xi_n^i \xi_\tau^j \xi_s^k f^T\right\rangle_{<0}=\sum_{m=0}^k \frac{k !}{m !(k-m) !}\left\langle\xi_n^i \xi_\tau^j c_s^k f^T\right\rangle_{<0} u_s^{k-m}, \\
		& \left\langle\xi_n^i \xi_\tau^j c_s^k f^T\right\rangle_{<0}=\sum_{m=0}^j \frac{j !}{m !(j-m) !}\left\langle\xi_n^i c_\tau^m c_s^k f^T\right\rangle_{<0} u_\tau^{j-m}.
	\end{aligned}
	\label{eq35}
\end{equation}

\noindent Here, $i$, $j$, and $k$ represent non-negative integers and $f^T$ can represent both $f_0^T$ and $f_1^T$. In accordance with the definition of the truncated distribution function, we can express $\left\langle\xi_n^i c_\tau^j c_s^k f_0^T\right\rangle_{<0}$ and $\left\langle\xi_n^i c_\tau^j c_s^k f_1^T\right\rangle_{<0}$ as the linear combinations of integral coefficients associated with the equilibrium distribution function as follows
\begin{equation}
	\begin{aligned}
		\left\langle\xi_n^i c_\tau^j c_s^k f_0^T\right\rangle_{<0}= & \left\langle\xi_n^i c_n^0\right\rangle_{<0}^{eq}\left\langle c_\tau^j\right\rangle^{eq}\left\langle c_s^k\right\rangle^{eq}+\sigma_{n n}^*\left\langle\xi_n^i c_n^2\right\rangle_{<0}^{eq}\left\langle c_\tau^j\right\rangle^{eq}\left\langle c_s^k\right\rangle^{eq} \\
		& +\sigma_{\tau \tau}^*\left\langle\xi_n^i c_n^0\right\rangle_{<0}^{eq}\left\langle c_\tau^{j+2}\right\rangle^{eq}\left\langle c_s^k\right\rangle^{eq}+\sigma_{s s}^*\left\langle\xi_n^i c_n^0\right\rangle_{<0}^{eq}\left\langle c_\tau^j\right\rangle^{eq}\left\langle c_s^{k+2}\right\rangle^{eq} \\
		& +2 \sigma_{n \tau}^*\left\langle\xi_n^i c_n^1\right\rangle_{<0}^{eq}\left\langle c_\tau^{j+1}\right\rangle^{eq}\left\langle c_s^k\right\rangle^{eq}+2 \sigma_{n s}^*\left\langle\xi_n^i c_n^1\right\rangle_{<0}^{eq}\left\langle c_\tau^j\right\rangle^{eq}\left\langle c_s^{k+1}\right\rangle^{eq} \\
		& +2 \sigma_{\tau s}^*\left\langle\xi_n^i c_n^0\right\rangle_{<0}^{eq}\left\langle c_\tau^{j+1}\right\rangle^{eq}\left\langle c_s^{k+1}\right\rangle^{eq}-q_{t, n}^*\left\langle\xi_n^i c_n^1\right\rangle_{<0}^{eq}\left\langle c_\tau^j\right\rangle^{eq}\left\langle c_s^k\right\rangle^{eq} \\
		& -q_{t, \tau}^*\left\langle\xi_n^i c_n^0\right\rangle_{<0}^{eq}\left\langle c_\tau^{j+1}\right\rangle^{eq}\left\langle c_s^k\right\rangle^{eq}-q_{t, s}^*\left\langle\xi_n^i c_n^0\right\rangle_{<0}^{eq}\left\langle c_\tau^j\right\rangle^{eq}\left\langle c_s^{k+1}\right\rangle^{eq} \\
		& +0.4 \lambda_t q_{t, n}^*\left\langle\xi_n^i c_n^3\right\rangle_{<0}^{eq}\left\langle c_\tau^j\right\rangle^{eq}\left\langle c_s^k\right\rangle^{eq}+0.4 \lambda_t q_{t, n}^*\left\langle\xi_n^i c_n^1\right\rangle_{<0}^{eq}\left\langle c_\tau^j\right\rangle^{eq}\left\langle c_s^{k+2}\right\rangle^{eq} \\
		& +0.4 \lambda_t q_{t, n}^*\left\langle\xi_n^i c_n^1\right\rangle_{<0}^{eq}\left\langle c_\tau^{j+2}\right\rangle^{eq}\left\langle c_s^k\right\rangle^{eq}+0.4 \lambda_t q_{t, \tau}^*\left\langle\xi_n^i c_n^2\right\rangle_{<0}^{eq}\left\langle c_\tau^{j+1}\right\rangle^{eq}\left\langle c_s^k\right\rangle^{eq} \\
		& +0.4 \lambda_t q_{t, \tau}^*\left\langle\xi_n^i c_n^0\right\rangle_{<0}^{eq}\left\langle c_\tau^{j+3}\right\rangle^{eq}\left\langle c_s^k\right\rangle^{eq}+0.4 \lambda_t q_{t, \tau}^*\left\langle\xi_n^i c_n^0\right\rangle_{<0}^{eq}\left\langle c_\tau^{j+1}\right\rangle^{eq}\left\langle c_s^{k+2}\right\rangle^{eq} \\
		& +0.4 \lambda_t q_{t, s}^*\left\langle\xi_n^i c_n^2\right\rangle_{<0}^{eq}\left\langle c_\tau^j\right\rangle^{eq}\left\langle c_s^{k+1}\right\rangle^{eq}+0.4 \lambda_t q_{t, s}^*\left\langle\xi_n^i c_n^0\right\rangle_{<0}^{eq}\left\langle c_\tau^{j+2}\right\rangle^{eq}\left\langle c_s^{k+1}\right\rangle^{eq} \\
		& +0.4 \lambda_t q_{t, s}^*\left\langle\xi_n^i c_n^0\right\rangle_{<0}^{eq}\left\langle c_\tau^j\right\rangle^{eq}\left\langle c_s^{k+3}\right\rangle^{eq},
	\end{aligned}
	\label{eq36}
\end{equation}

\noindent and
\begin{equation}
	\begin{aligned}
		\left\langle\xi_n^i c_\tau^j c_s^k f_1^T\right\rangle_{<0}= & \frac{d_r}{2} T_r\left\langle\xi_n^i c_\tau^j c_s^k f_0^T\right\rangle_{<0}+q_{r, n}^*\left\langle\xi_n^i c_n^1\right\rangle_{<0}^{e q}\left\langle c_\tau^j\right\rangle^{e q}\left\langle c_s^k\right\rangle^{e q} \\
		& +q_{r, \tau}^*\left\langle\xi_n^i c_n^0\right\rangle_{<0}^{e q}\left\langle c_\tau^{j+1}\right\rangle^{e q}\left\langle c_s^k\right\rangle^{e q}+q_{r, s}^*\left\langle\xi_n^i c_n^0\right\rangle_{<0}^{e q}\left\langle c_\tau^j\right\rangle^{e q}\left\langle c_s^{k+1}\right\rangle^{e q},
	\end{aligned}
	\label{eq37}
\end{equation}

\noindent where $\lambda_t=1 / 2 T_t$, and the stress and heat flux marked with an asterisk superscript are given by
\begin{equation}
	\begin{gathered}
		\sigma_{n n}^*=\sigma_{n n} /\left(2 p_t T_t\right), \quad \sigma_{n \tau}^*=\sigma_{n \tau} /\left(2 p_t T_t\right), \quad \sigma_{n s}^*=\sigma_{n s} /\left(2 p_t T_t\right), \\
		\sigma_{\tau \tau}^*=\sigma_{\tau \tau} /\left(2 p_t T_t\right), 
  \quad 
  \sigma_{\tau s}^*=\sigma_{\tau s} /\left(2 p_t T_t\right), \\
		q_{t, n}^*=q_{t, n} /\left(p_t T_t\right), \quad q_{t, \tau}^*=q_{t, \tau} /\left(p_t T_t\right), \quad q_{t, s}^*=q_{t, s} /\left(p_t T_t\right), \\
		q_{r, n}^*=q_{r, n} / p_t, \quad q_{r, \tau}^*=q_{r, \tau} / p_t, \quad q_{r, s}^*=q_{r, s} / p_t.
	\end{gathered}
	\label{eq38}
\end{equation}

The integral coefficients associated with the equilibrium distribution function are defined as
\begin{equation}
	\begin{aligned}
		& \langle\cdots\rangle^{e q}=\frac{1}{\rho}\int_{-\infty}^{+\infty} \int_{-\infty}^{+\infty} \int_{-\infty}^{+\infty}(\cdots) f_0^{e q} d \xi_n d \xi_\tau d \xi_s \\
		& \langle\cdots\rangle_{<0}^{e q}=\frac{1}{\rho}\int_{-\infty}^0 \int_{-\infty}^{+\infty} \int_{-\infty}^{+\infty}(\cdots) f_0^{e q} d \xi_n d \xi_\tau d \xi_s \\
		& \langle\cdots\rangle_{>0}^{e q}=\frac{1}{\rho}\int_0^{+\infty} \int_{-\infty}^{+\infty} \int_{-\infty}^{+\infty}(\cdots) f_0^{e q} d \xi_n d \xi_\tau d \xi_s .
	\end{aligned}
	\label{eq39}
\end{equation}

To obtain the coefficient of $\left\langle\xi_n^i c_n^j\right\rangle_{<0}^{eq}$, we have the following equation
\begin{equation}
	\left\langle\xi_n^i c_n^j\right\rangle_{<0}^{eq}=(-1)^{j-m} \sum_{m=0}^j \frac{j !}{m !(j-m) !}\left\langle\xi_n^{m+i}\right\rangle_{<0}^{e q}\left(u_n\right)^{j-m}.
	\label{eq40}
\end{equation}

While our previous description only provides the calculation procedure for the integration coefficients on the right side of interface, i.e., $\left\langle\xi_n^i \xi_\tau^j \xi_s^k f^T\right\rangle_{<0}$, the coefficients on the left side follows a similar procedure. The only difference is that the expressions $\langle\cdot\cdot\cdot\rangle_{<0}$ in Eqs.~(\ref{eq35}), (\ref{eq36}) and (\ref{eq37}) should be replaced  by $\langle\cdot\cdot\cdot\rangle_{>0}$ and the macroscopic variables in the right side of the interface should be replaced by that in the left side. Taking the diffuse reflection boundary condition as an example, the integration coefficients on the left side of the interface corresponding to Eqs.~(\ref{eq36}) and (\ref{eq37}) can be simplified to
\begin{equation}
	\begin{aligned}
		\left\langle\xi_n^i c_\tau^j c_s^k f_0^T\right\rangle_{>0}= & \left\langle\xi_n^i c_n^0\right\rangle_{>0}^{eq}\left\langle c_\tau^j\right\rangle^{eq}\left\langle c_s^k\right\rangle^{eq},
	\end{aligned}
	\label{eq41}
\end{equation}
and
\begin{equation}
	\begin{aligned}
		\left\langle\xi_n^i c_\tau^j c_s^k f_1^T\right\rangle_{>0}= & \frac{d_r}{2} T_r\left\langle\xi_n^i c_\tau^j c_s^k f_0^T\right\rangle_{>0}.
	\end{aligned}
	\label{eq42}
\end{equation}

\noindent The remaining task is to calculate the terms $\left\langle\xi_{n}^{k}\right\rangle_{>0}^{eq}$, $\left\langle\xi_{n}^{k}\right\rangle_{<0}^{eq}$, $\left\langle c_{\tau}^{k}\right\rangle^{eq}$ and $\left\langle c_{s}^{k}\right\rangle^{eq}$, which are related to the moment integration of equilibrium state, see \ref{C:1} for details.

Prior to computing the macroscopic numerical flux, it is essential to obtain the macroscopic variables on both sides of the interface $\mathbf{W}$. It is worth noting that the expression of macroscopic numerical flux presented here is of a general form, without specifying whether the wall is located to the left or right of the interface. We continue to use the example where the wall is located to the left of the interface. The values $\mathbf{W}^R$ on the right side of the interface can be interpolated to the interface by the values at the centers of the nearest cells adjacent to the wall as
\begin{equation}
	\mathbf{W}^R\left(\mathbf{x}_{i j}\right)=\mathbf{W}\left(\mathbf{x}_j\right)+L\left(\mathbf{W}_j\right) \nabla \mathbf{W}\left(\mathbf{x}_j\right) \cdot\left(\mathbf{x}_{i j}-\mathbf{x}_j\right),
	\label{eq43}
\end{equation}
where $L\left(\mathbf{W}\right)$ is the Venkatakrishnan’s limiter~\cite{venkatakrishnan_convergence_1995} functions of the variable $\mathbf{W}$. The gradient $\nabla \mathbf{W}(\mathbf{x})^{n}$ can be computed by the the least squares method. 

In addition, the numerical integration based on the truncated distribution function requires the determination of stress and heat flux. Since the boundary conditions here are intended for the macroscopic synthetic equation, the stress and heat flux should be obtained explicitly based on the constitutive relations and taking into account the high-order corrections applied in GSIS as
\begin{equation}
	\begin{gathered}
		\boldsymbol{\sigma}^R=-\mu\left(\nabla \mathbf{u}+\nabla \mathbf{u}^{\mathrm{T}}-\frac{2}{3} \nabla \cdot u \mathrm{I}\right)+\Delta \boldsymbol{\sigma}_j, \\
		\mathbf{q}_t^R=-\kappa_t \nabla T_t+\Delta \mathbf{q}_{t, j}, \\
		\mathbf{q}_r^R=-\kappa_r \nabla T_r+\Delta \mathbf{q}_{r, j},
	\end{gathered}
	\label{eq44}
\end{equation}

\noindent where the explicit constitutive relations facilitate real-time updates of the macroscopic boundary conditions during inner iterations, boosting faster information exchange in the flow field, and consequently achieving quicker convergence. Furthermore, the introduction of high-order corrections ensures that the macroscopic numerical flux at the boundary remains accurate and stable even under strong rarefied flow conditions. 

The macroscopic variables on the left side of the interface, denoted as $\mathbf{W}^L$, are determined based on the boundary conditions, i.e., $\mathbf{W}^L = \mathbf{W}^W$, and the stress and heat flux on the left side of the wall interface need to be set to zero. Taking isothermal wall boundary conditions as an example, the velocity at the interface is set to $\mathbf{u}^L = \mathbf{u}^W = 0$, while the temperature is set to the wall temperature $T^L = T^W$. The density is determined according to the non-penetration boundary condition as follows:
\begin{equation}
	\rho_W=-\sqrt{\frac{2\pi}{T_t^W}}\left\langle\xi_n^1 \xi_\tau^0 \xi_s^0 f_{0, i j}^T\right\rangle_{<0}.
	\label{eq45}
\end{equation}


\section{Numerical experiments}
\label{S:4}

In this section, we assess the performance of the GSIS-GBT. Our tests focus on two key issues: (i) whether the GBT can yield stable and accurate results, especially in hypersonic and strong rarefied flow scenarios, and (ii) whether the GBT can assist the further acceleration of GSIS. The planar Fourier flow, hypersonic flow past a cylinder, pressure-driven flow in a variable-diameter circular pipe, and the multiscale flow around a hypersonic technology vehicle, are considered.

The viscosity index $\omega$ in Eq.~\eqref{eq2.1.3} are chosen as 0.81, 0.75 and 0.71 for the monatomic gas, air and nitrogen, respectively. Following the previous study~\cite{li_uncertainty_2021}, the thermal relaxation rates in Eq.~\eqref{eq2.1.8} are given as $A_{t r}=-0.059$, $A_{t t}=0.786$, $A_{r t}=-0.059$ and $A_{r r}=0.842$ for polyatomic gas. The convergence criterion is determined based on Eq.~\eqref{eq2.2.11} to ensure the convergence of all macroscopic quantities.

\subsection{Planar Fourier flow}
\label{sec4-1}

Consider two stationary parallel plates separated by a distance $H$, with temperatures $T_0+\Delta T$ and $T_0+\Delta T$ on the upper and lower plates, respectively. Due to the temperature difference, the gas temperature and density distribution exhibit gradients. Furthermore, with the enhancement of rarefaction effects, a notable temperature jump will be observed. 

The computational settings are as follows: the plate spacing $H$ is set to 1.0 as the reference length, and a mesh of 50 points is employed to discretize the computational domain of $y \in[0, 1]$ uniformly. The monatomic gas is initially set to a state with the reference density $\rho_0$ and temperature $T_0$ both equal to 1.0. The temperature difference at the wall is set to $\Delta T = 0.25$, resulting in temperatures of 1.25 and 0.75 for the upper and lower surfaces, respectively. The wall boundary condition is subjected to complete diffuse reflection. The Gauss–Hermite quadrature with $28 \times 28$ points is adopted in the discretization of velocity domain. The collision time $\tau$ and shear viscosity $\mu$ are determined according to 
Eqs.~\eqref{eq2.1.2} and \eqref{eq2.1.3}, with the rotational collision number set to 0.001 to recover the monatomic gas model. To expedite computations and facilitate rapid convergence to steady state, the number of inner iterations in Eq.~\eqref{eq2.2.5} is set to $m=100$, and the CFL numbers for the mesoscopic and macroscopic levels are designated as 1000 and 500, respectively.

\begin{figure}[ht]
	\centering
	\includegraphics[width=0.35\textwidth]{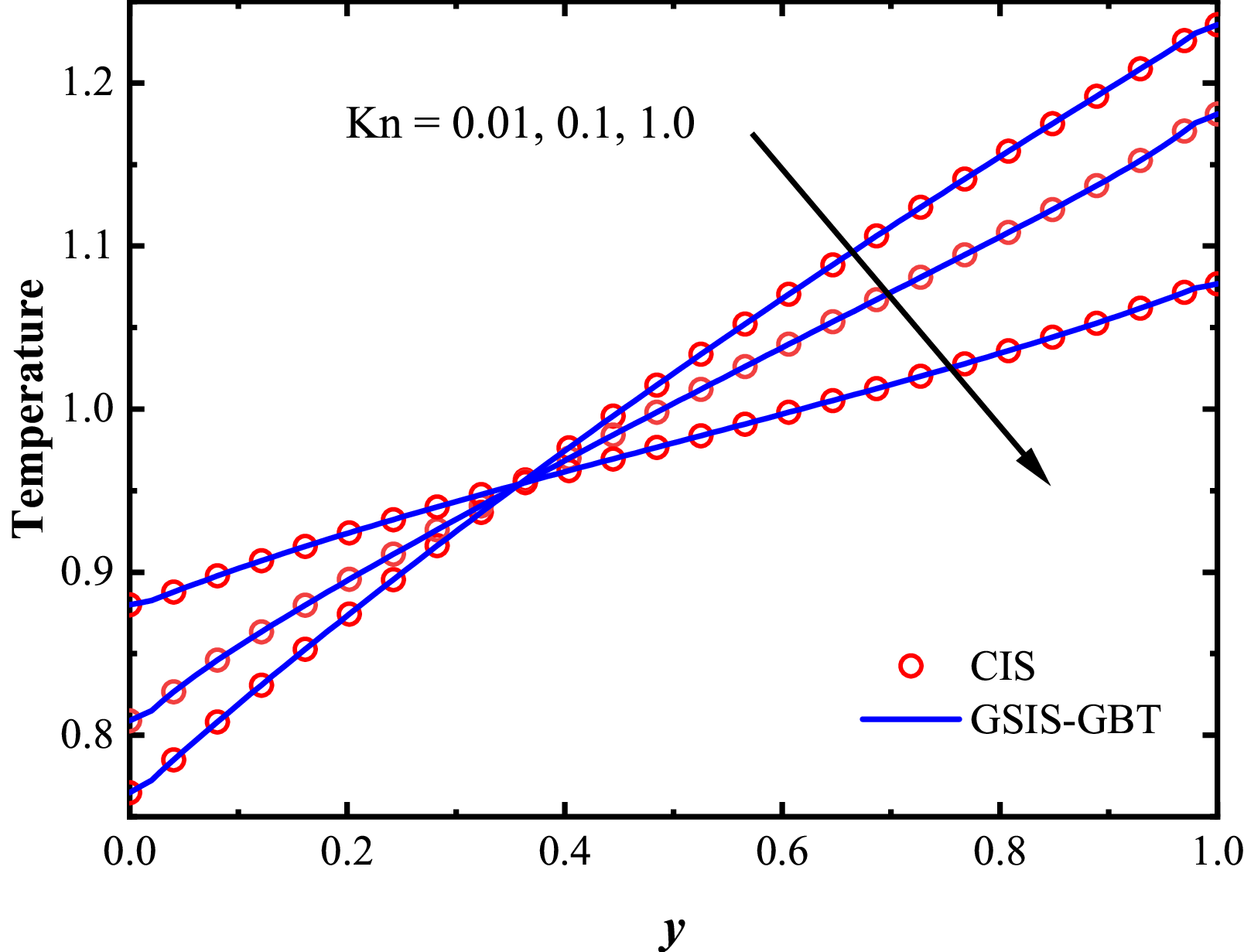}
	\quad
    \includegraphics[width=0.35\textwidth]{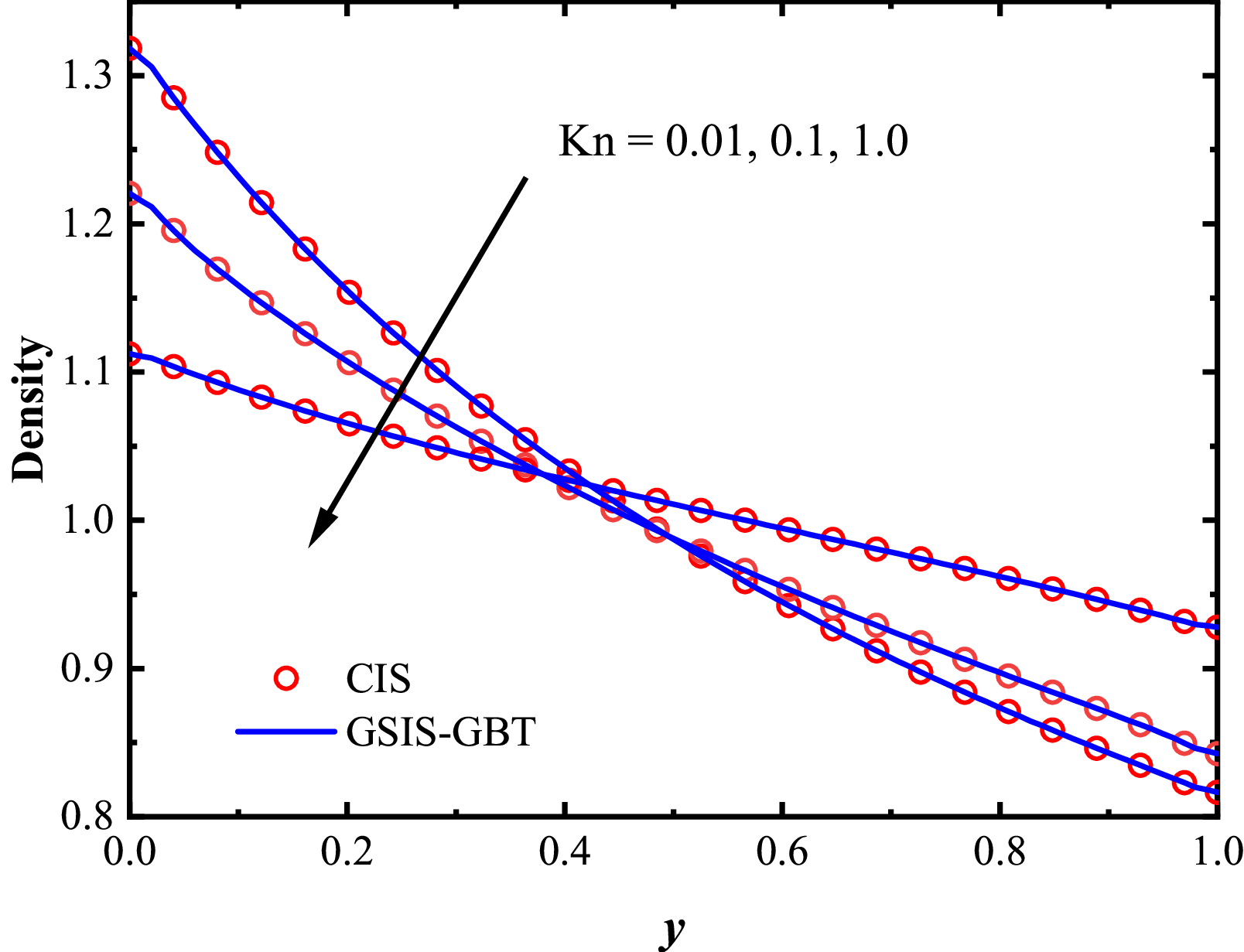}
	\caption{Temperature and density profiles of the heat transfer between two plates at different Kn.}
	\label{Fig1}
\end{figure}

\begin{figure}[th]
	\centering
	\includegraphics[width=0.35\textwidth]{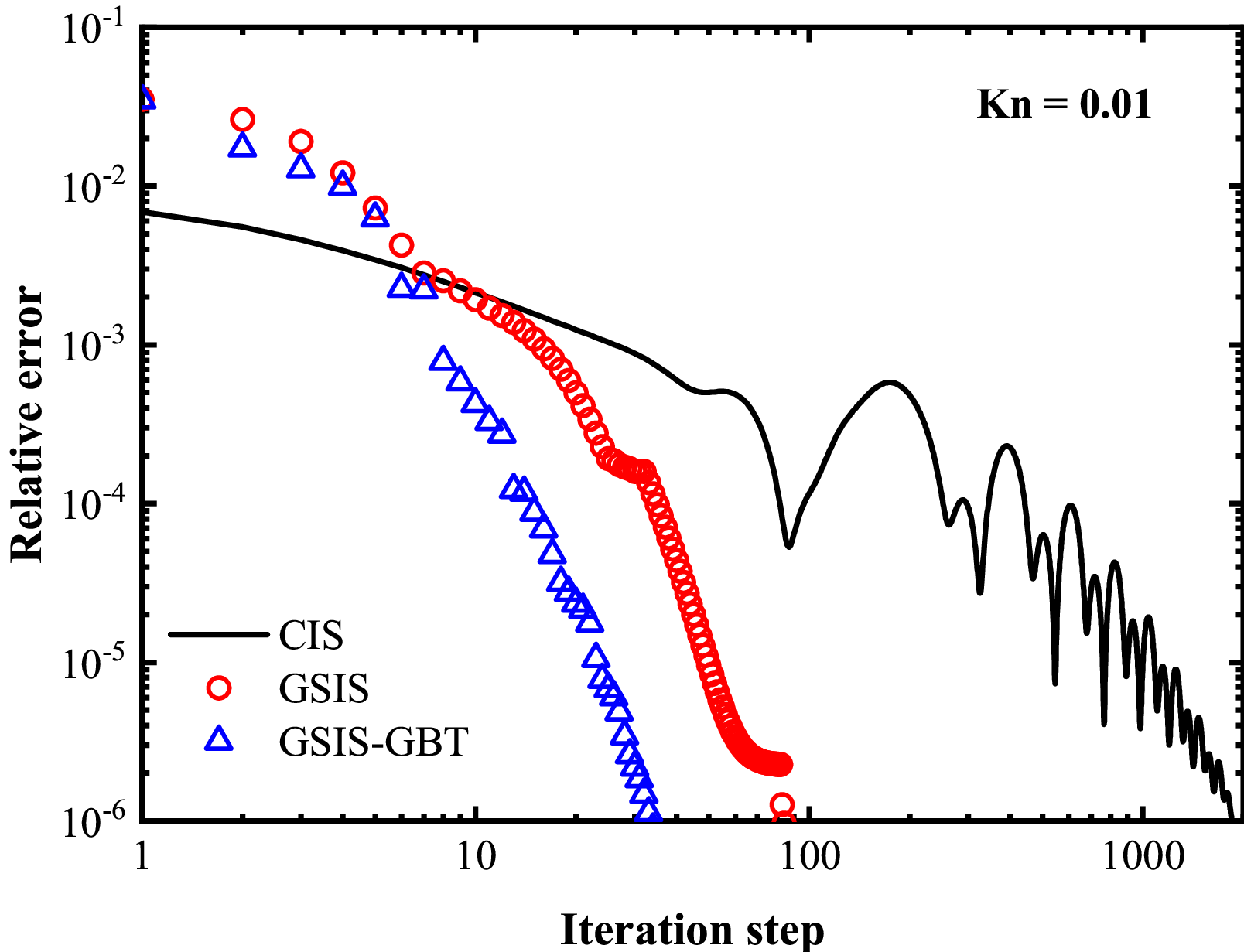}\quad
	\includegraphics[width=0.35\textwidth]{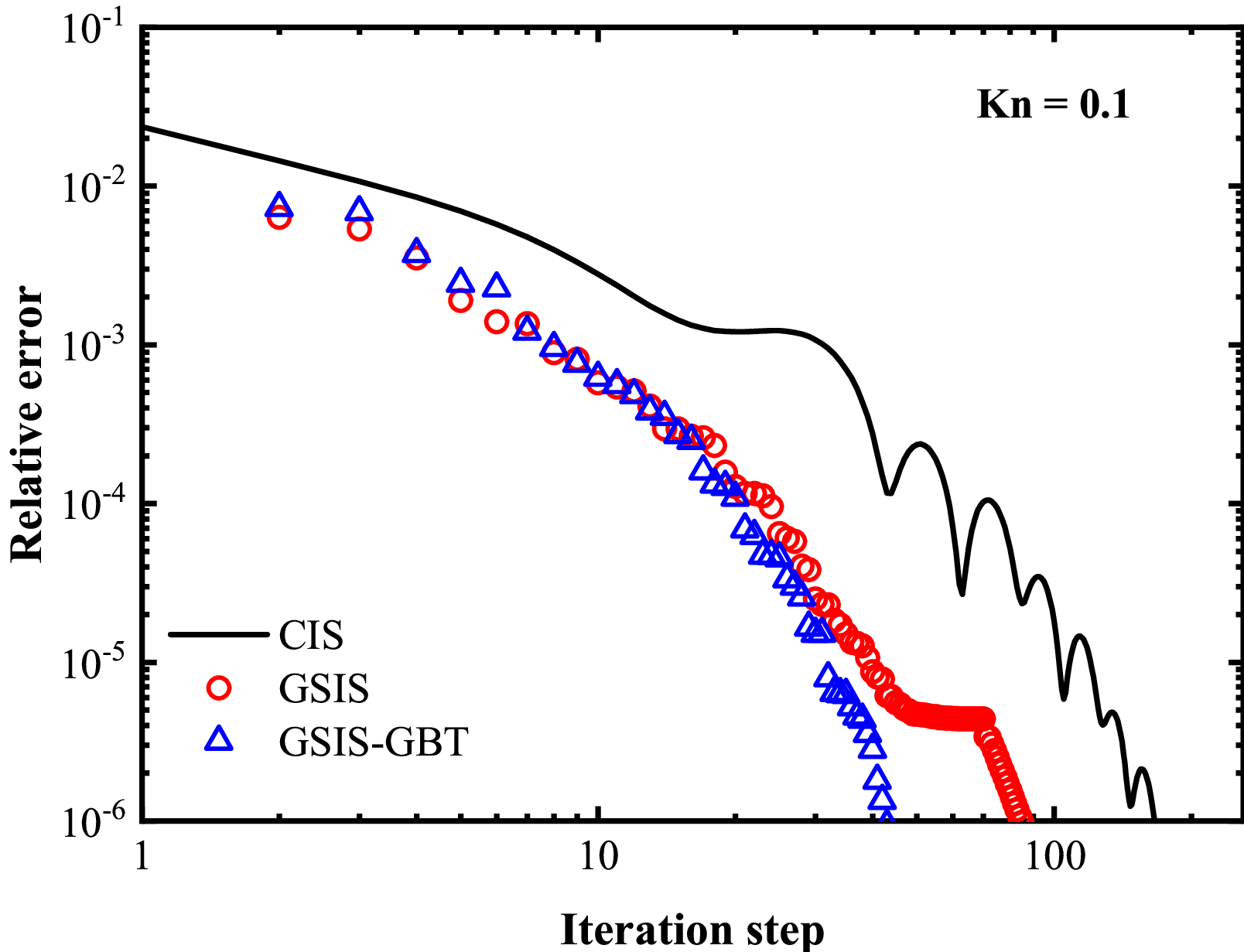}\\
 \vspace{0.5cm}
	\includegraphics[width=0.35\textwidth]{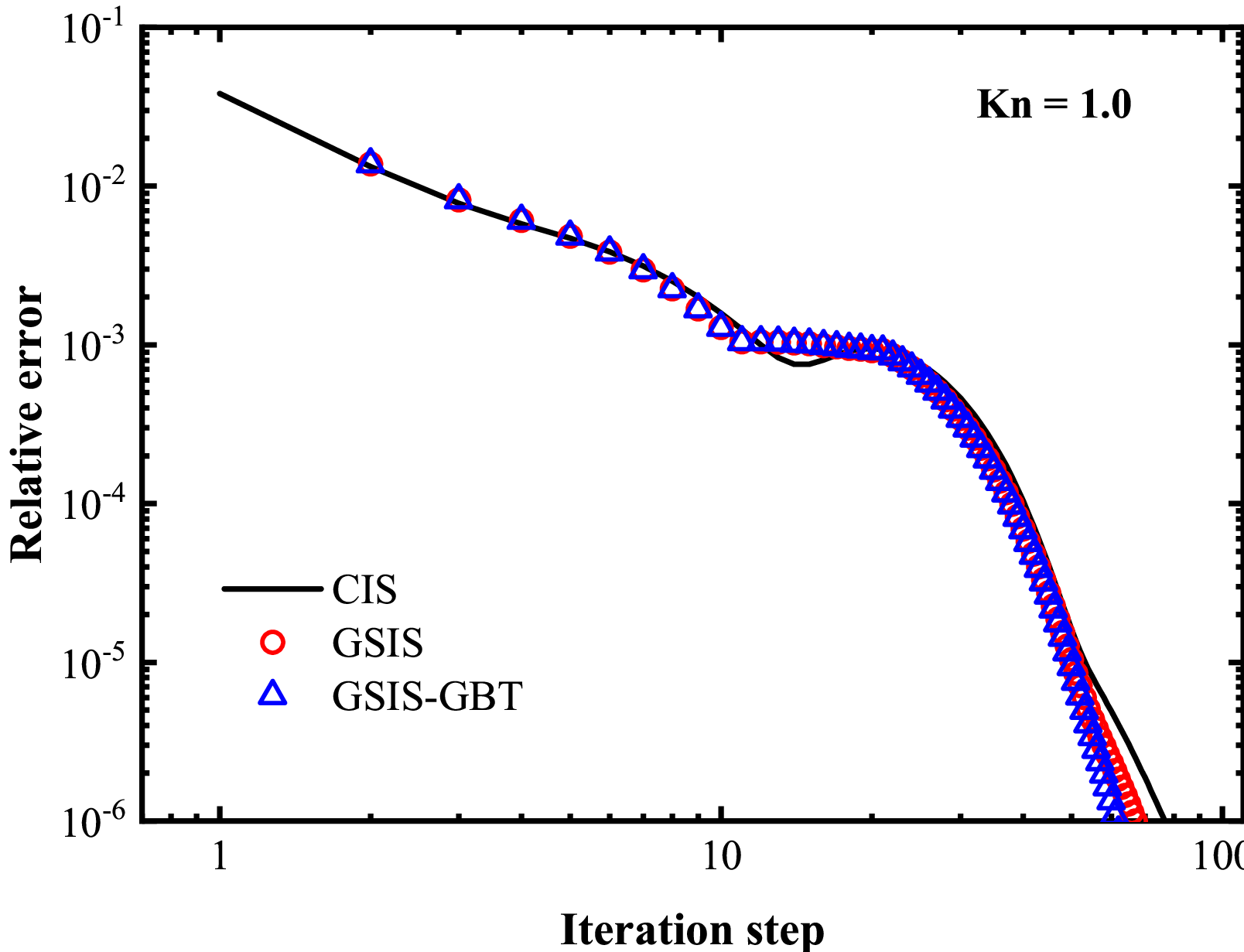}   
	\caption{Convergence history of the CIS, GSIS and GSIS-GBT in the simulation of heat transfer between two plates at different Kn. The dashed lines represent the density and temperature profiles at different iteration steps, while the circles indicate the accurate results serving as references.
 }
	\label{Fig2}
\end{figure}

To validate the accuracy of GSIS-GBT, the temperature and velocity profiles between two plates at Kn = 0.01, 0.1, and 1.0 are compared to those obtained from the CIS~\cite{zhu_general_2021}. Figure~\ref{Fig1} shows that, as the Knudsen number increases, the temperature profiles become fatter, and the temperature jump near the wall increases. The agreement between the results obtained from GSIS-GBT and CIS at different Knudsen number confirms the accuracy of GBT in different flow regimes.

Figure~\ref{Fig2} compares the iteration steps used in the CIS, GSIS~\cite{zeng_general_2023}, and GSIS-GBT. When Kn=1.0, the three methods complete the computation within 100 steps. However, as the Knudsen number decreases, the iteration steps in CIS significantly increase. At Kn = 0.01, CIS requires more than 1800 steps; the GSIS, due to the coupled macroscopic inner iterations, significantly expedites the information exchange within the flow field, achieving convergence in as few as 85 steps; the GSIS-GBT further reduces the steps to 34, where the additional acceleration is attributed to the GBT on the evolution of the flow field.

\begin{figure}[th]
	\centering
	\includegraphics[width=0.35\textwidth]{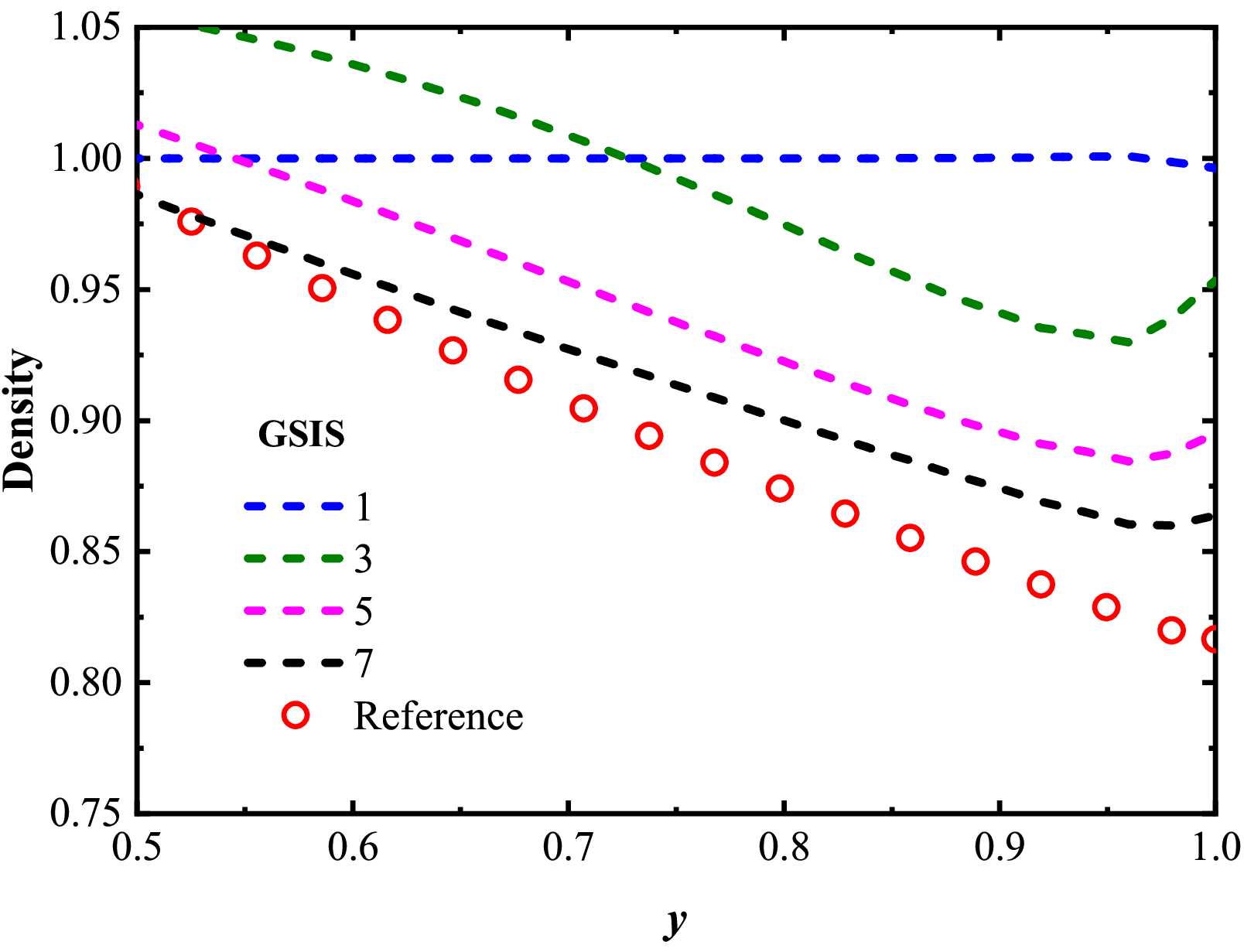}
 \quad
	\includegraphics[width=0.35\textwidth]{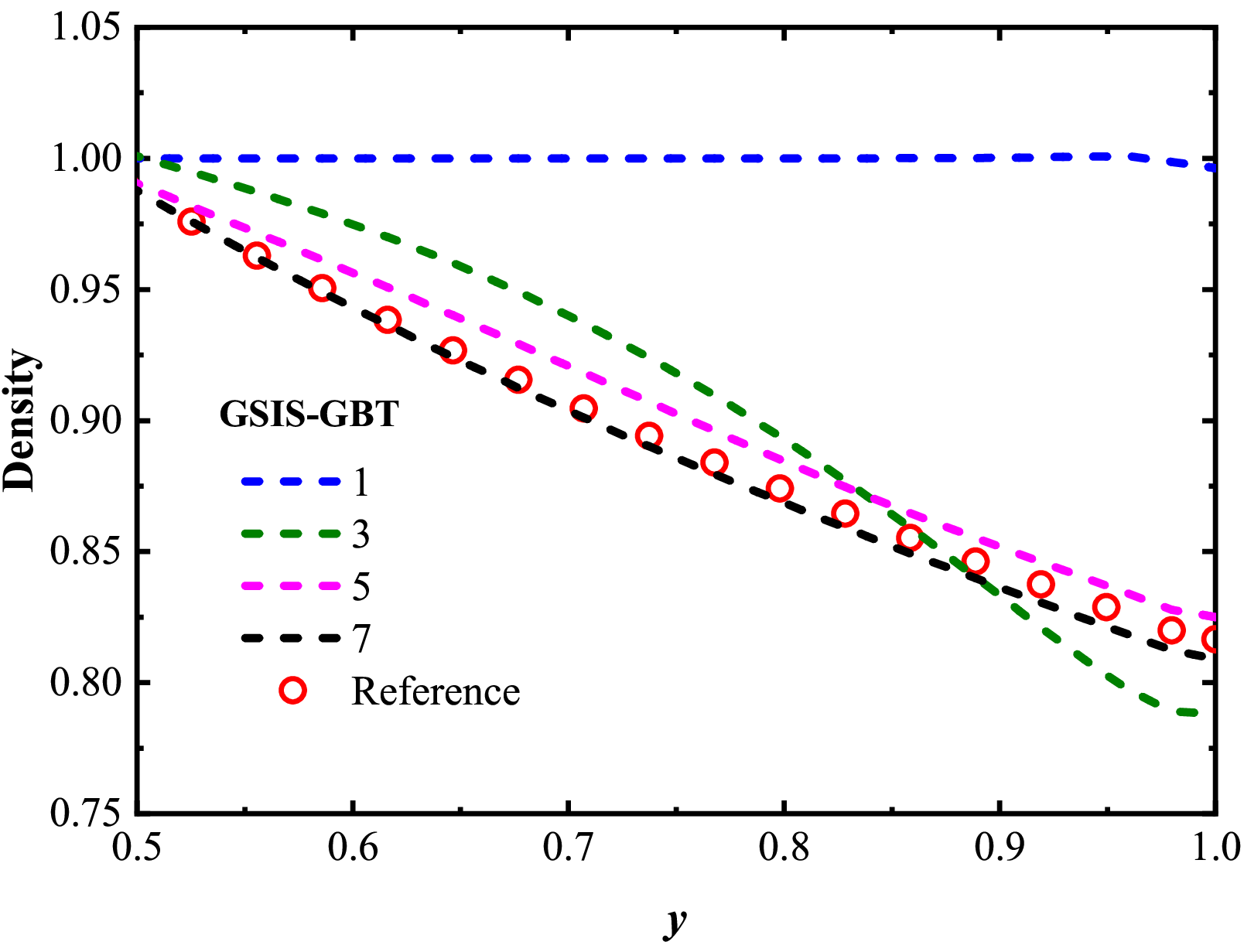}\\
	\vspace{0.5cm}
 \includegraphics[width=0.35\textwidth]{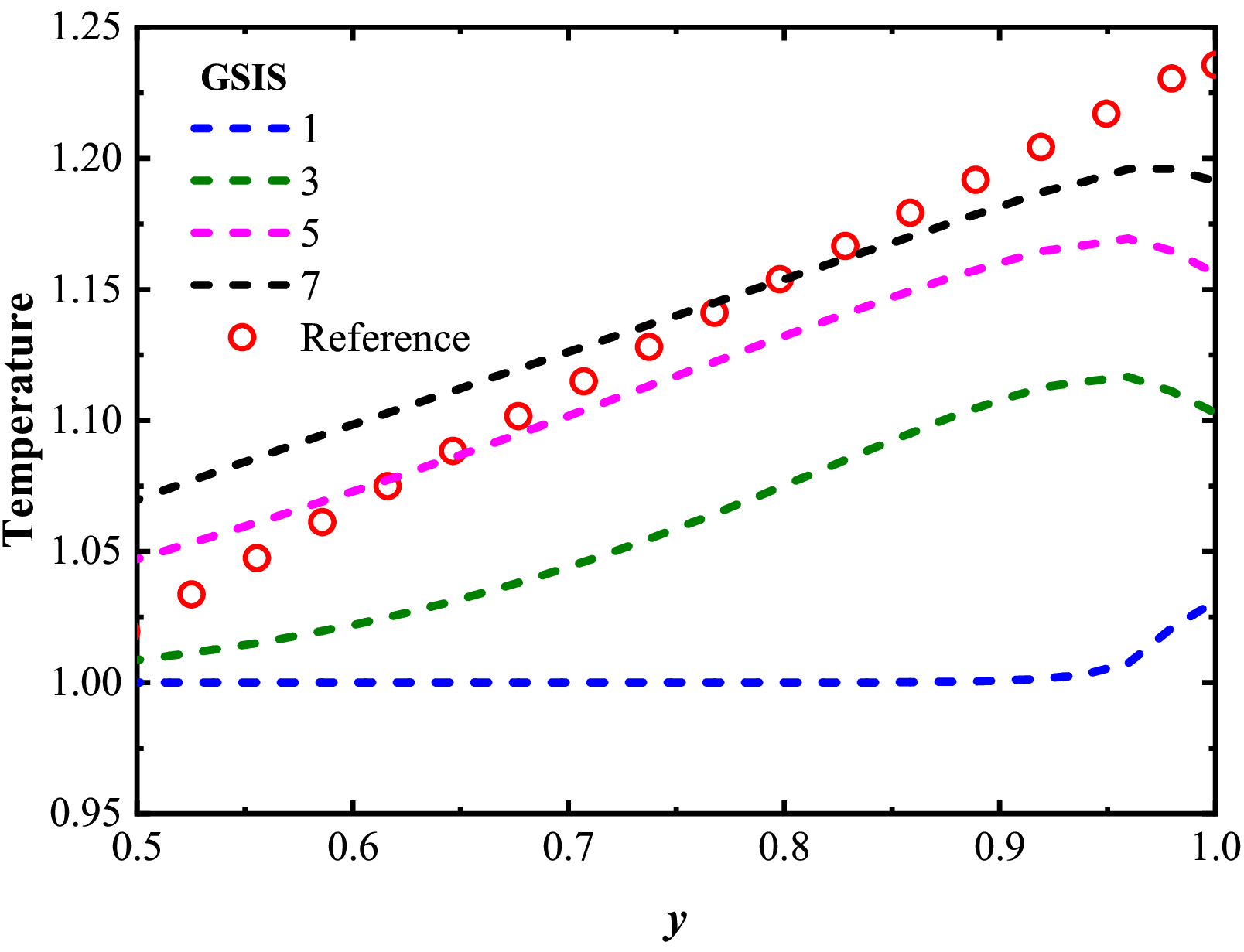}
 \quad
	\includegraphics[width=0.35\textwidth]{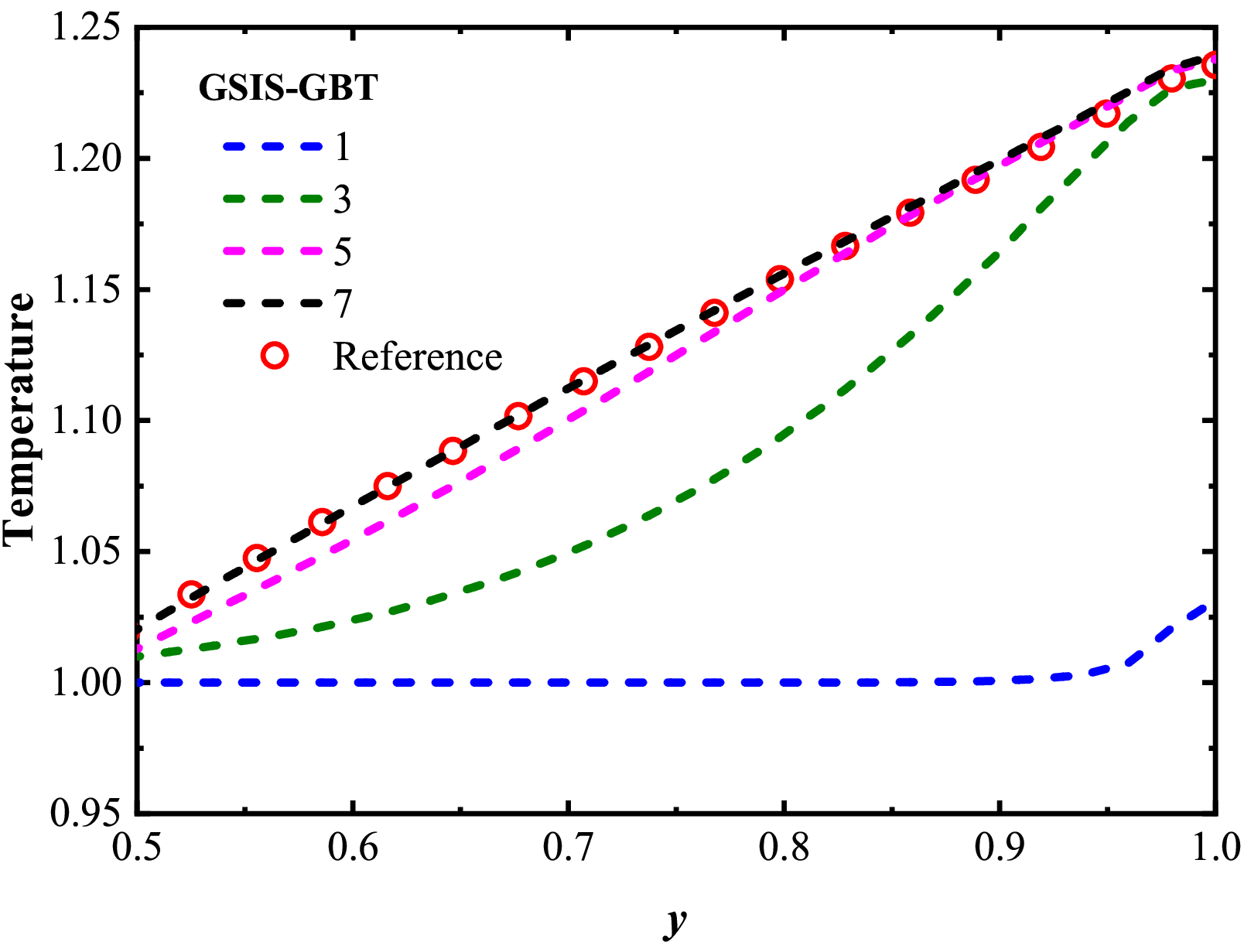} 
	\caption{Density and temperature at different iteration steps from the GSIS (left column) and GSIS-GBT (right column) for the planar Fourier heat transfer when Kn = 0.01. Reference solutions from~\cite{zhu_general_2021}.
 }
	\label{Fig3}
\end{figure}

To {confirm} that the GSIS-GBT further accelerates the evolution of the flow field at the boundaries, we compare in Fig.~\ref{Fig3} the density and temperature profiles for the GSIS~\cite{zeng_general_2023} and GSIS-GBT at intermediate iteration steps. After the first iteration, both methods produce the density and temperature close to their initial states. However, after just 3 iterations, the density and temperature in GSIS-GBT quickly converge towards the steady-state solutions, while GSIS evolves obviously slower. After the seventh iteration, the GSIS-GBT results closely align with the reference ones, while GSIS still has discrepancies with the reference solution, mostly near the wall. Hence, the relative error in GSIS-GBT exhibits a rapid decrease after the seventh iteration, whereas that of the GSIS performs relatively slower. This example provides a strong support that the GBT aids in accelerating the flow field evolution near the boundaries, thereby expediting the overall convergence.


\subsection{Supersonic flow passing through the cylinder}
\label{sec4-2}

Compared to the flat-plate heat transfer, the hypersonic flow around a cylinder poses greater challenges. The primary challenge arises from that the shock waves can be generated by the interaction between the hypersonic incoming flow and the cylinder surface. The computational parameters are configured as follows: the cylinder radius \(R = 0.5\) is served as the reference length, and the flow domain is defined as a circular region with a radius of 6, discretized by 128 radial and 100 axial structural grid points. The gas is selected as nitrogen with the rotational collision number set to 2.226.
The grid height near the cylinder surface is set to 0.002. 
The reduced VDFs are introduced to further reduce the molecular velocity space $\xi_z$~\cite{zeng_general_2023} and a grid of 64 $\times$ 64 velocity points is applied, discretized uniformly over the range of $\left[-12, 12 \right]^{2}$. The reference temperature for the flow is 273 K. At the initial state, both the dimensionless temperature of the flow field and the wall are set to 1.0. The CFL numbers for the mesoscopic and macroscopic equations are set to 10,000 and 1000, respectively. 


\begin{figure}[th]
	\centering
	\includegraphics[width=0.31\textwidth]{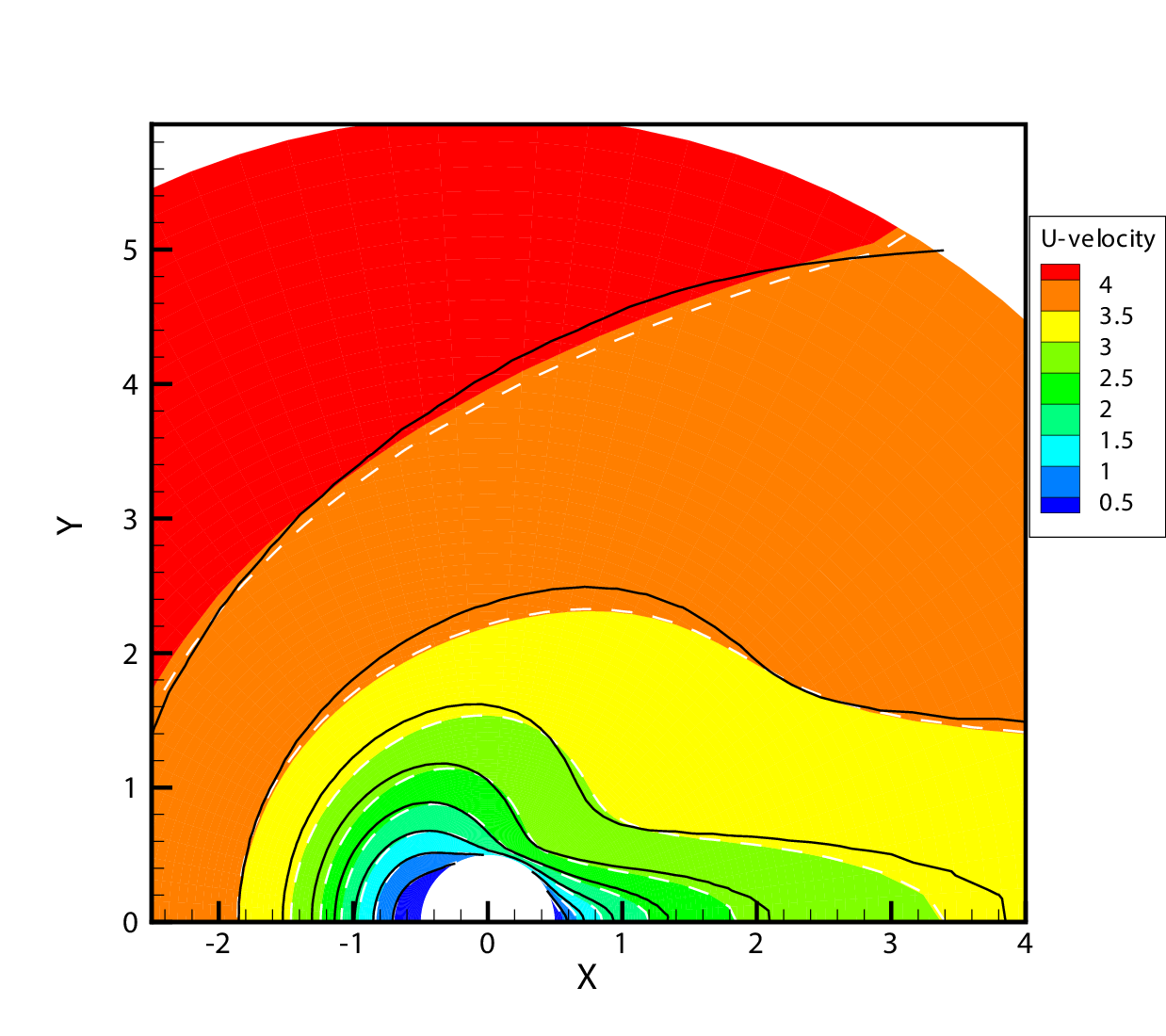}
	\includegraphics[width=0.31\textwidth]{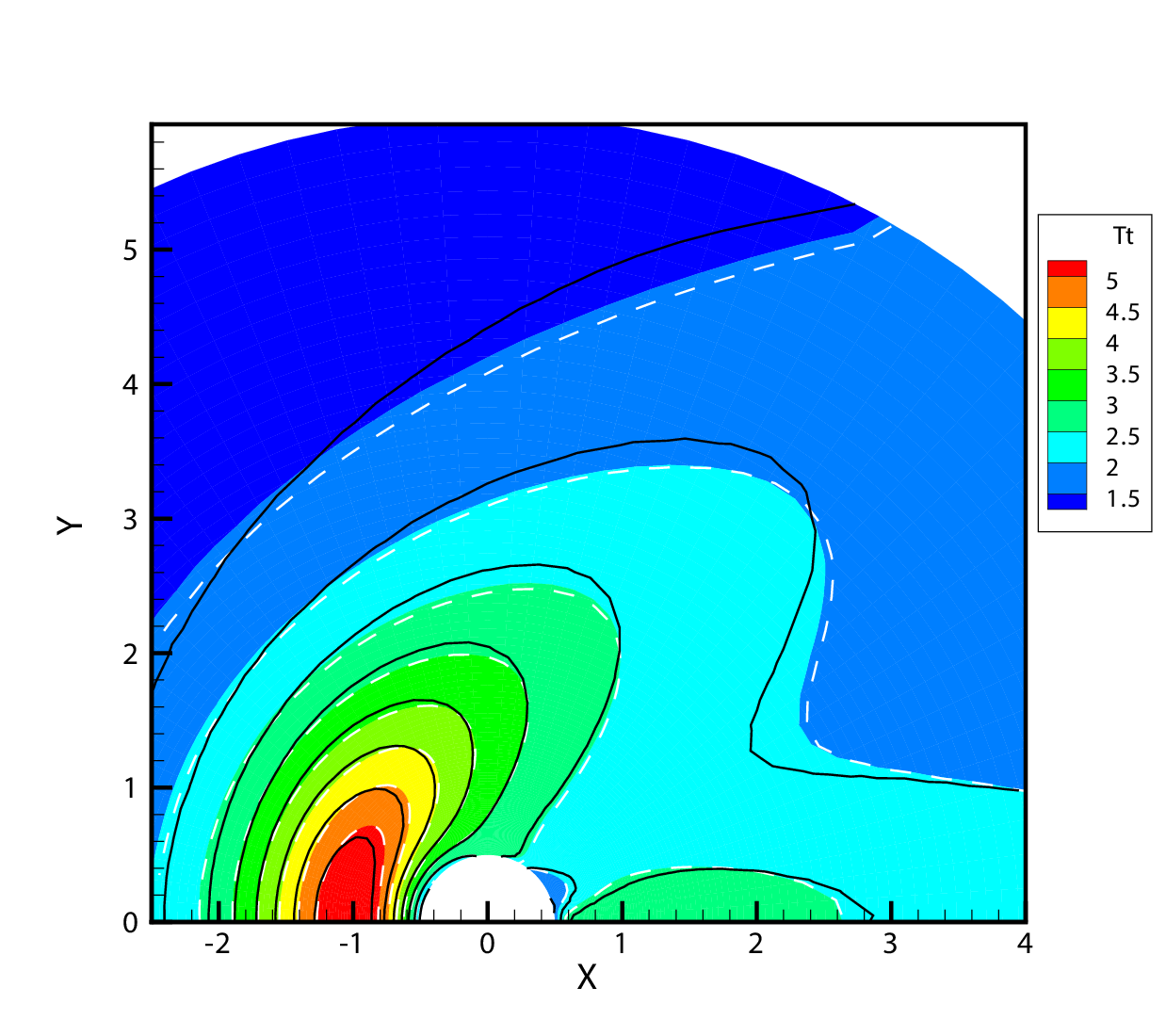}
	\includegraphics[width=0.31\textwidth]{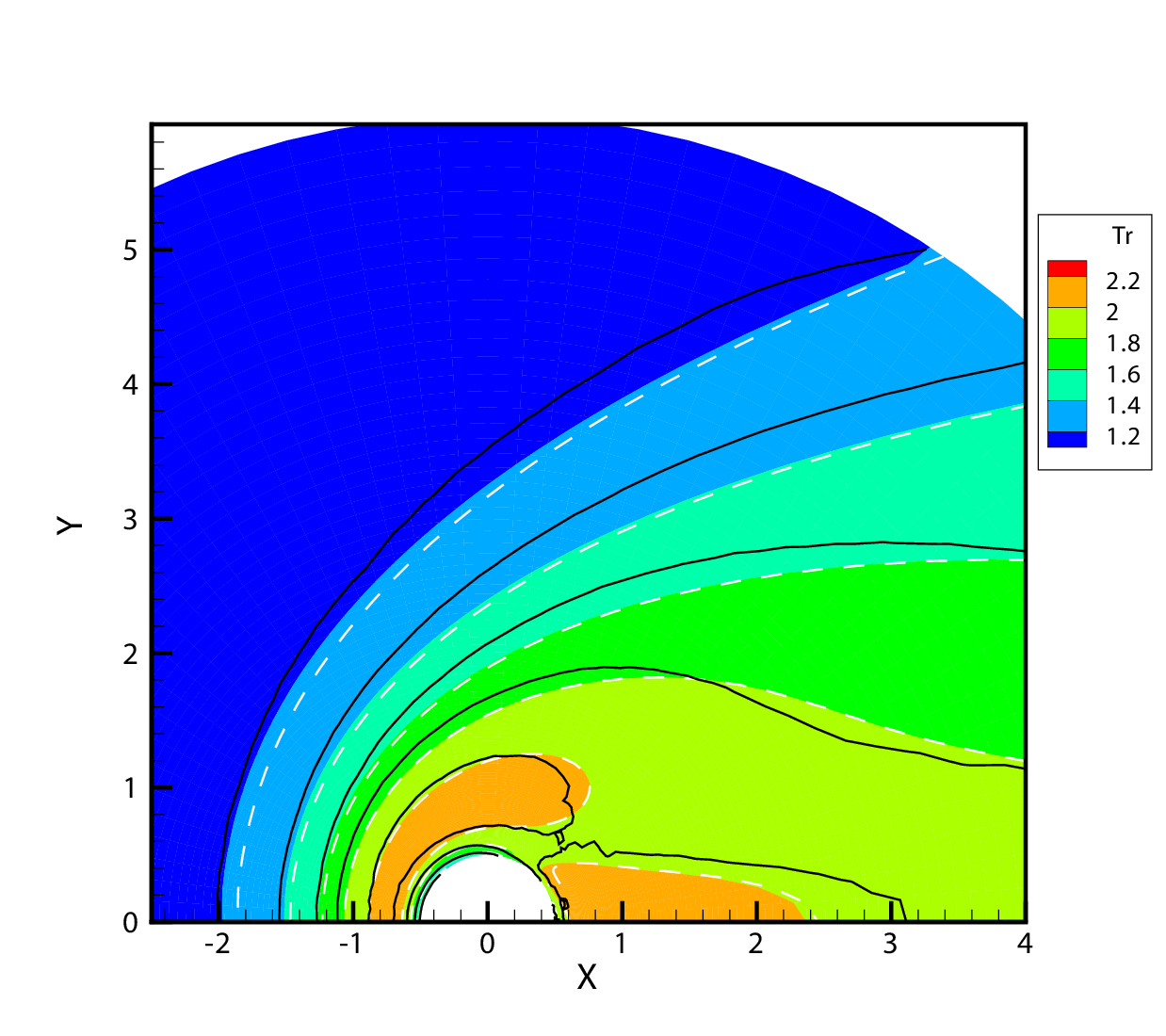}
	\includegraphics[width=0.31\textwidth]{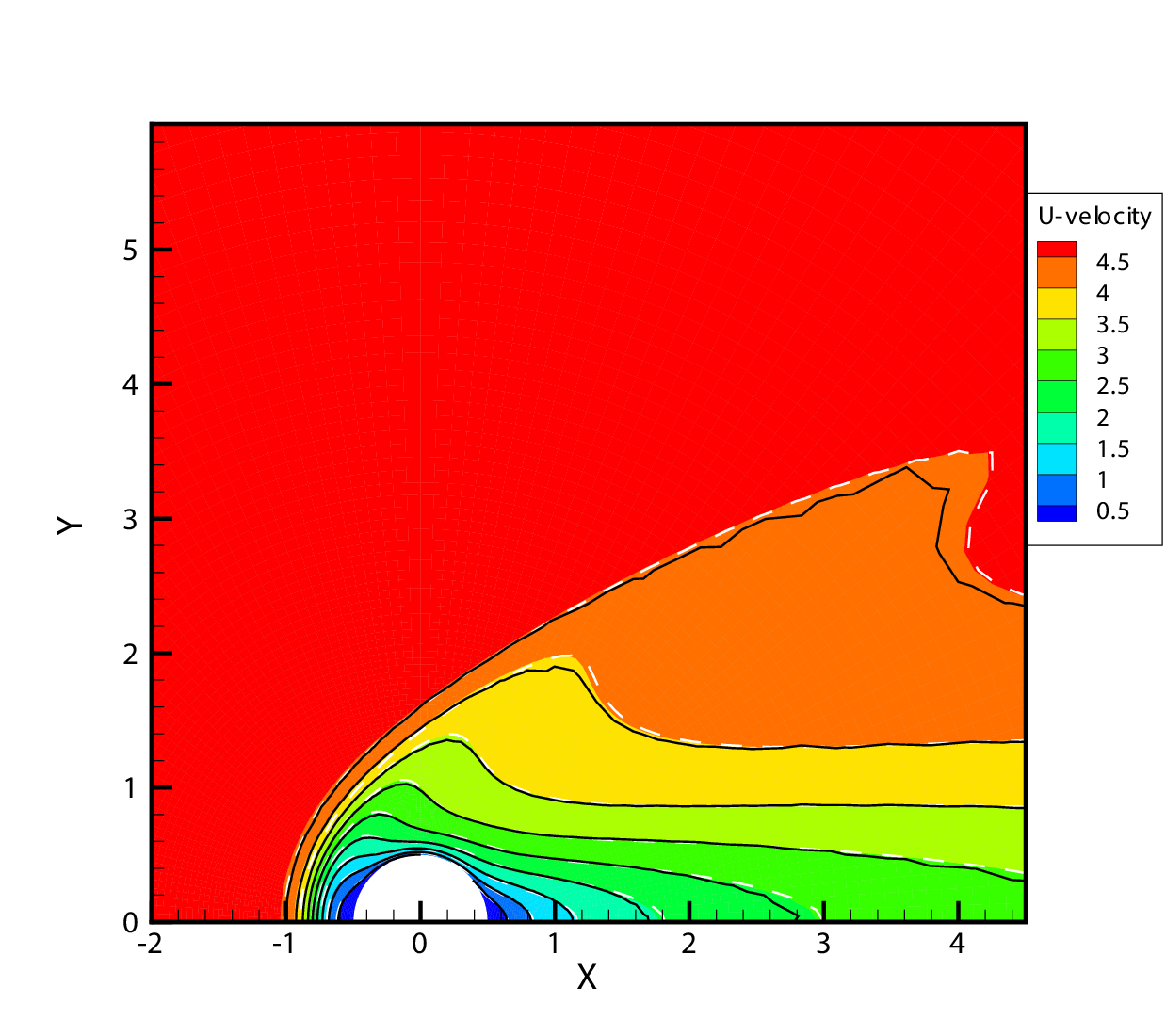}
	\includegraphics[width=0.31\textwidth]{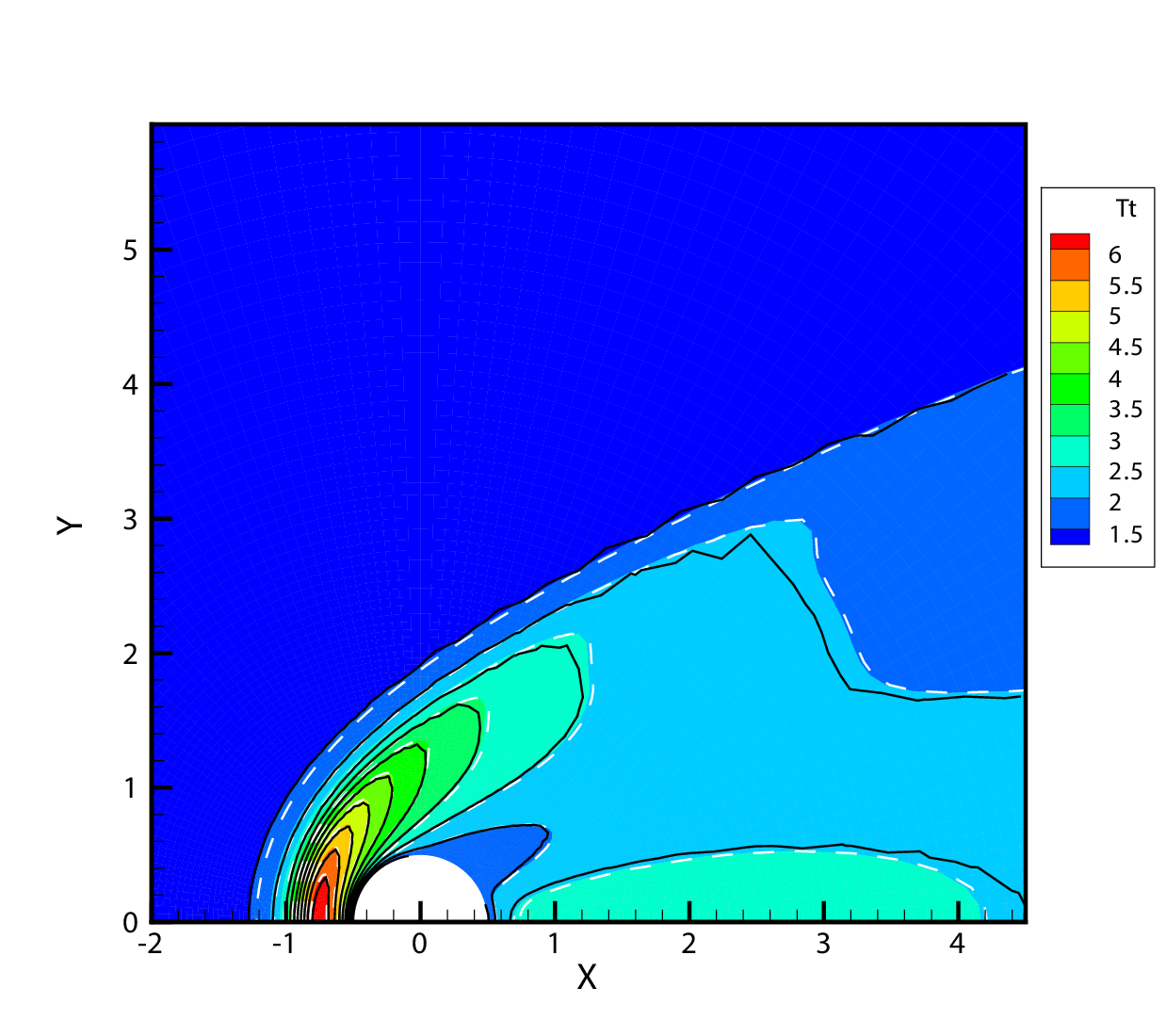}
	\includegraphics[width=0.31\textwidth]{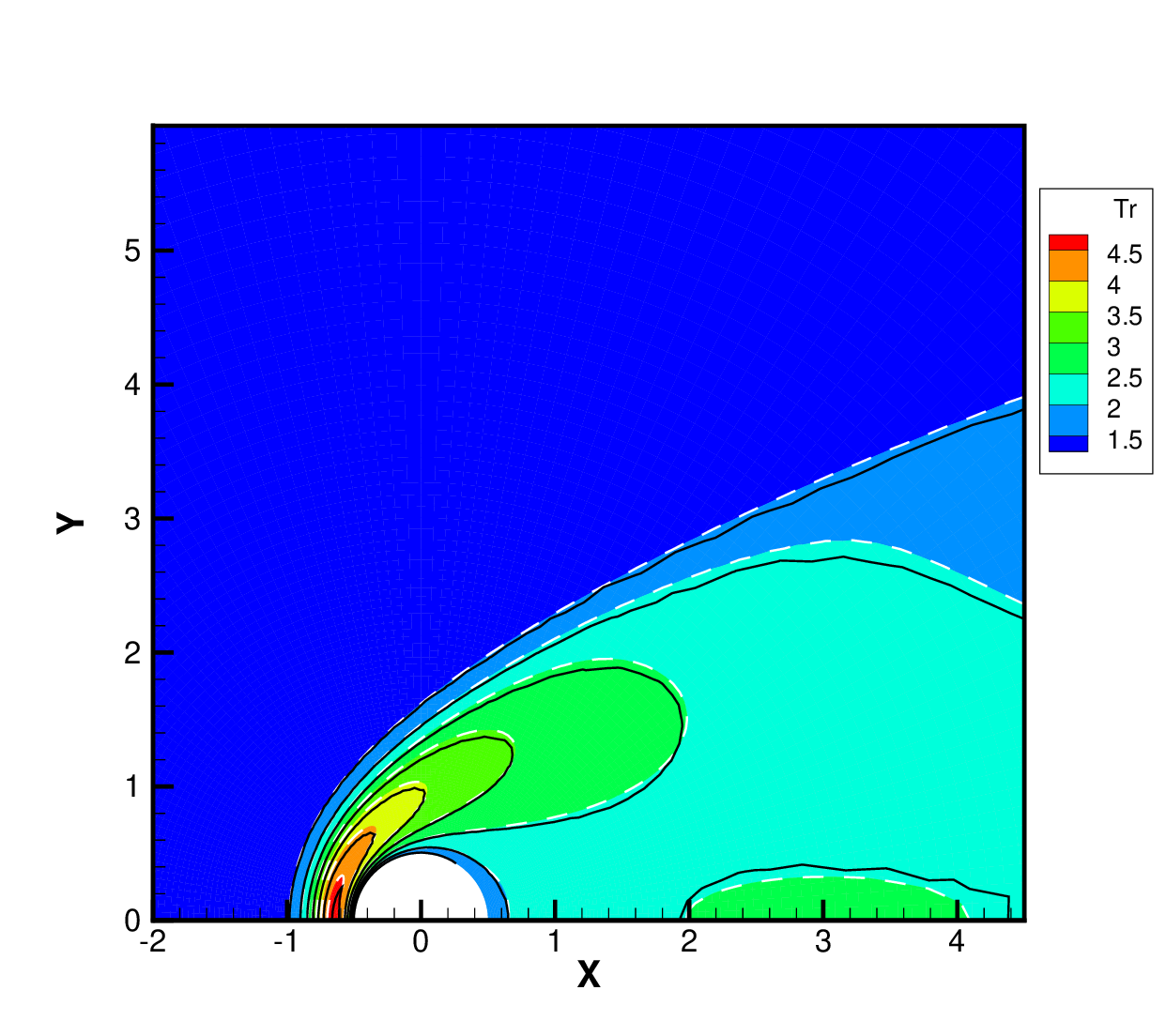}
	\caption{Comparison of the velocity (1st column), translational temperature (2nd column) and rotational temperature (3rd column) for supersonic flows around cylinder at Ma = 5.0 and Kn = 1 (1st row) and 0.1 (2nd row). Background contour: GSIS-GBT; White dashed line: GSIS; Black solid line: DSMC.}
	\label{Fig5}
\end{figure}

For the inflow boundary condition, the macroscopic quantities $\mathbf{W}_{\infty}=\left[\rho_{\infty}, \mathbf{u}_{\infty}, T_{\infty}\right]$ in the far field is prescribed, and the corresponding equilibrium VDFs in Eq.~\eqref{eq2.1.5} are used. 
For the outflow boundary condition, the gradients of macroscopic quantities along the flow direction are set to zero.

\begin{figure}[t]
	\centering
	\includegraphics[width=0.31\textwidth]{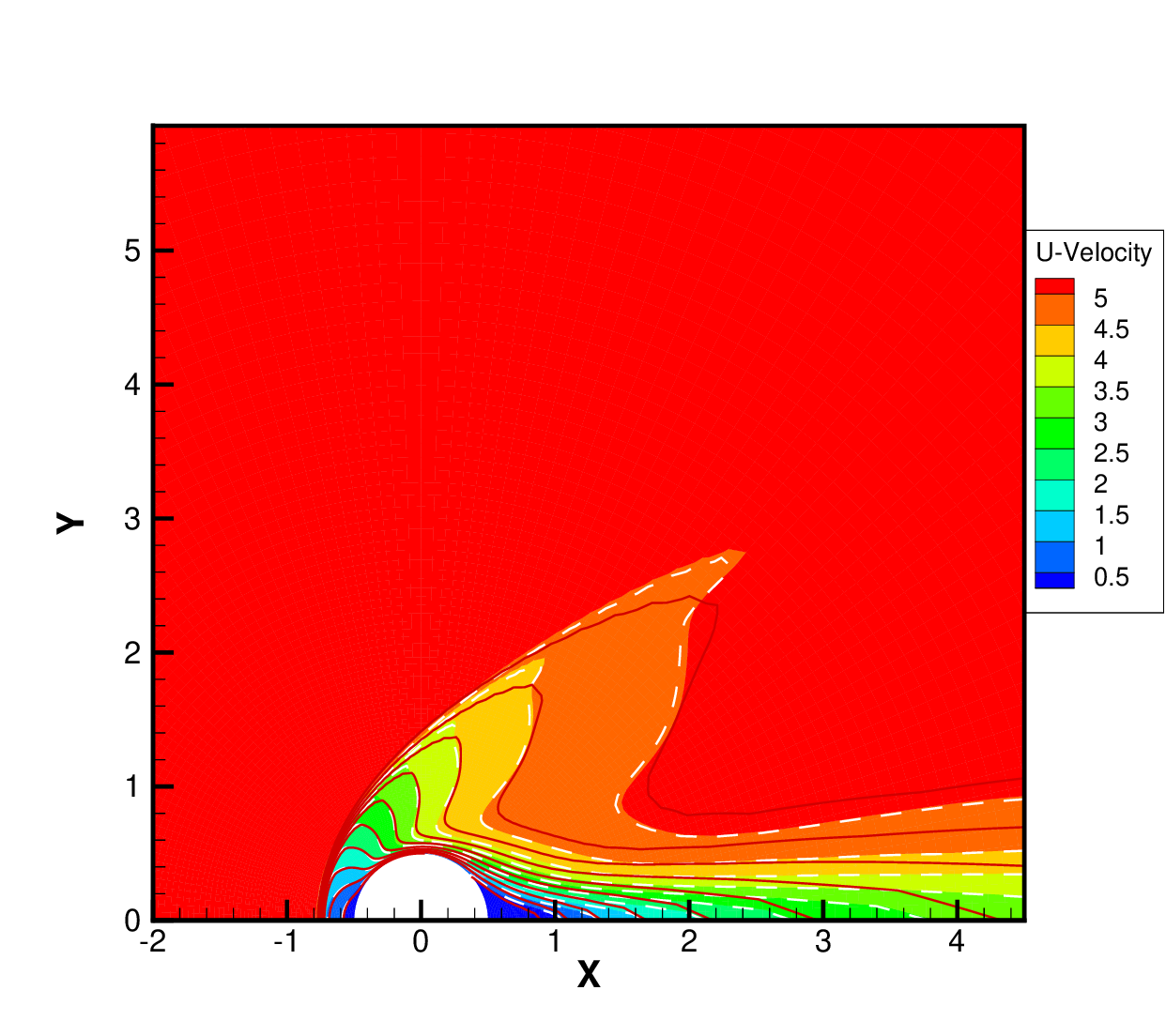}
	\includegraphics[width=0.31\textwidth]{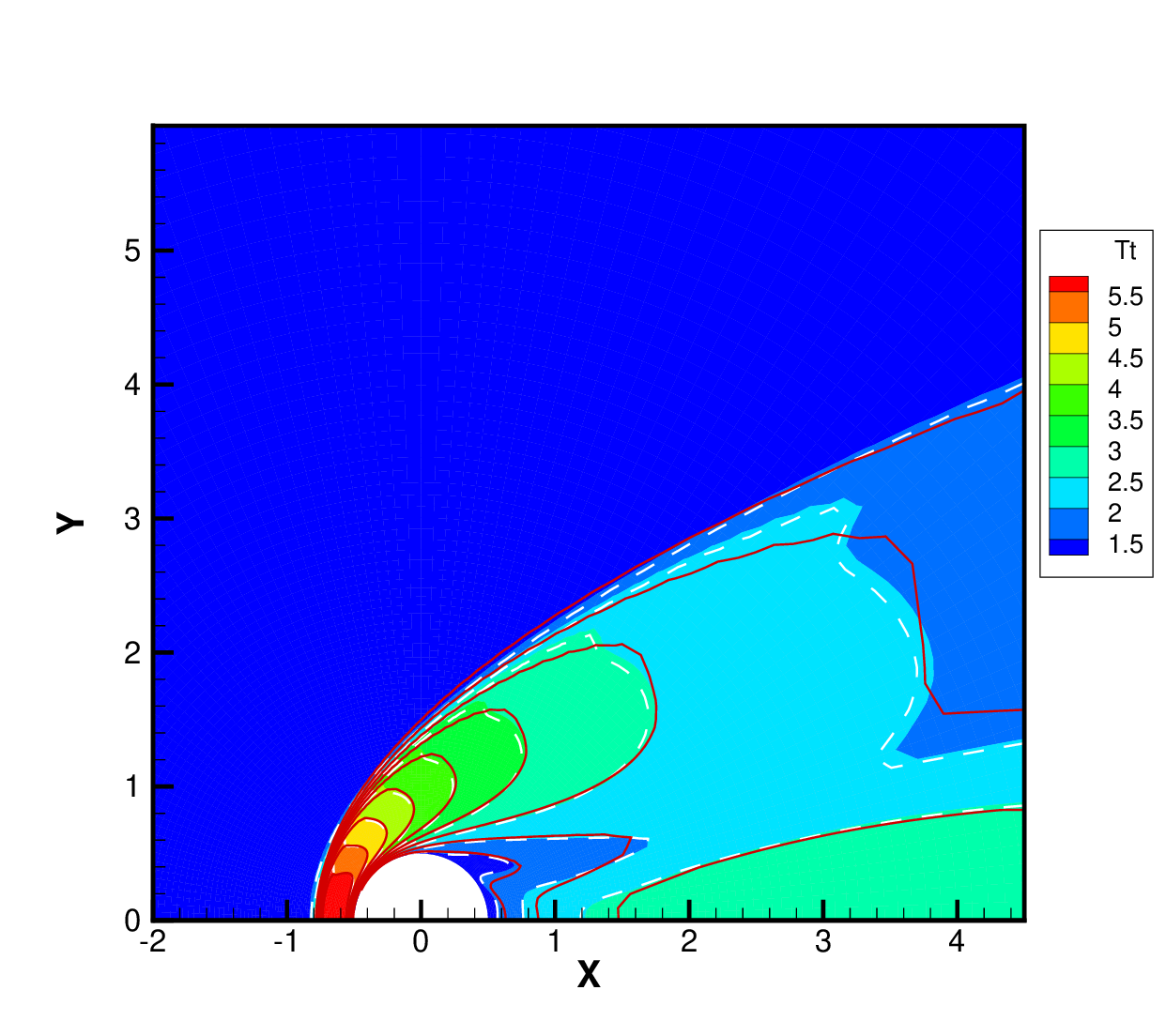}
	\includegraphics[width=0.31\textwidth]{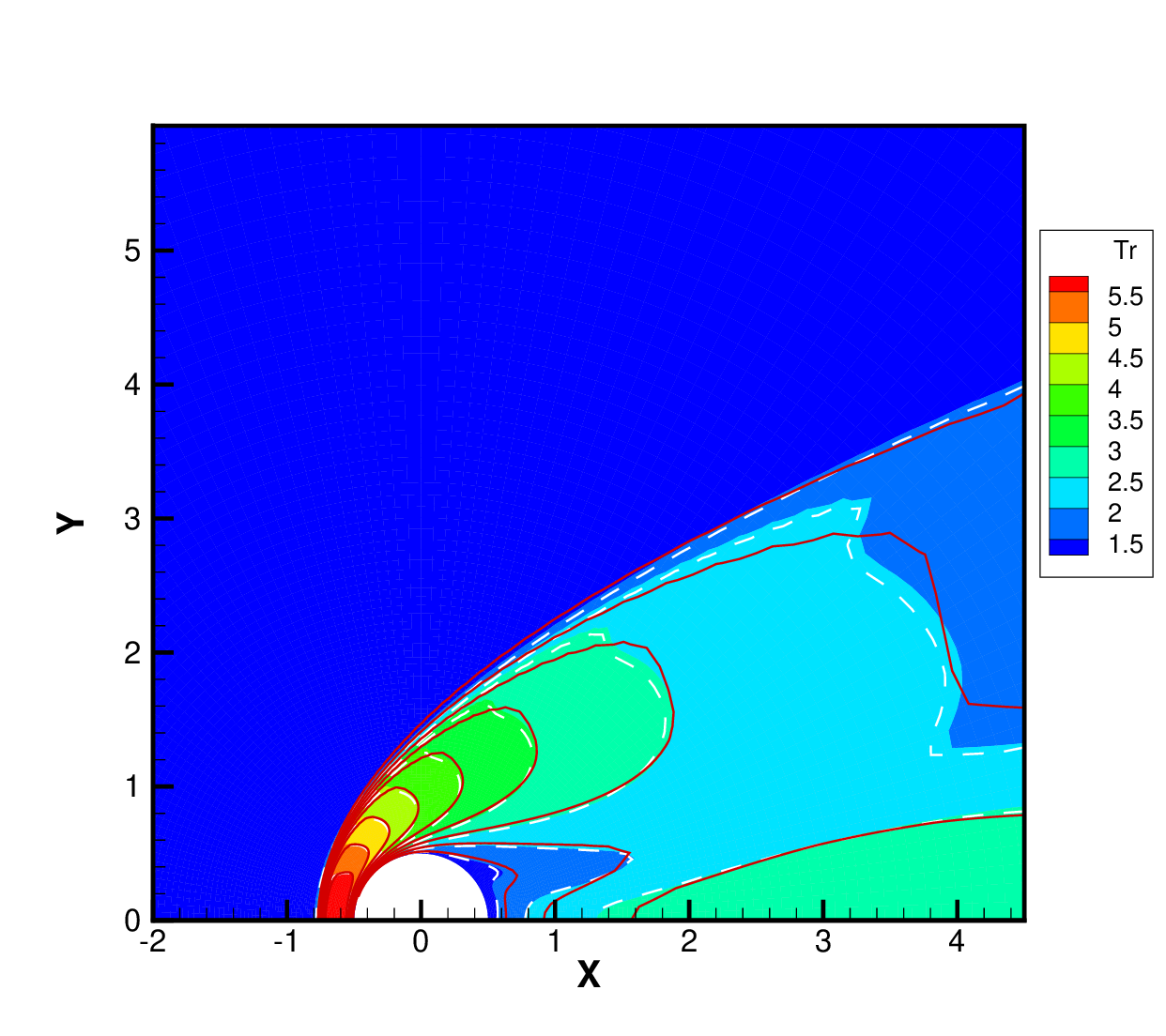}   
	\caption{Comparison of the (left) velocity, (middle) translational temperature and (right) rotational temperature for the supersonic flow around cylinder at Ma = 5.0 and Kn = 0.1. Background contour: GSIS-GBT; White dashed lines: GSIS; Red solid lines: NS-GBT.}
	\label{Fig6}
\end{figure}

We first consider the flow with the Mach number 5.0 and Knudsen numbers 1.0, 0.1 and 0.01. Figure~\ref{Fig5} compare the density, velocity, and temperature profiles obtained from the GSIS-GBT, GSIS~\cite{zeng_general_2023}, and DSMC. The shown velocity are normalized by the sound speed
\begin{equation}
    c_s=\sqrt{\gamma RT_0}, \quad 
    \text{with} \quad 
    \gamma=\frac{5+d_r}{3+d_r}.
\end{equation}
Results from GSIS and GSIS-GBT align well with those obtained from DSMC, confirming the accuracy of GSIS in rarefied gas flows. Furthermore, the GSIS-GBT exhibits comparable performance to the GSIS in highly rarefied flows, supporting the assertion that GBT accurately provides boundary conditions. When the Knudsen number decreases to 0.01, DSMC would necessitate an exceedingly dense meshes for accurate calculations, rendering the computational cost high. Given that the flow is approaching the slip regime and the rarefaction effects are less pronounced, we present the contours of GSIS-GBT in Fig.~\ref{Fig6}, along with the results from the Navier-Stokes equations (NS-GBT)\footnote{Note that the NS-GBT utilizes the governing equations~(\ref{eq2.1.9}) and \eqref{eq2.1.10} to compute the internal interfaces of the flow field, while employing GBT to simulate the velocity slip and temperature jump boundary conditions. In contrast to the macroscopic inner iterations in GSIS-GBT, NS-GBT does not introduce higher-order corrections for stress and heat flux, relying solely on the constitutive relations. As a result, it is applicable primarily to rarefied flows near the slip regime.}. 
The results in Fig.~\ref{Fig6} indicate a close agreement between the results from the GSIS-GBT and NS-GBT, particularly in the frontal region of the cylinder. 

\begin{figure}[p]
	\centering
	\subfigure[]{
		\includegraphics[width=4.5cm]{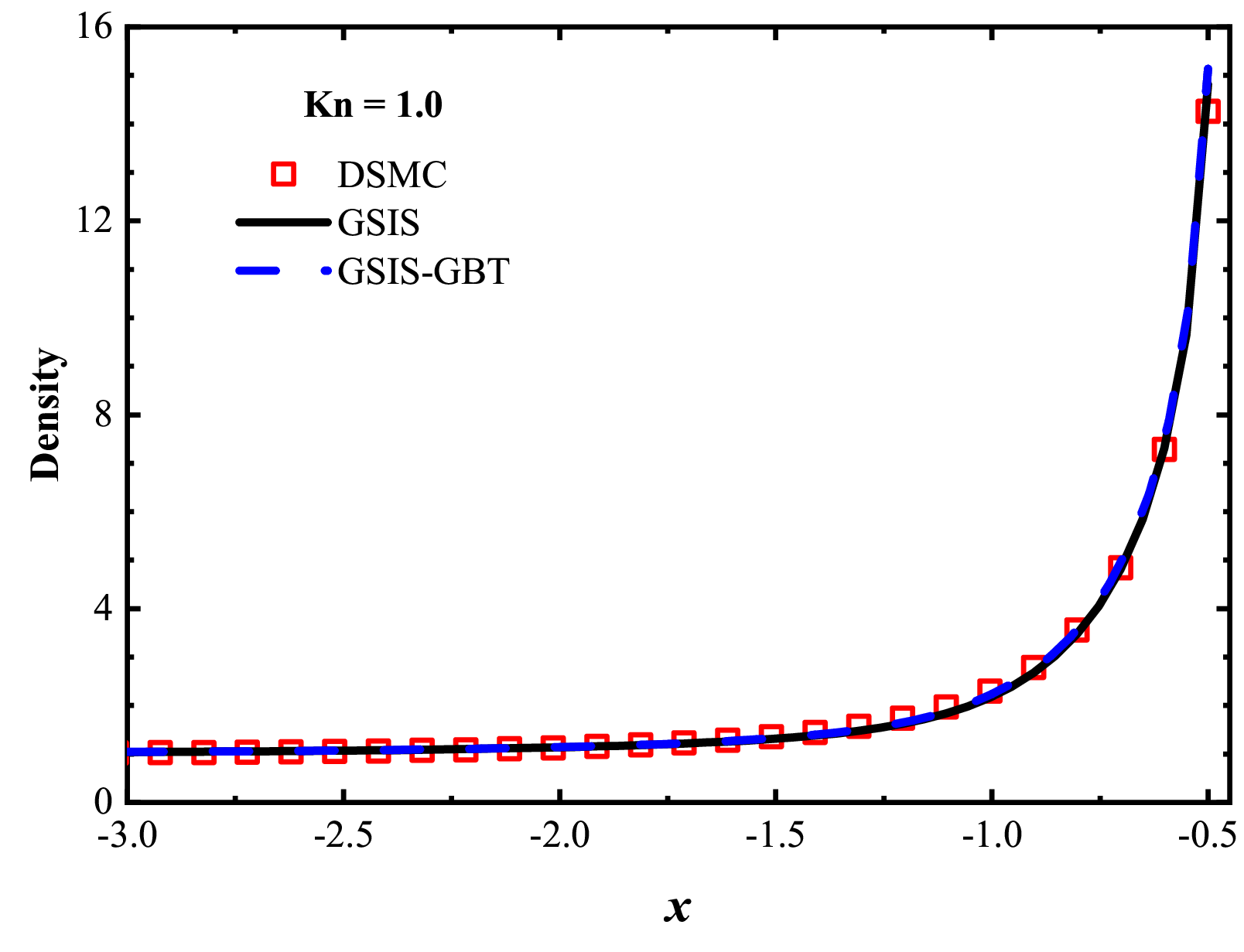}
	}
	\quad
	\subfigure[]{
		\includegraphics[width=4.5cm]{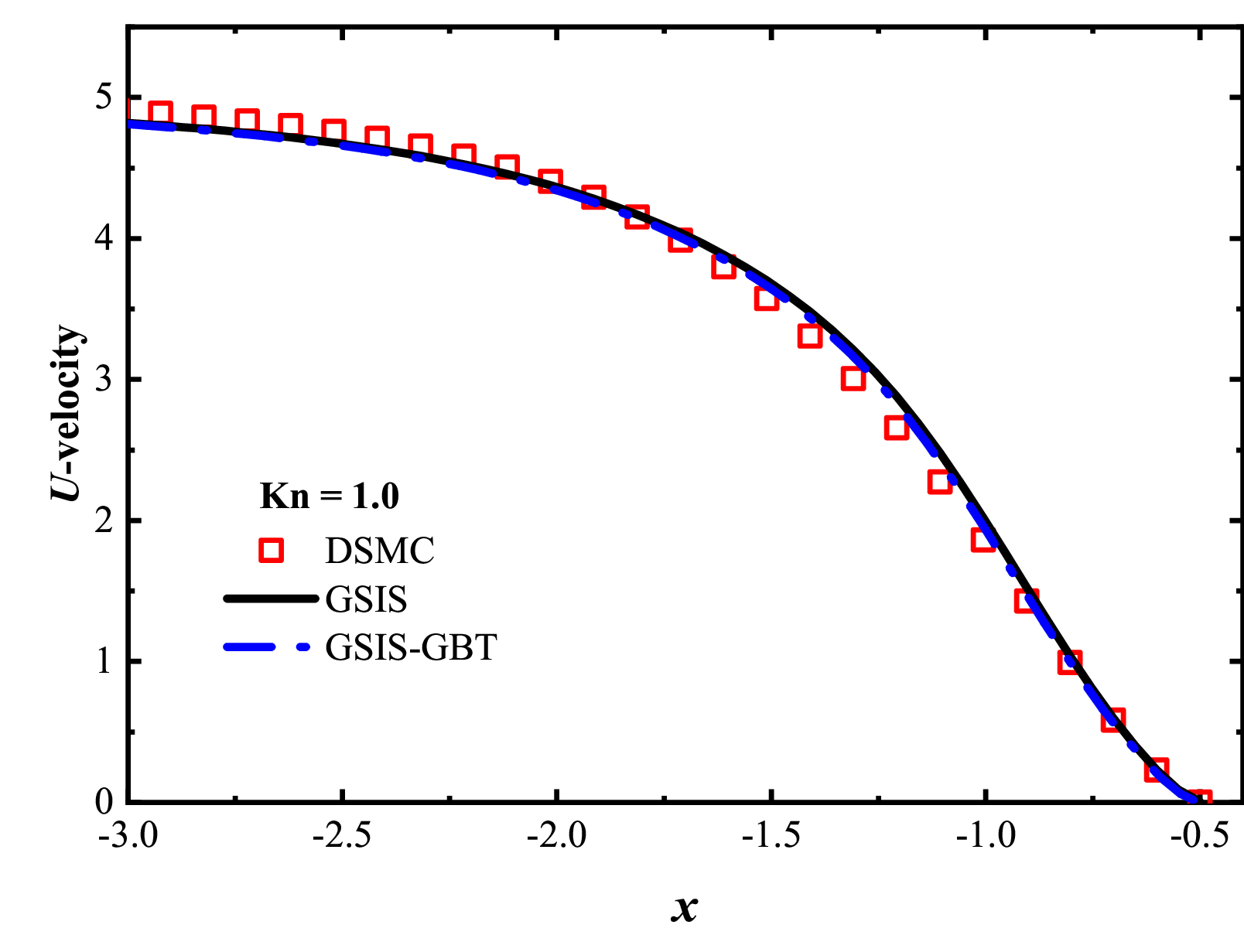}
	}
	\quad
	\subfigure[]{
		\includegraphics[width=4.5cm]{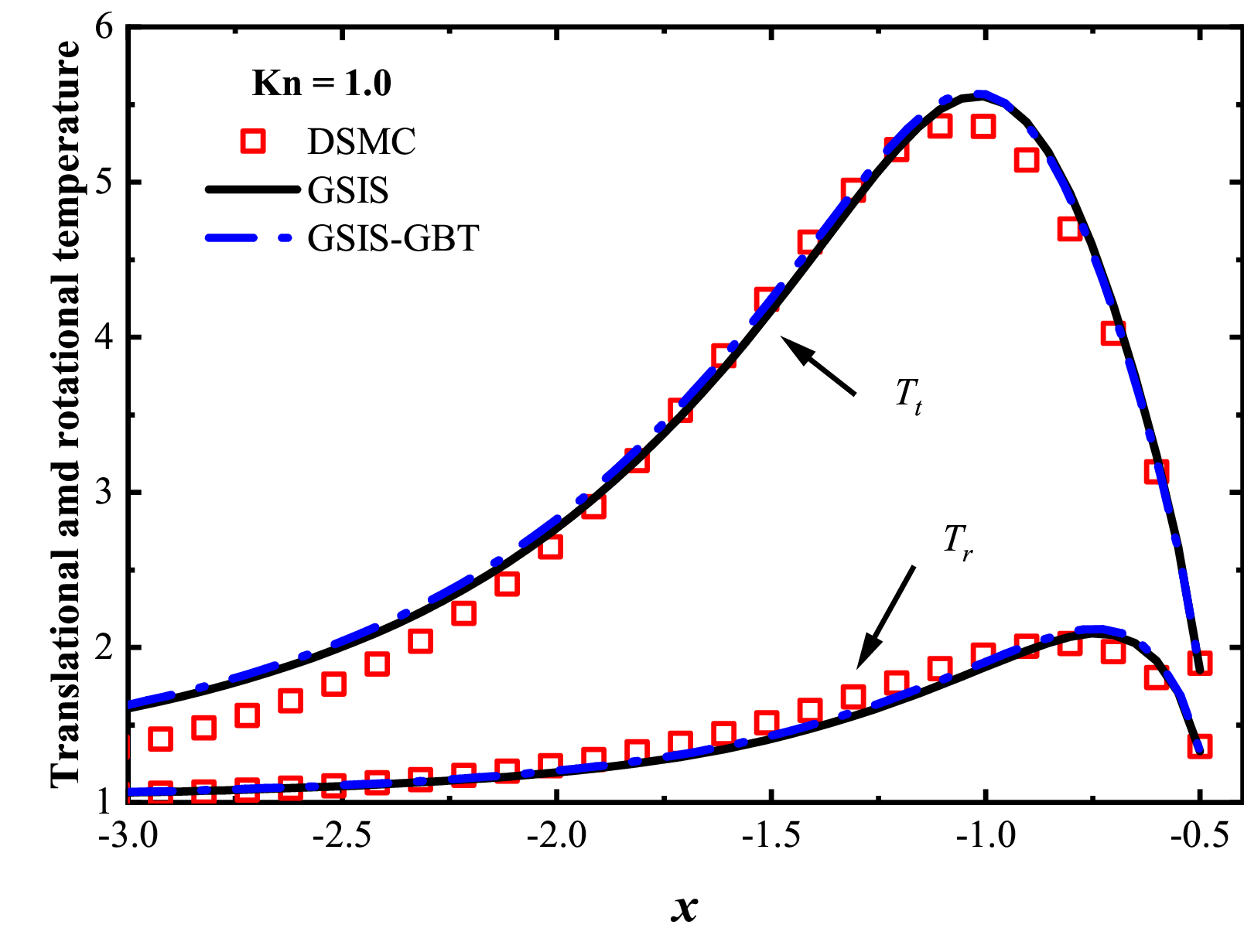}
	}
	\subfigure[]{
		\includegraphics[width=4.5cm]{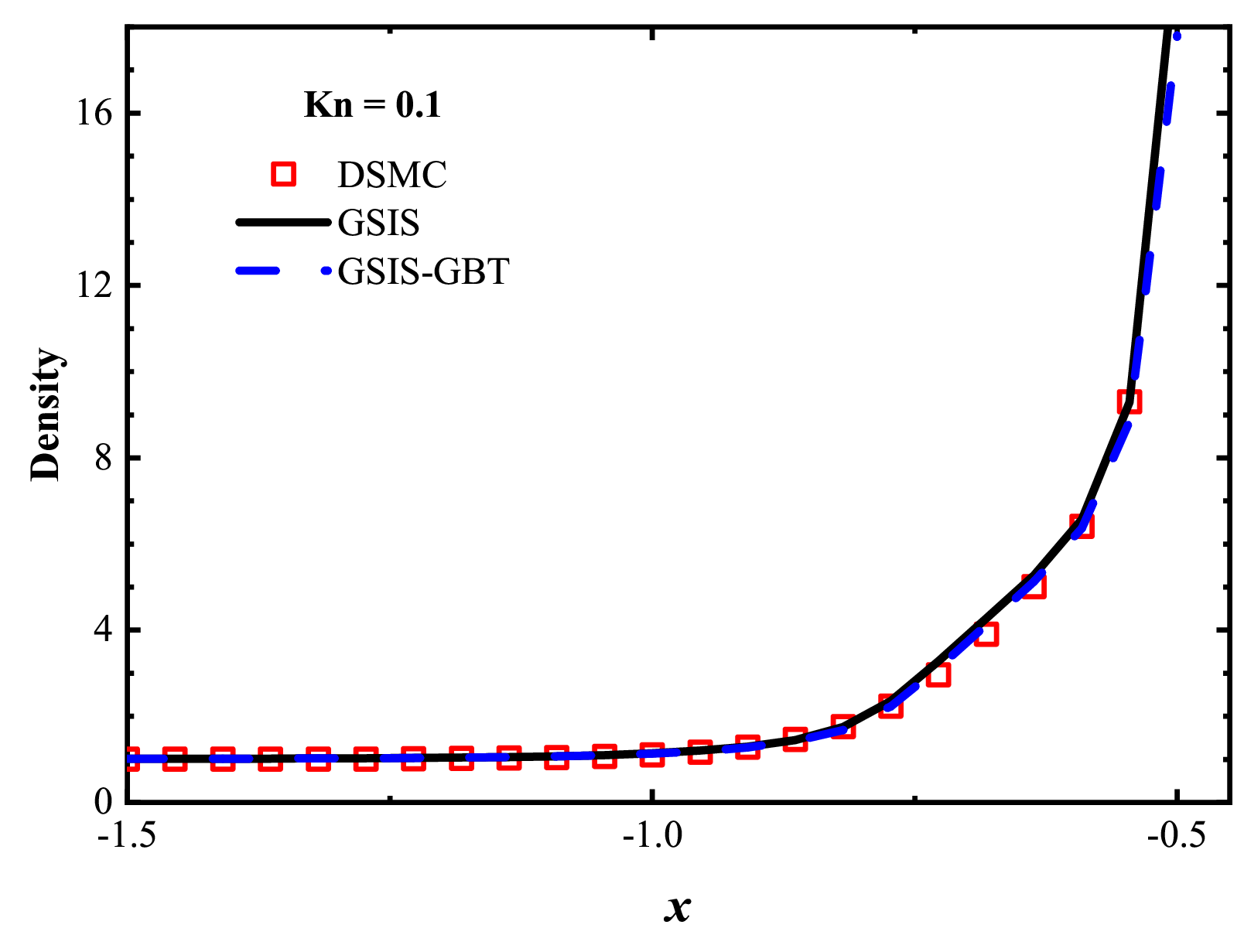}
	}
	\quad
	\subfigure[]{
		\includegraphics[width=4.5cm]{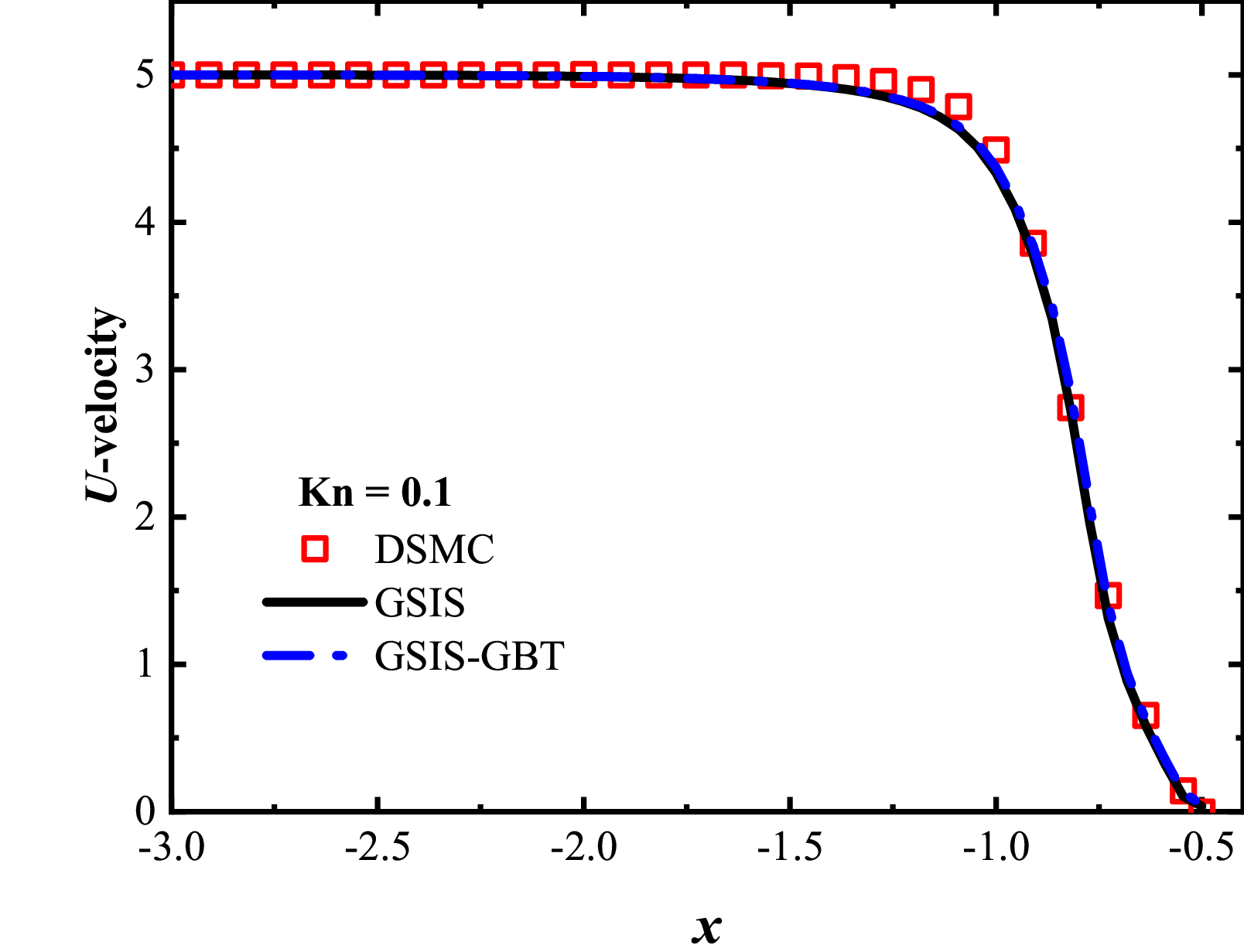}
	}
	\quad
	\subfigure[]{
		\includegraphics[width=4.5cm]{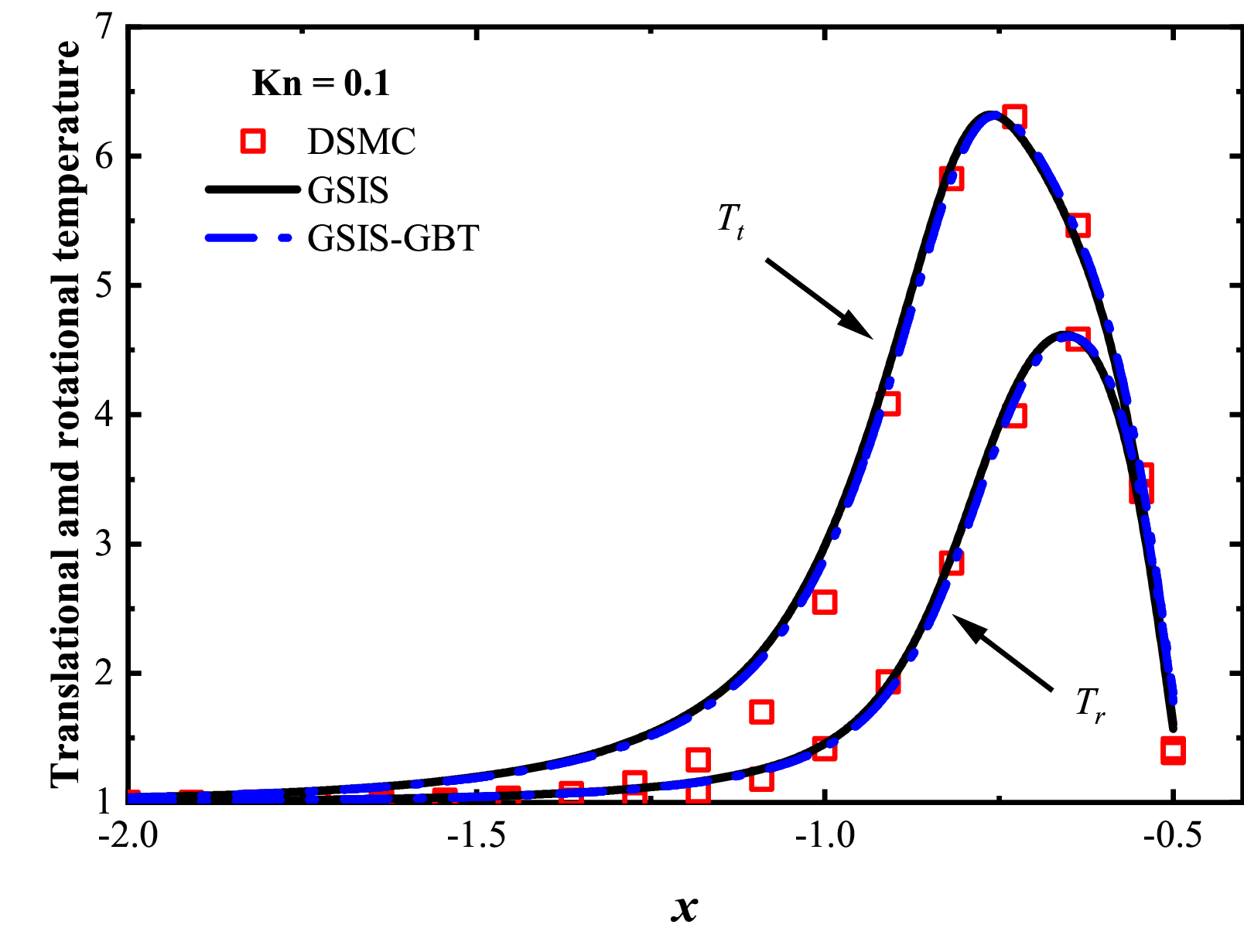}
	}    
	\subfigure[]{
		\includegraphics[width=4.5cm]{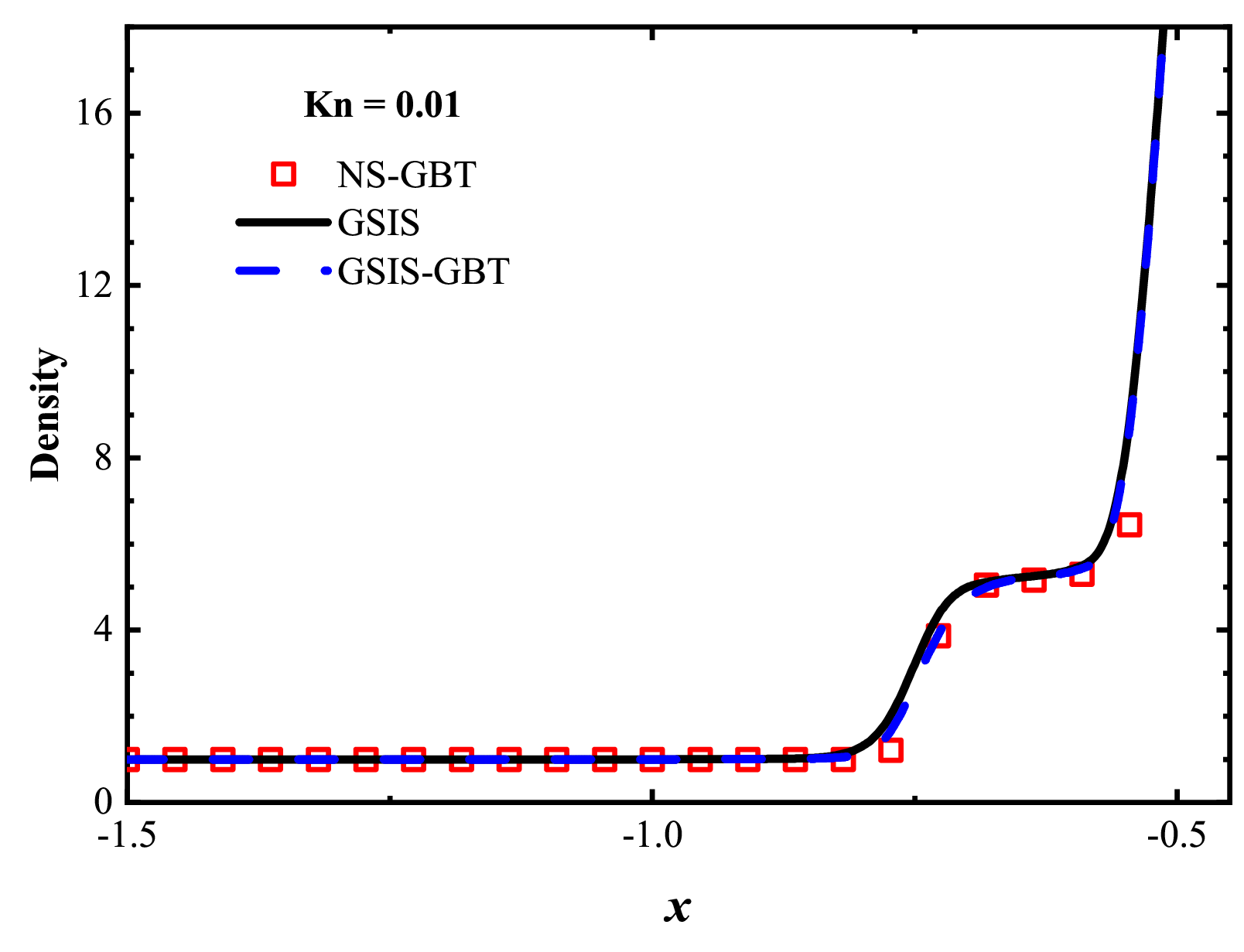}
	}
	\quad
	\subfigure[]{
		\includegraphics[width=4.5cm]{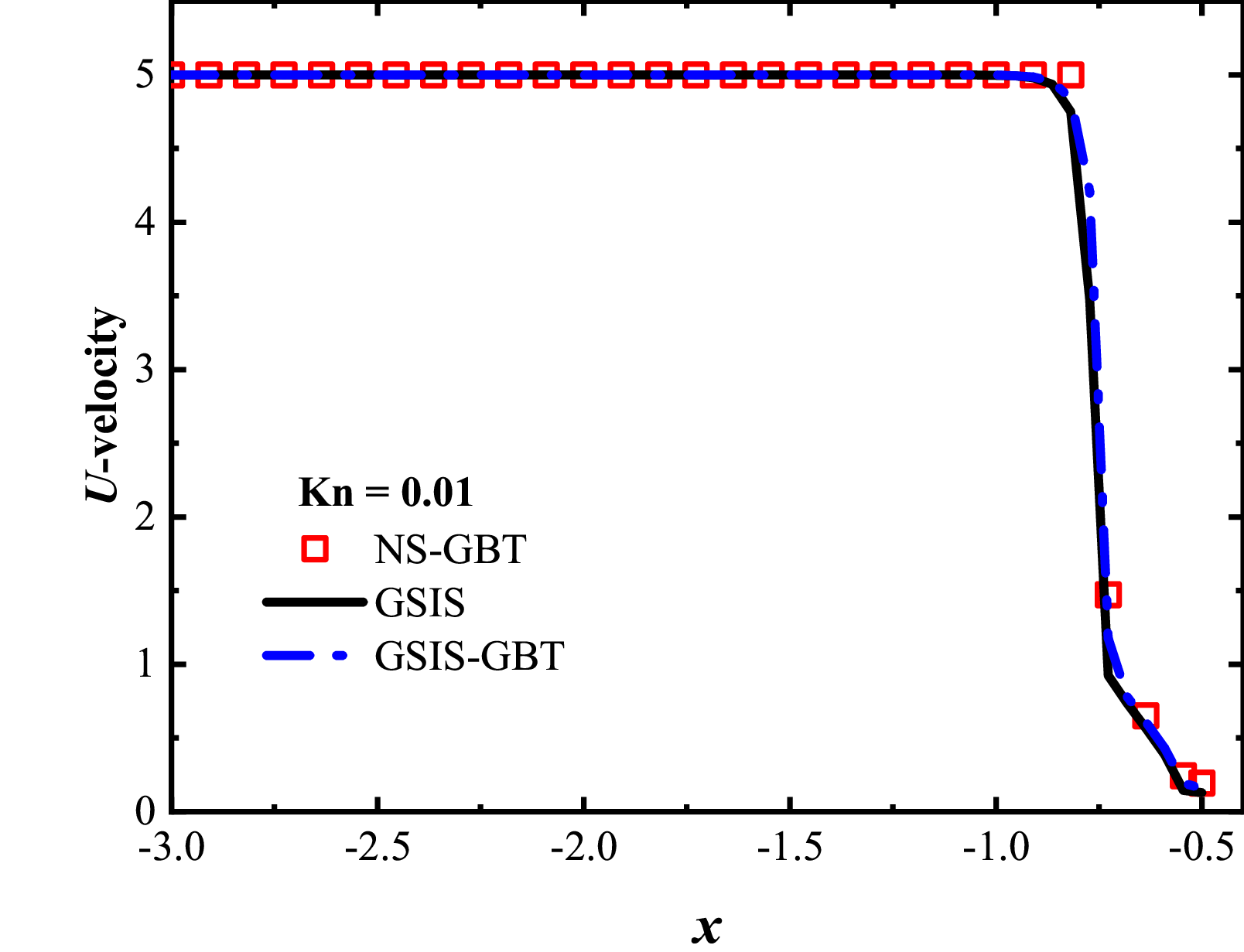}
	}
	\quad
	\subfigure[]{
		\includegraphics[width=4.5cm]{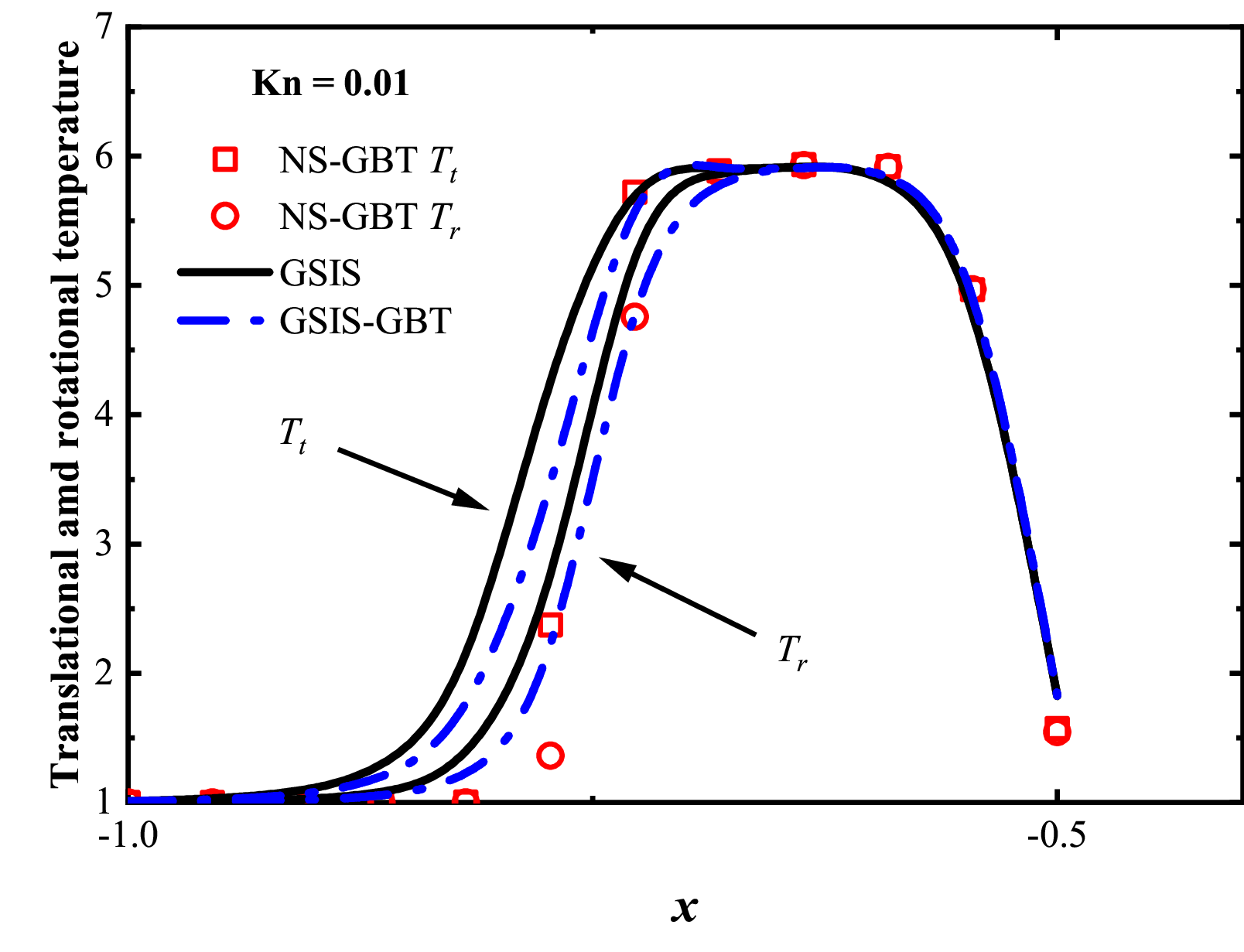}
	} 
	\subfigure[]{
		\includegraphics[width=4.5cm]{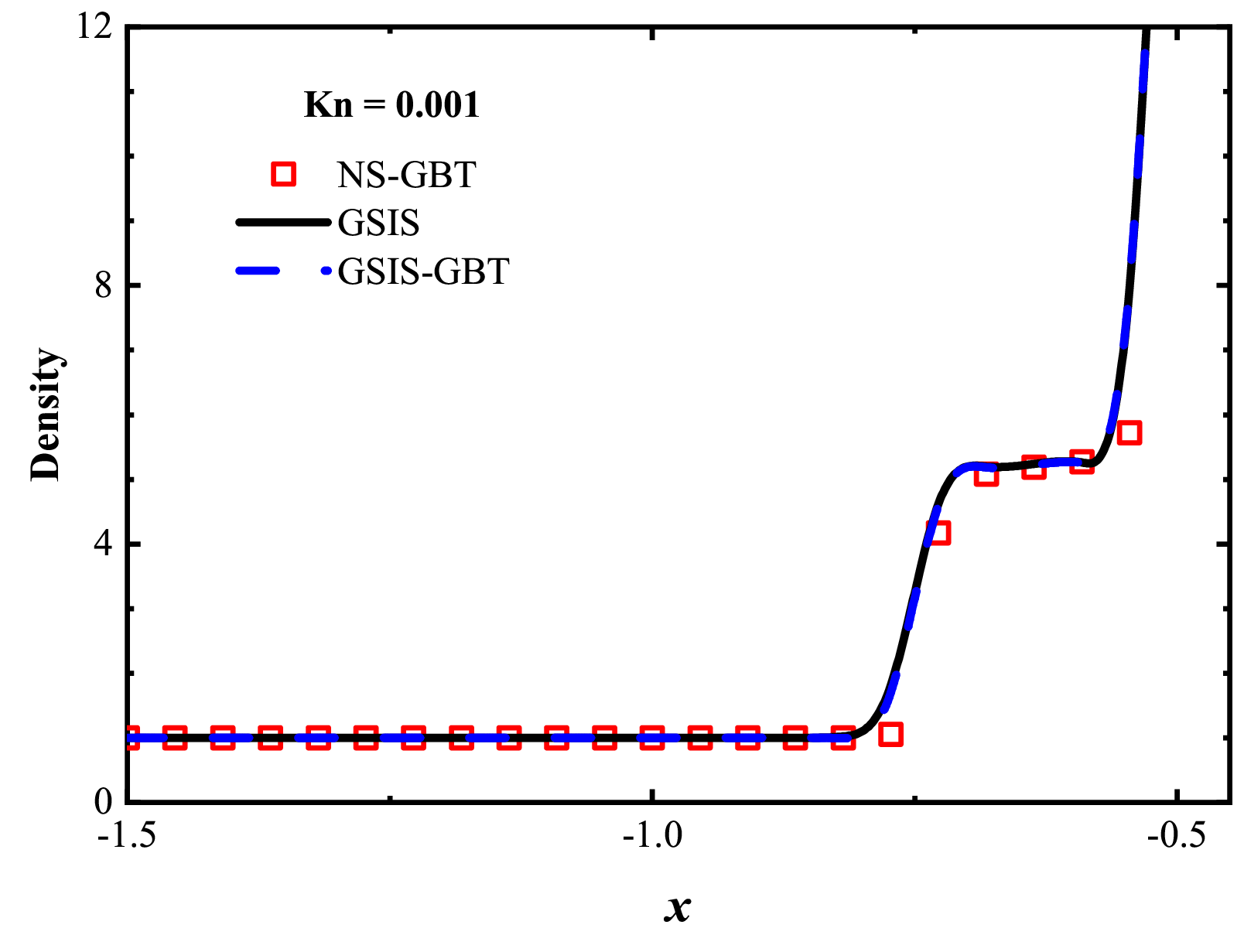}
	}
	\quad
	\subfigure[]{
		\includegraphics[width=4.5cm]{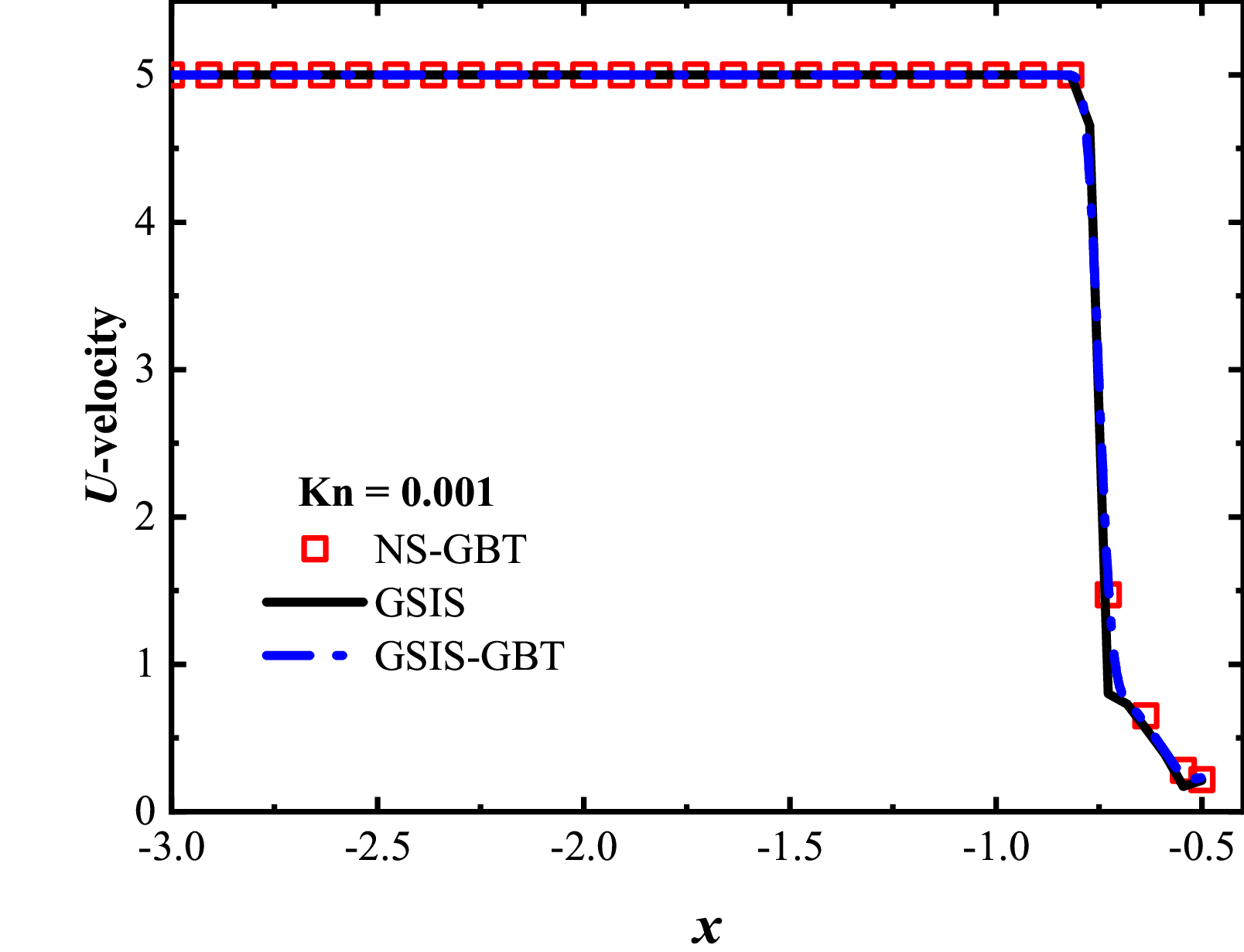}
	}
	\quad
	\subfigure[]{
		\includegraphics[width=4.5cm]{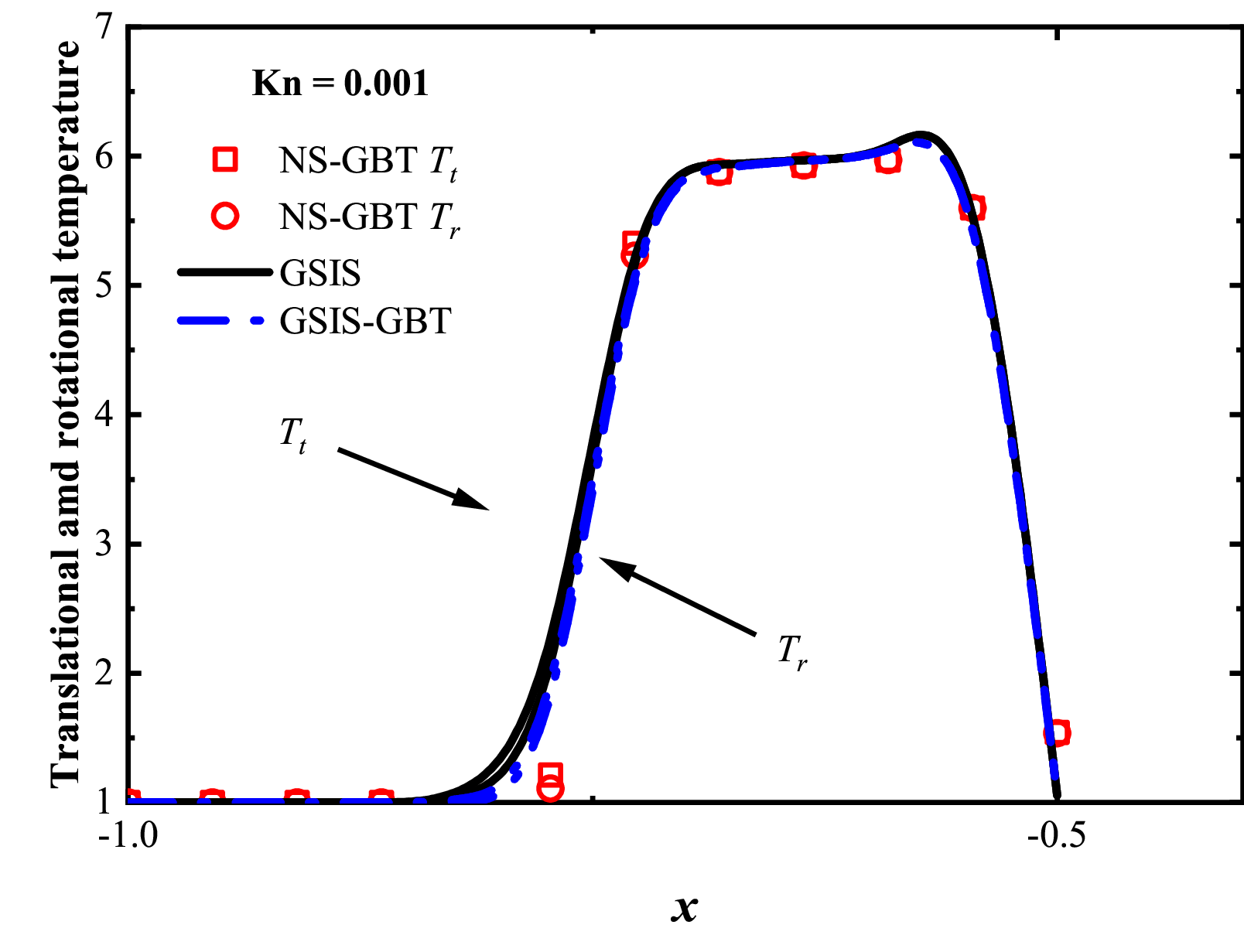}
	}       
	\caption{Comparison of density, velocity, translational temperature and rotational temperature along the stagnation line for the supersonic cylinder flow at Ma = 5.0 and different Kn.}
	\label{Fig7}
\end{figure}

To validate the performance of GSIS-GBT in hypersonic flows, results along the stagnation line obtained from the GSIS~\cite{zeng_general_2023}, GSIS-GBT, DSMC, and NS-GBT are compared. Figures~\ref{Fig7}(a-f) shows that the results from the GSIS and GSIS-GBT exhibit a favorable agreement with DSMC. When the Knudsen number decreases to 0.01 and 0.001, slight discrepancies emerge in the results of GSIS and NS equations, especially in the temperature results presented in Figs.~\ref{Fig7}(i-l):  the temperature profile obtained from the GSIS at the shock wave is slightly higher than the that from the NS equations. This is due to the stronger local rarefaction effects in the shock wave, making the NS equations loss the accuracy. However, outside the shock wave region, results obtained from the two versions of GSIS align very well with those from the NS equations.

\begin{table}[t]
	\centering
	\setlength{\abovecaptionskip}{10pt}
	\setlength{\belowcaptionskip}{10pt}
	\caption{Convergence step and computational time in CIS, GSIS and GSIS-GBT for the supersonic cylinder flow at Ma = 5.0 and various Kn.
 The simulations are conducted on a parallel computer of AMD Epyc 7742 processor with 20 cores.
 }
	\setlength{\tabcolsep}{3.6mm}
	\begin{threeparttable}
		\begin{tabular}{cccccccc}
			\hline \hline & \multicolumn{2}{c}{ CIS } & \multicolumn{2}{c}{ GSIS } & \multicolumn{2}{c}{ GSIS-GBT } & Speedup \\
			\cline { 2 - 7 } $\mathrm{Kn}$ & Steps & Times $(\mathrm{s})$ & Steps$^{\ast}$ & Times$^{\dagger}$ $(\mathrm{s})$ & Steps$^{\ast}$ & Times$^{\dagger}$ $(\mathrm{s})$ & Ratio$^{\ddagger}$ \\
			\hline 0.1 & 462 & 356 & 91 & 138.2 & 73 & 109.3 & 3.3/1.3 \\
			0.01 & 3952 & 3082 & 133 & 199.6 & 43 & 64.5 & 47/3.1 \\
			0.001 & 7865 & 5876 & 184 & 280.5 & 34 & 61.1 & 98/4.6 \\
			\hline \hline
		\end{tabular}
		\begin{tablenotes}    
			\footnotesize     
			\item[$^{\ast}$]The steps contains the total steps for the 10-step CIS iterations and the subsequent GSIS iterations to convergence. Each iteration of GSIS comprises solving one time mesoscopic equation and 200 times of macroscopic inner iterations.
			\item[$^{\dagger}$] The computational time contains the total time for the 10-step CIS iterations and the subsequent GSIS iterations to convergence.
			\item[$^{\ddagger}$] The speed-up ratio comprises the acceleration ratio of GSIS-GBT relative to CIS and the acceleration ratio relative to the original  GSIS.
		\end{tablenotes}
		\label{tb1}
	\end{threeparttable}
\end{table}

\begin{figure}[t]
	\centering
	\includegraphics[width=0.38\textwidth]{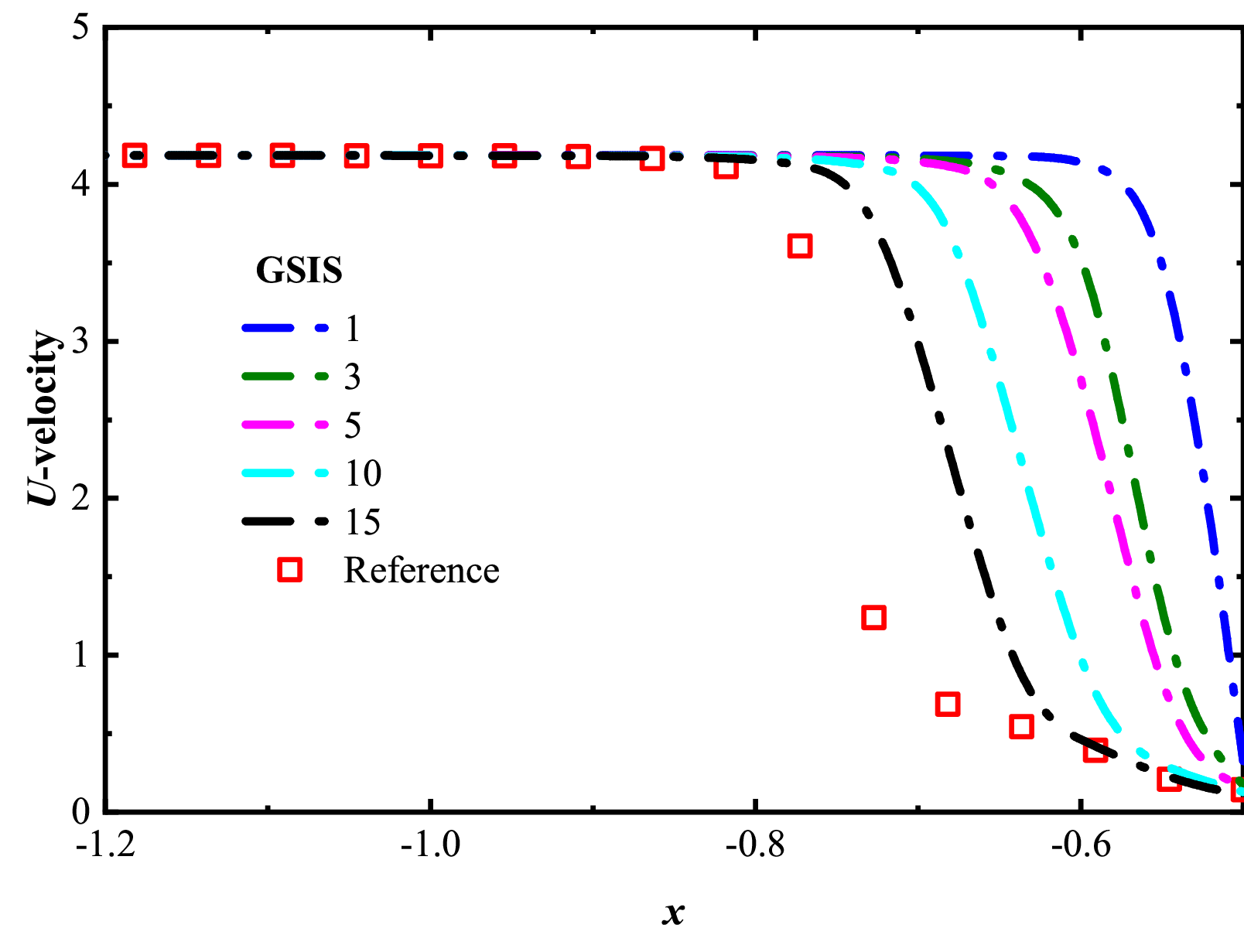}
	\quad 
    \includegraphics[width=0.38\textwidth]{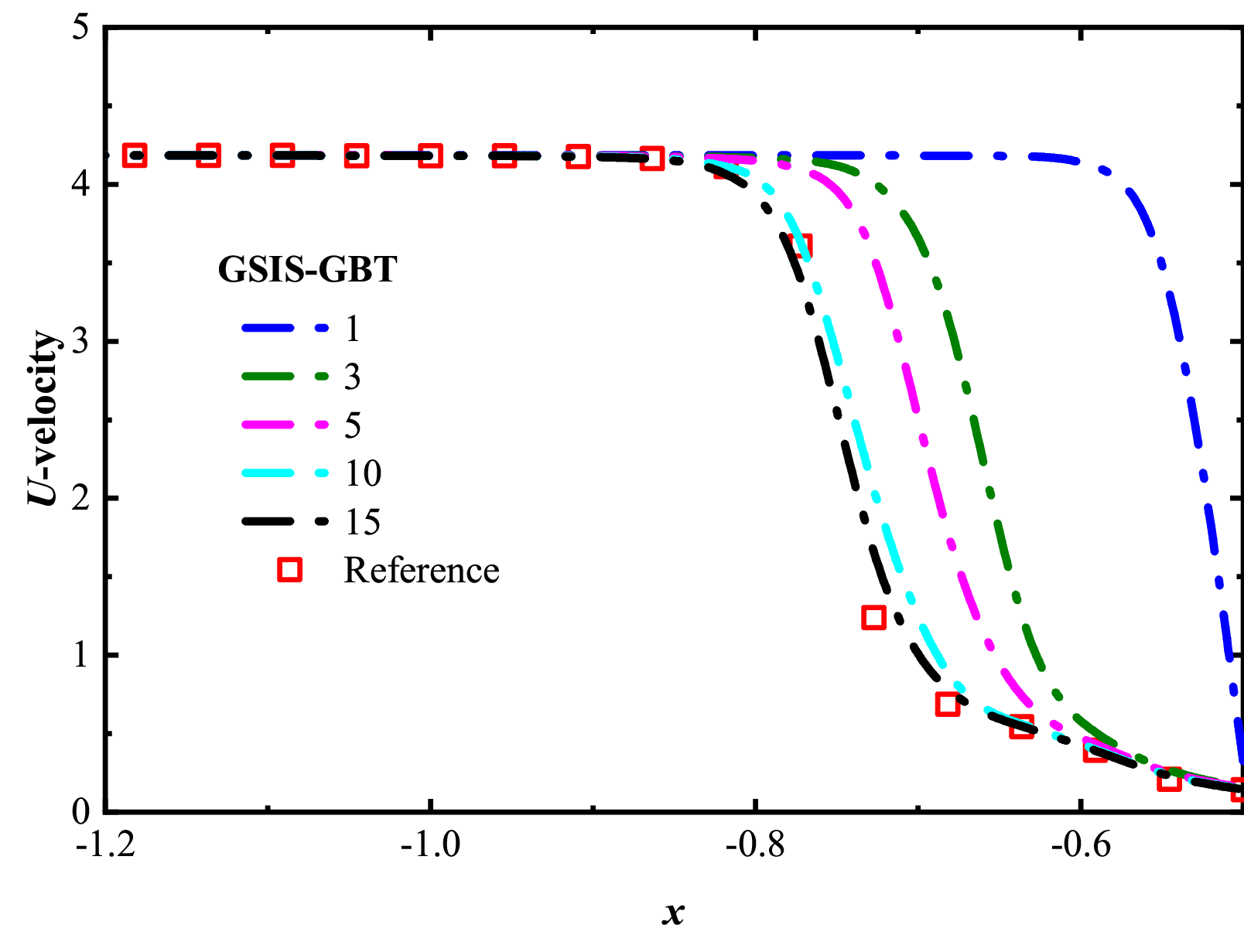}\\
	\includegraphics[width=0.38\textwidth]{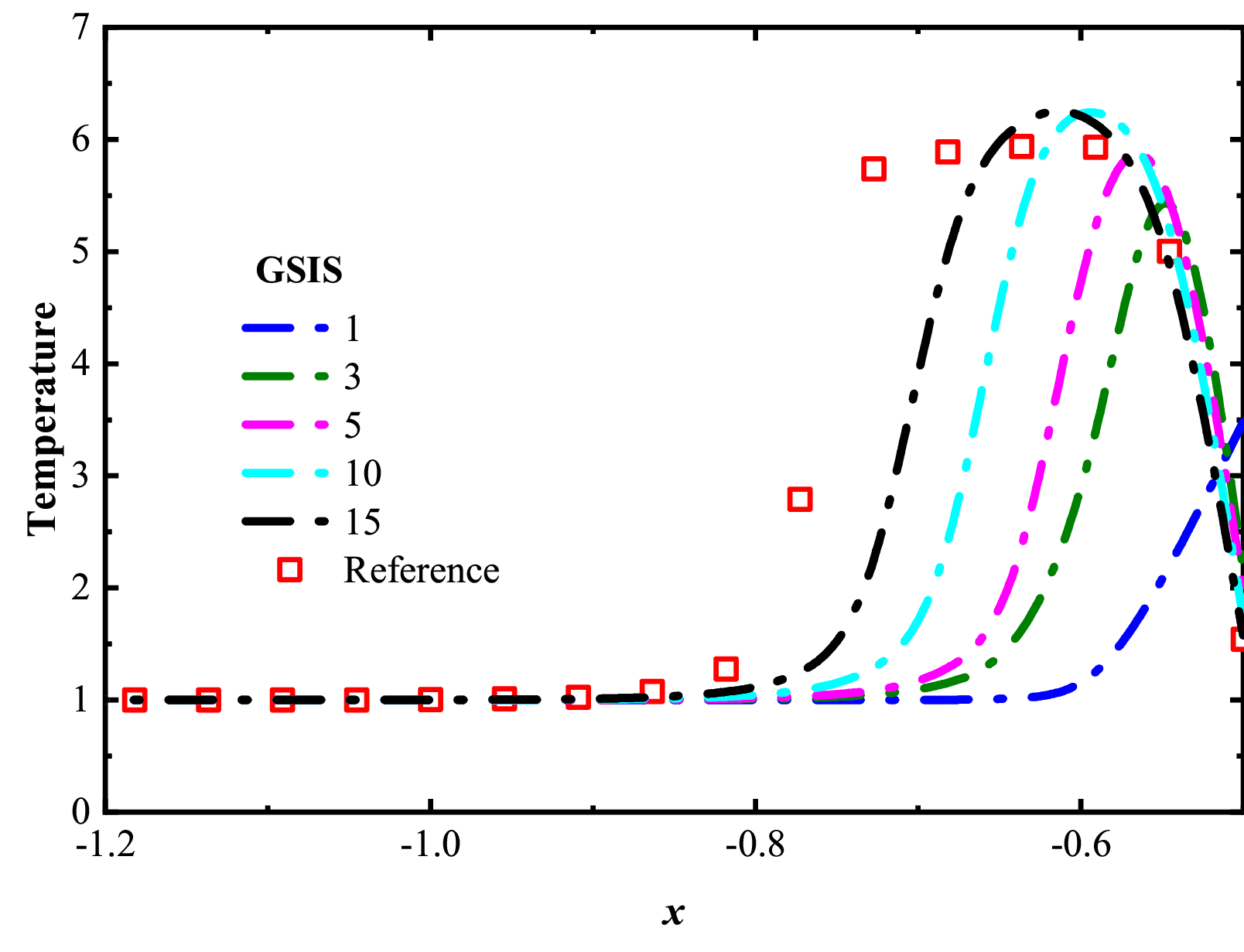}
	\quad
    \includegraphics[width=0.38\textwidth]{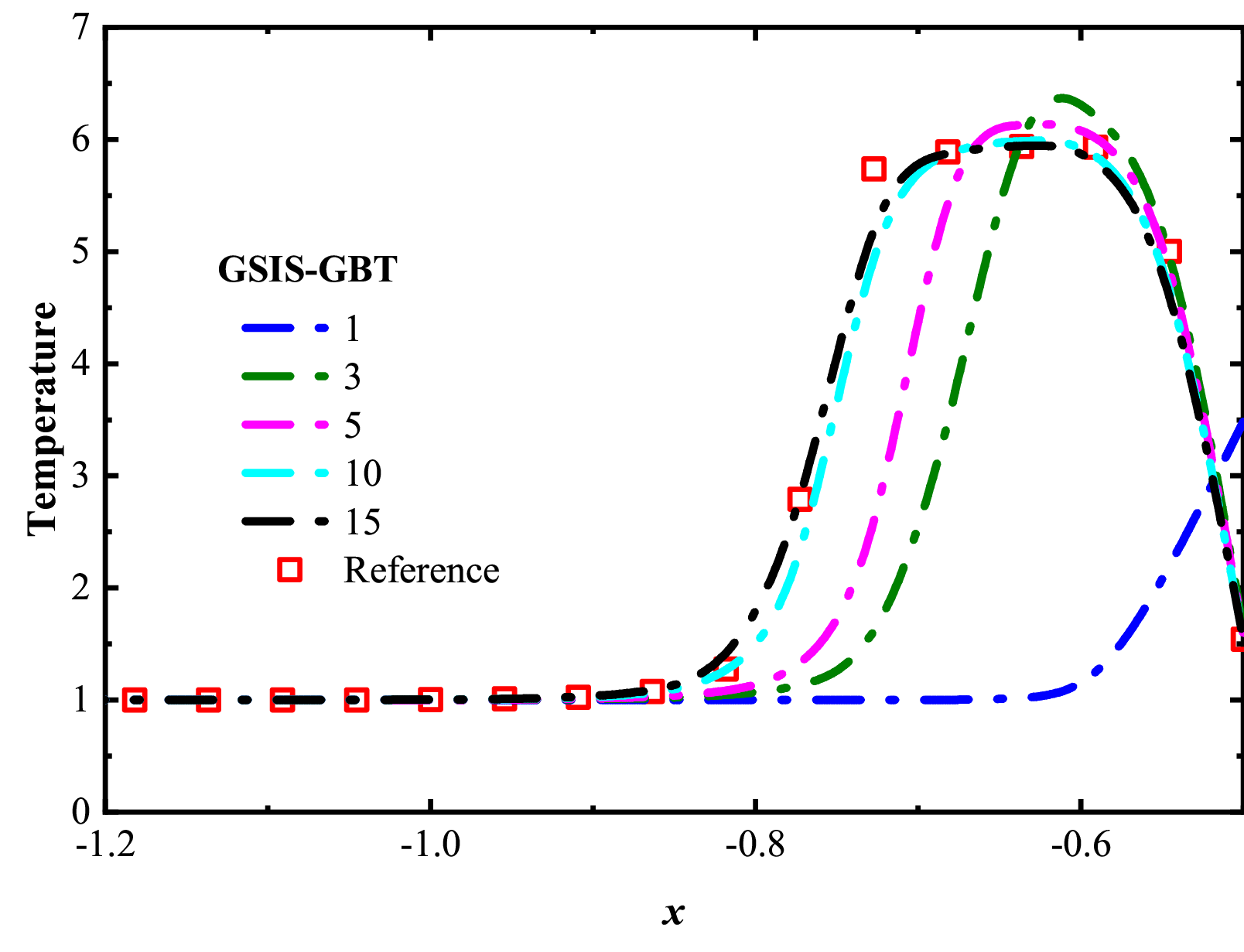}   
	\caption{Density and temperature profiles at different iteration steps, obtained from the GSIS (lst column) and GSIS-GBT (2nd column) for the supersonic cylinder flow when Ma = 5.0 and Kn = 0.01.}
	\label{Fig8}
\end{figure}

Table~\ref{tb1} presents the convergence steps for hypersonic cylinder flows. Both GSIS~\cite{zeng_general_2023} and GSIS-GBT exhibit accelerated convergence compared to the CIS across various Knudsen numbers, where the acceleration becomes more pronounced as the Knudsen number decreases. For instance, at Kn = 0.01, CIS require nearly an hour of computational time, whereas GSIS-GBT converges in approximately one minute. This acceleration can be attributed to two factors: (i) the macroscopic inner iterations facilitates the fast exchange of information within the flow field, and (ii) the GBT in the macroscopic equations accelerates the evolution of the flow field at the boundaries, resulting in a nearly five-fold acceleration over the GSIS. In Fig.~\ref{Fig8}, we present a comparison of velocity and temperature profiles between the GSIS and GSIS-GBT at the same iteration steps. It is seen that the flow evolution near the wall ($x = -0.5$) is more rapid in GSIS-GBT. Specifically, after the 10-th iteration, results from GSIS-GBT closely approach the reference solution, while GSIS exhibits a larger discrepancy to the reference results even after 15 iterations. 

\begin{figure}[th]
	\centering
	\includegraphics[width=0.4\textwidth]{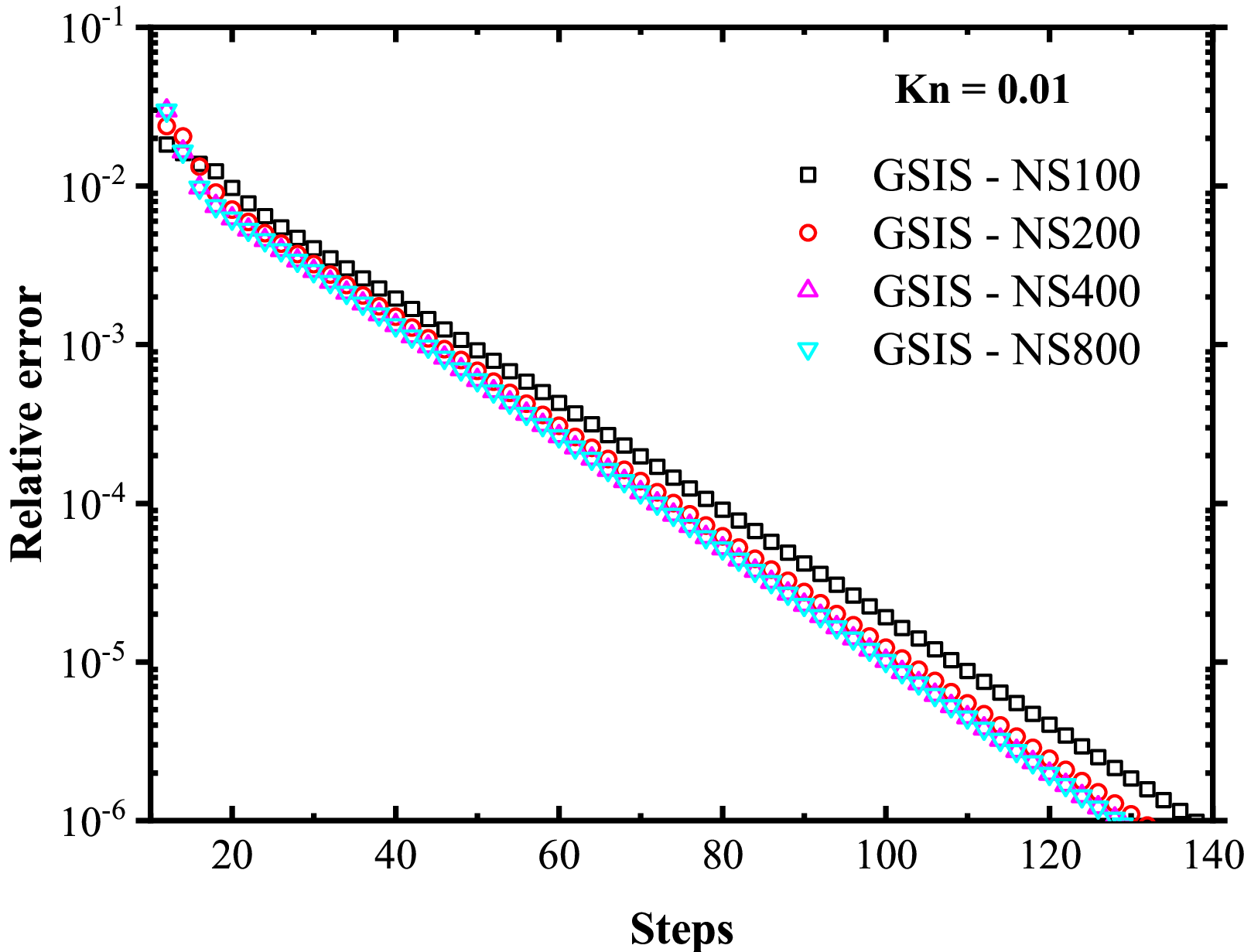}
 \quad
	\includegraphics[width=0.4\textwidth]{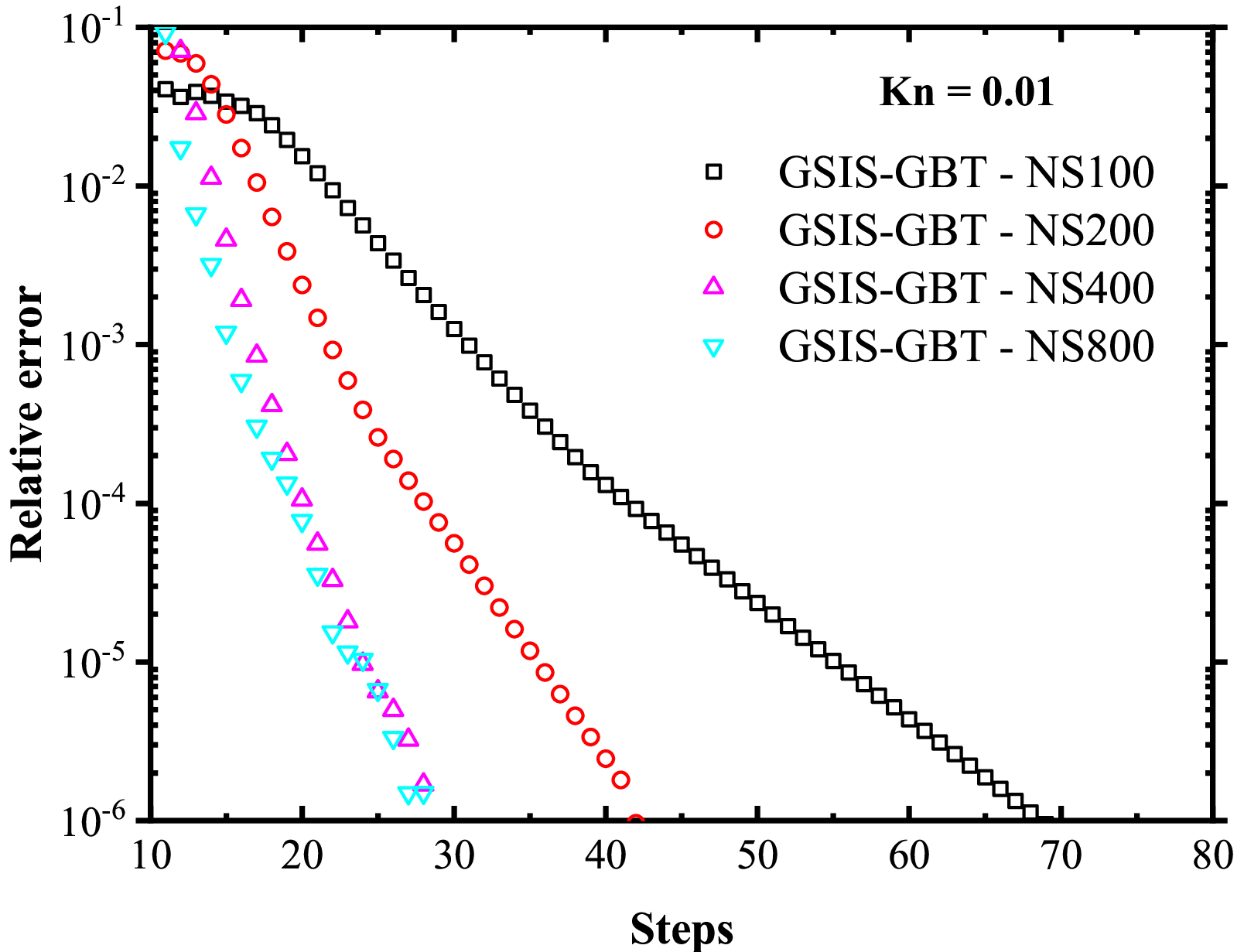}
	\caption{Convergence history of GSIS (left) and GSIS-GBT (right) in the simulation of planar Fourier heat transfer at Kn = 0.01 with different numbers of macroscopic inner iterations, e.g., NS800 means that the maximum value of $m$ in Eq.~\eqref{eq2.2.5} is 800.}
	\label{Fig9}
\end{figure}

In each iteration of {the kinetic equation in GSIS, the macroscopic synthetic equations need to be solved $m$ times} to accelerate the exchange of information within the flow field, see Eq.~\eqref{eq2.2.5}. Intuitively, increasing the value of $m$ boosts the convergence of GSIS. However, this is not be the case in practice. Taking the case {of $\text{Kn} =0.01$} as an example, we compare the convergence steps of the GSIS and GSIS-GBT under different numbers of macroscopic iterations. Figure~\ref{Fig9} shows that, increasing the solution steps of the NS equations in GSIS~\cite{zeng_general_2023} has a negligible effect on accelerating convergence: by increasing the inner iterations from 100 to 400 steps, the overall convergence steps decrease only from 138 to 129. 
However, when GBT is applied to the boundaries, increasing the inner iteration steps significantly reduces the convergence steps. To be specific, when the inner iterations increases from 100 to 400 steps, the convergence steps of GSIS-GBT decrease from 68 to 28, resulting in an acceleration of convergence by nearly 2.5 times. 
We believe this phenomenon stems from the constrained convergence of the macroscopic equations in the GSIS due to the limitations imposed by integral boundary conditions. With the application of GBT, the macroscopic equations can acquire more real-time information from the boundary, thus reducing the number of outer iteration, see $n$ in Eq.~\eqref{eq2.2.2}, which is the most time-consuming part.

\begin{figure}[p]
\centering
\includegraphics[width=0.4\textwidth]{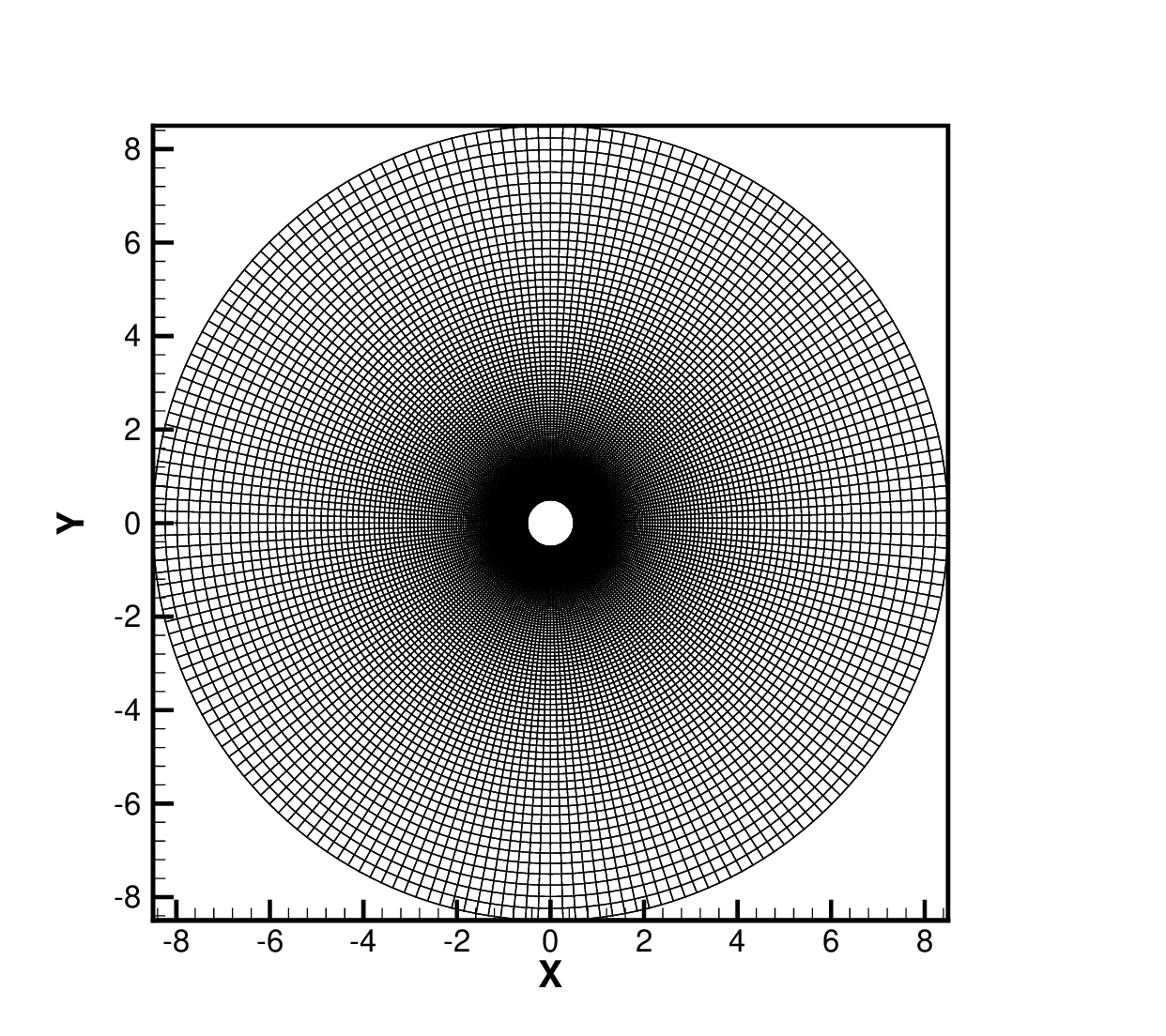}
\includegraphics[width=0.4\textwidth]{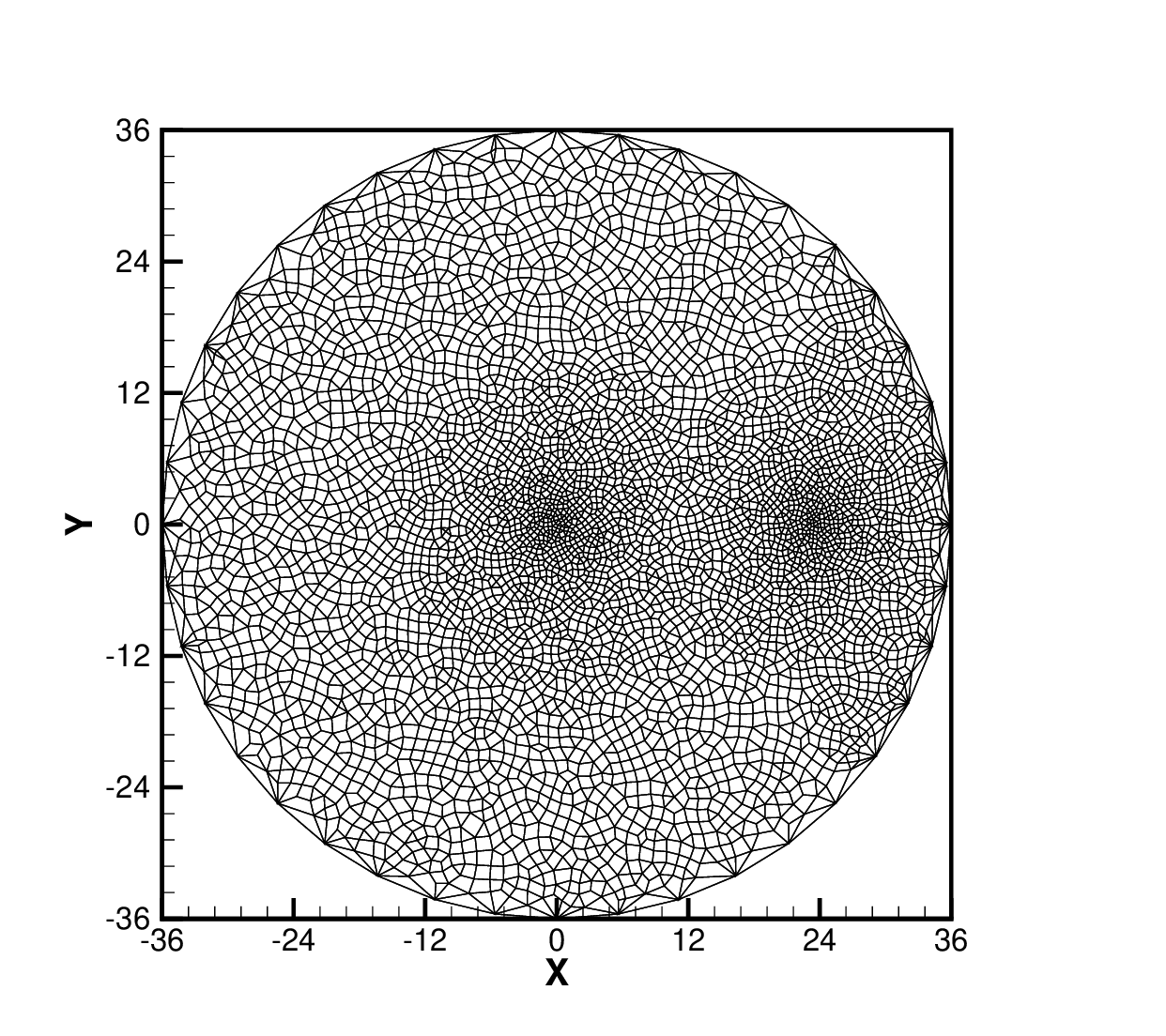}\\
	\includegraphics[width=0.32\textwidth]{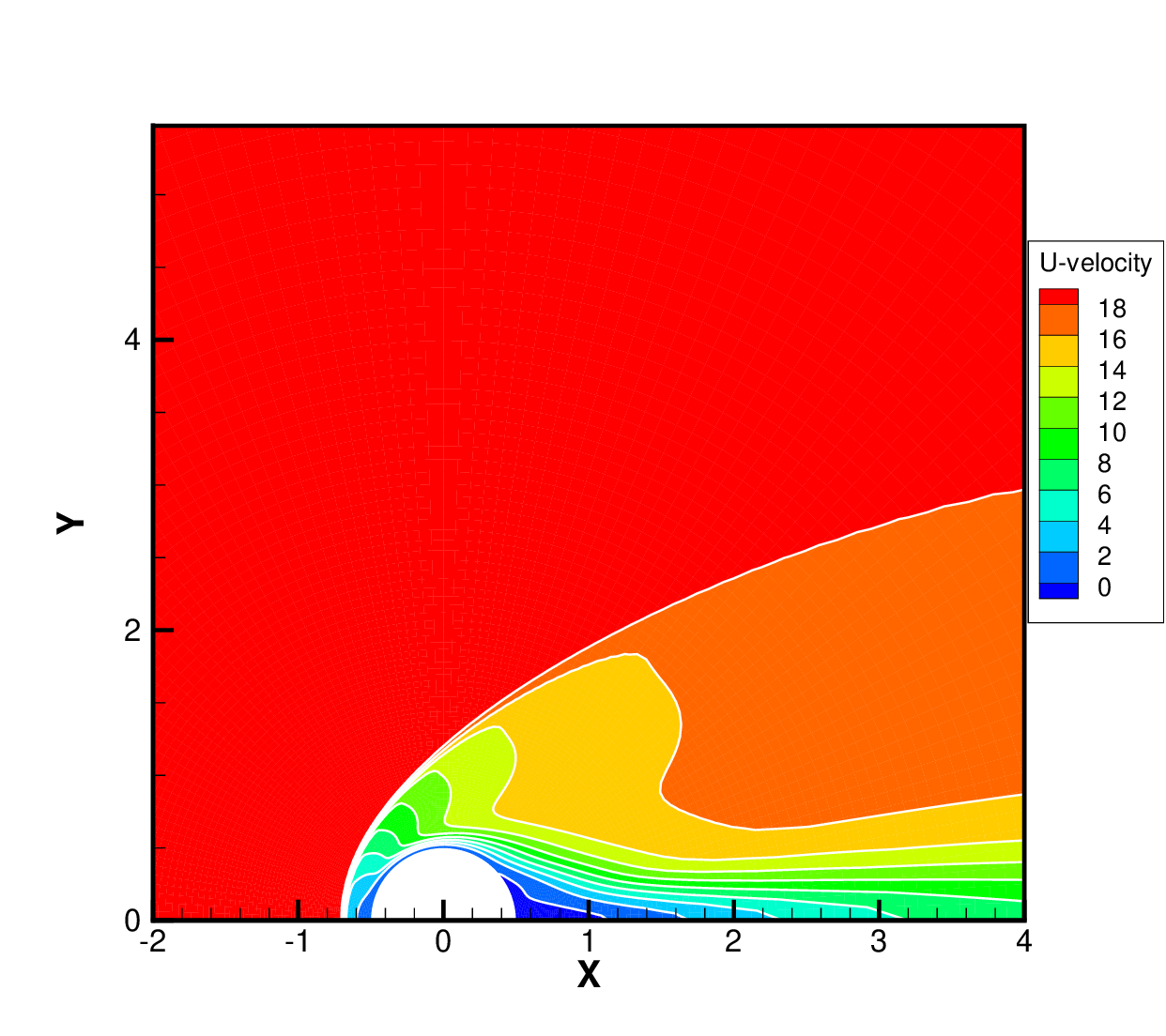}
	\includegraphics[width=0.32\textwidth]{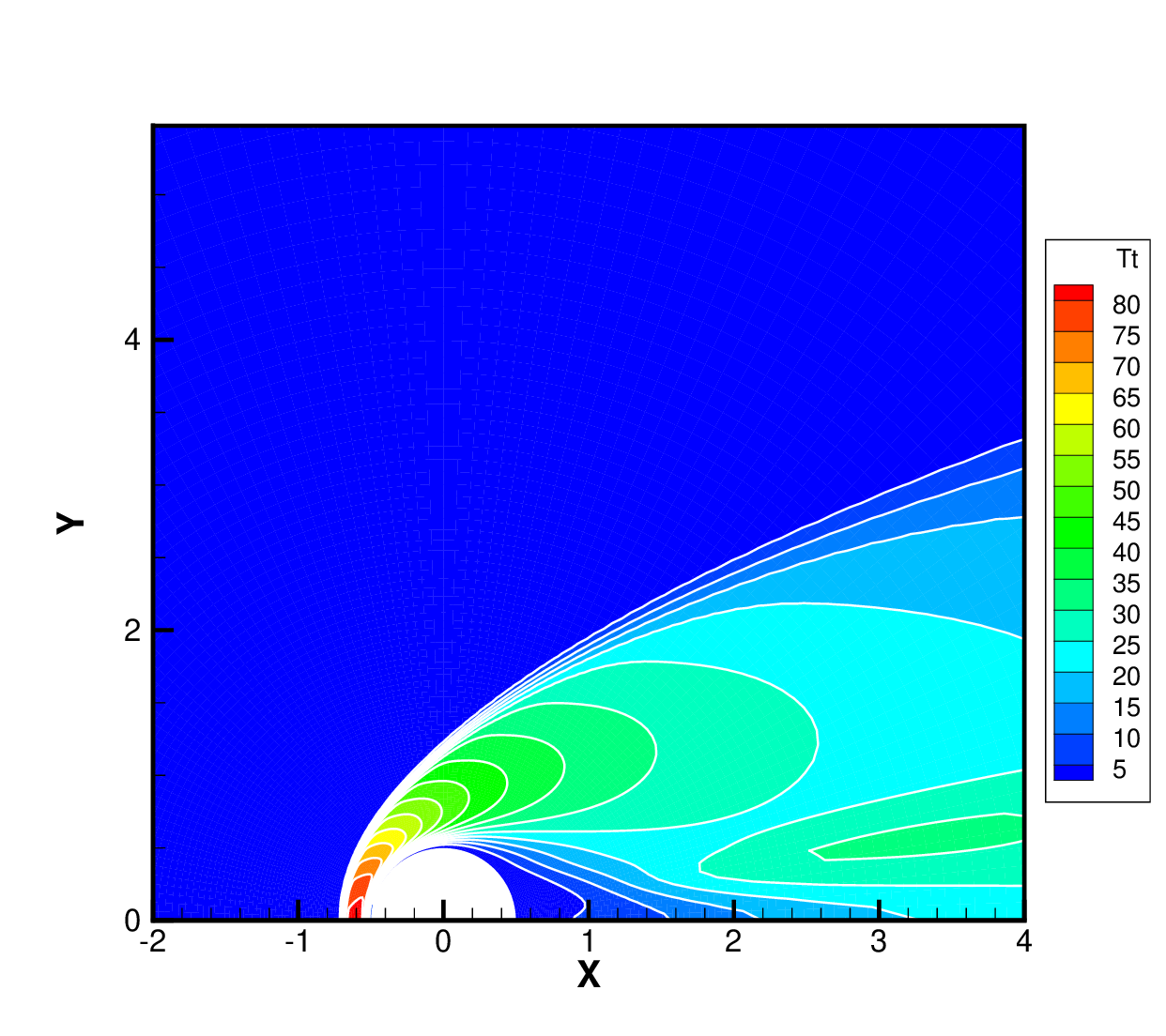}
	\includegraphics[width=0.32\textwidth]{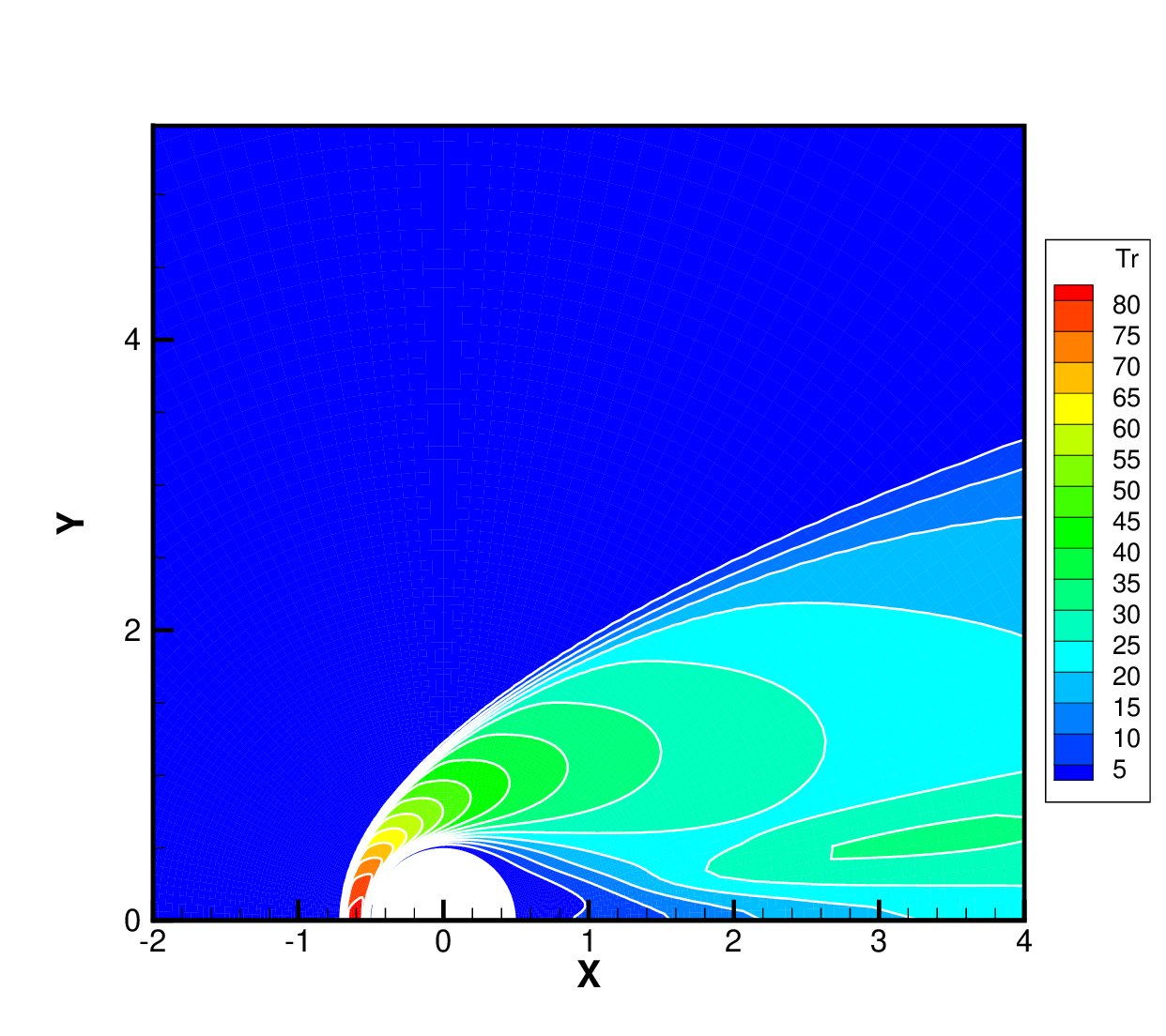}
	\includegraphics[width=0.32\textwidth]{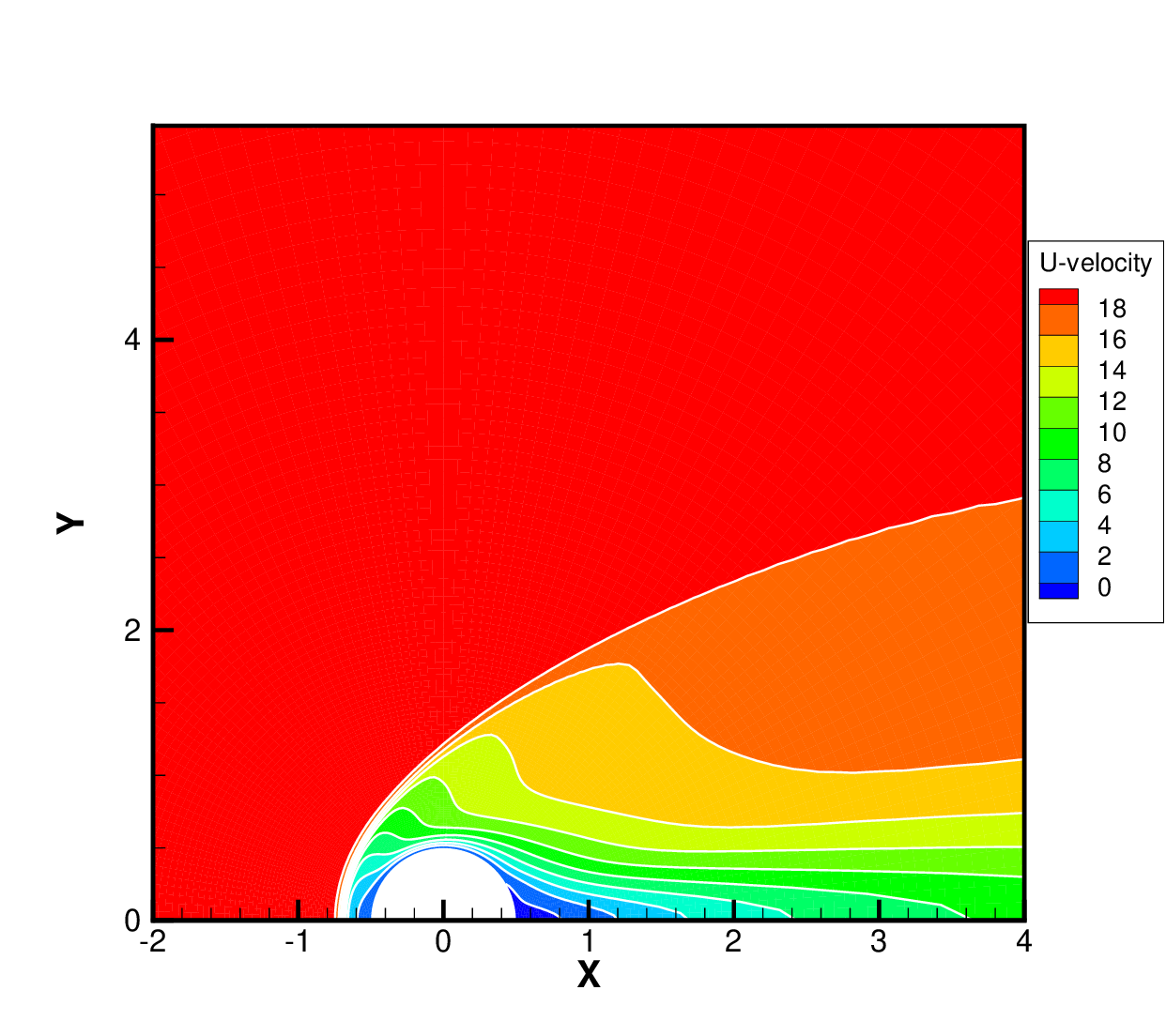}
	\includegraphics[width=0.32\textwidth]{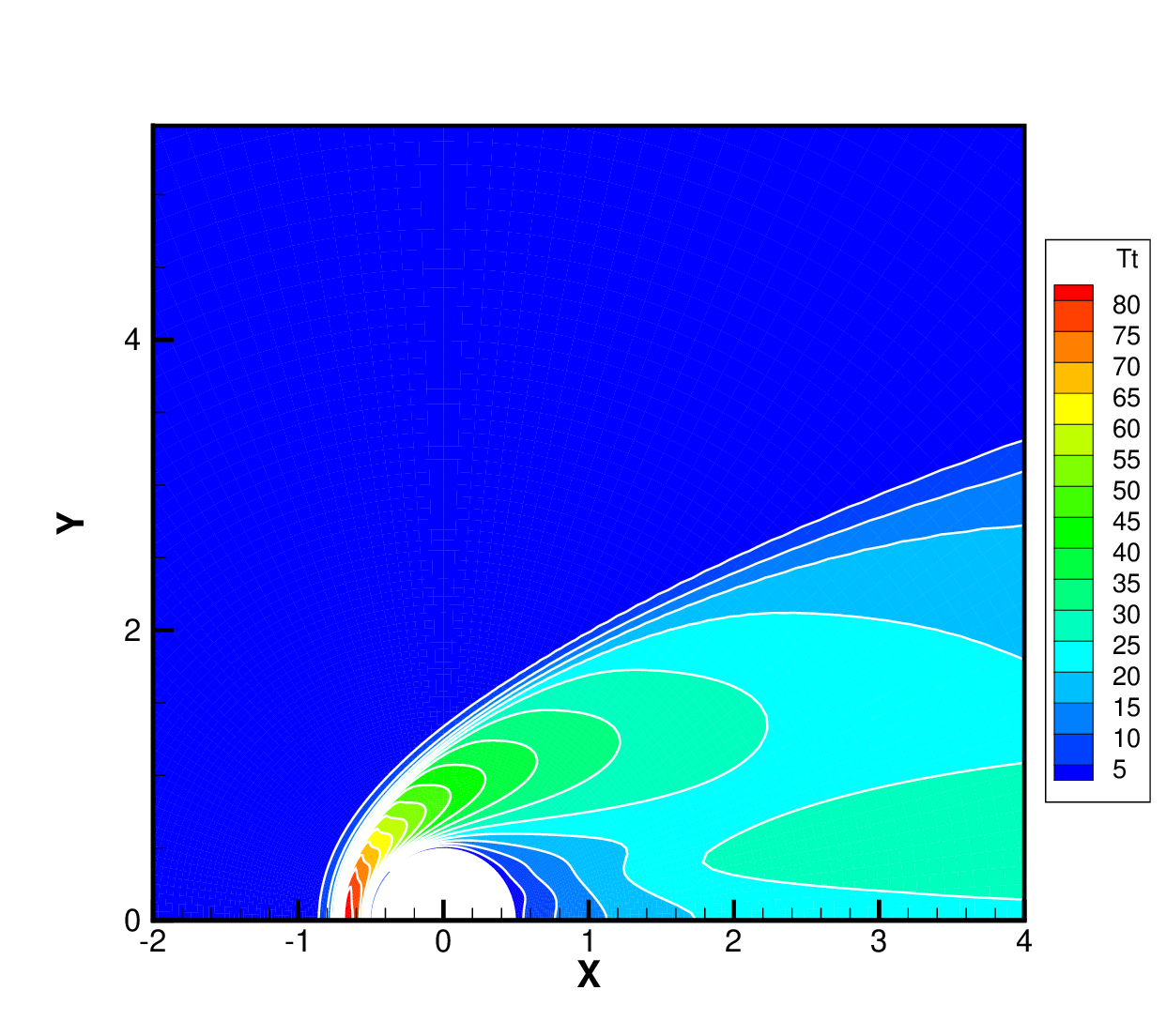}
	\includegraphics[width=0.32\textwidth]{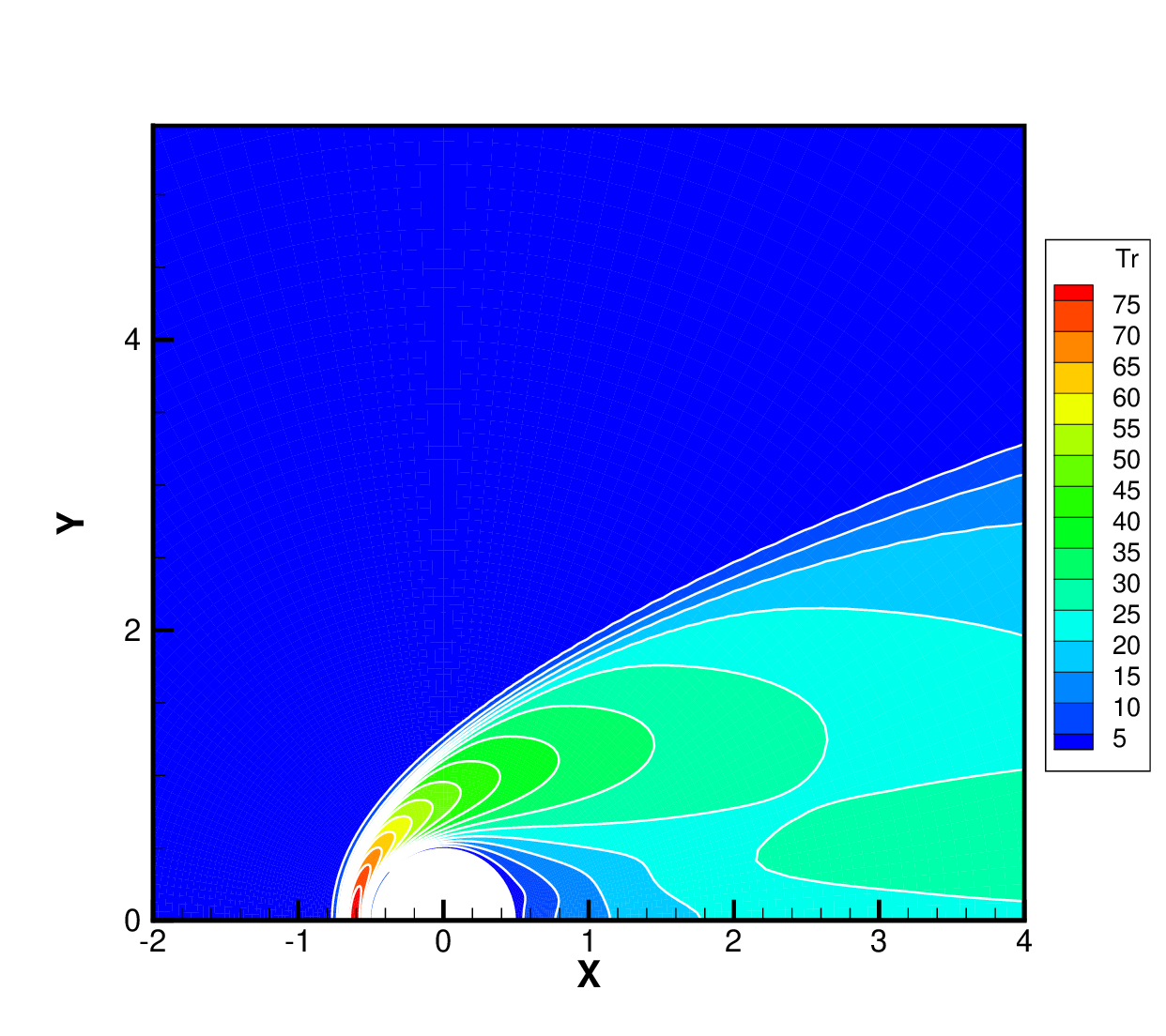}
	\includegraphics[width=0.33\textwidth]{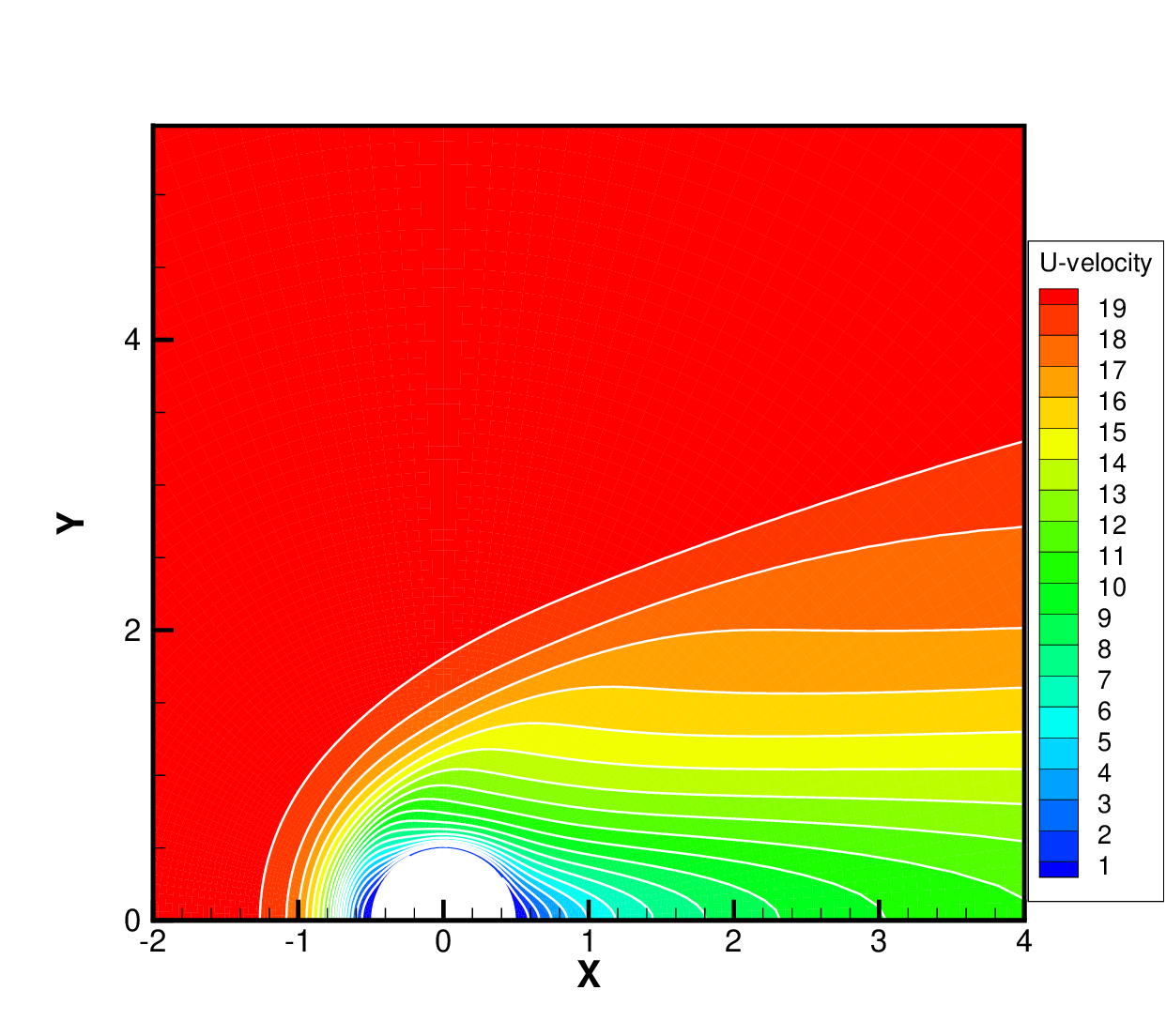}
	\includegraphics[width=0.33\textwidth]{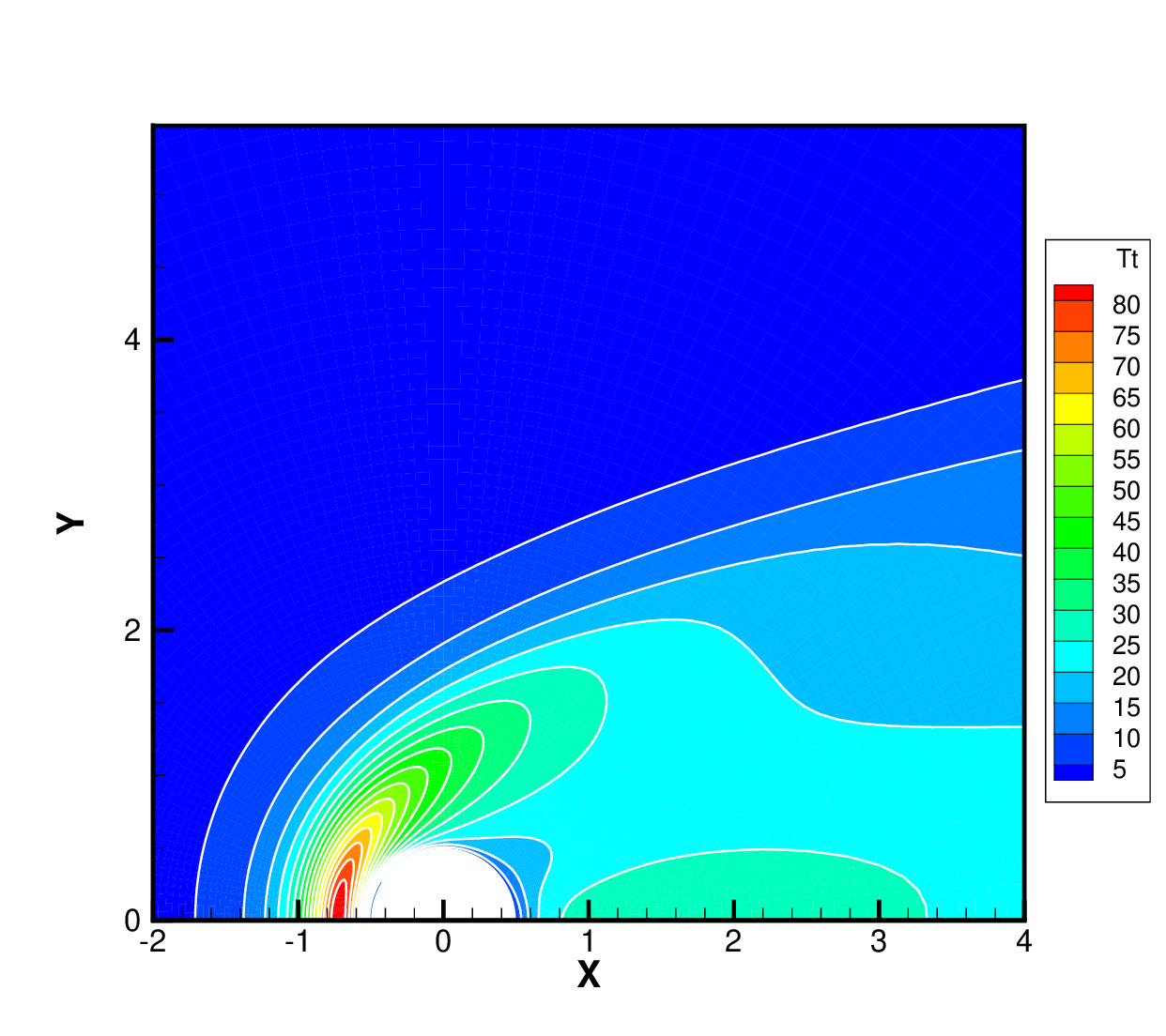}
	\includegraphics[width=0.32\textwidth]{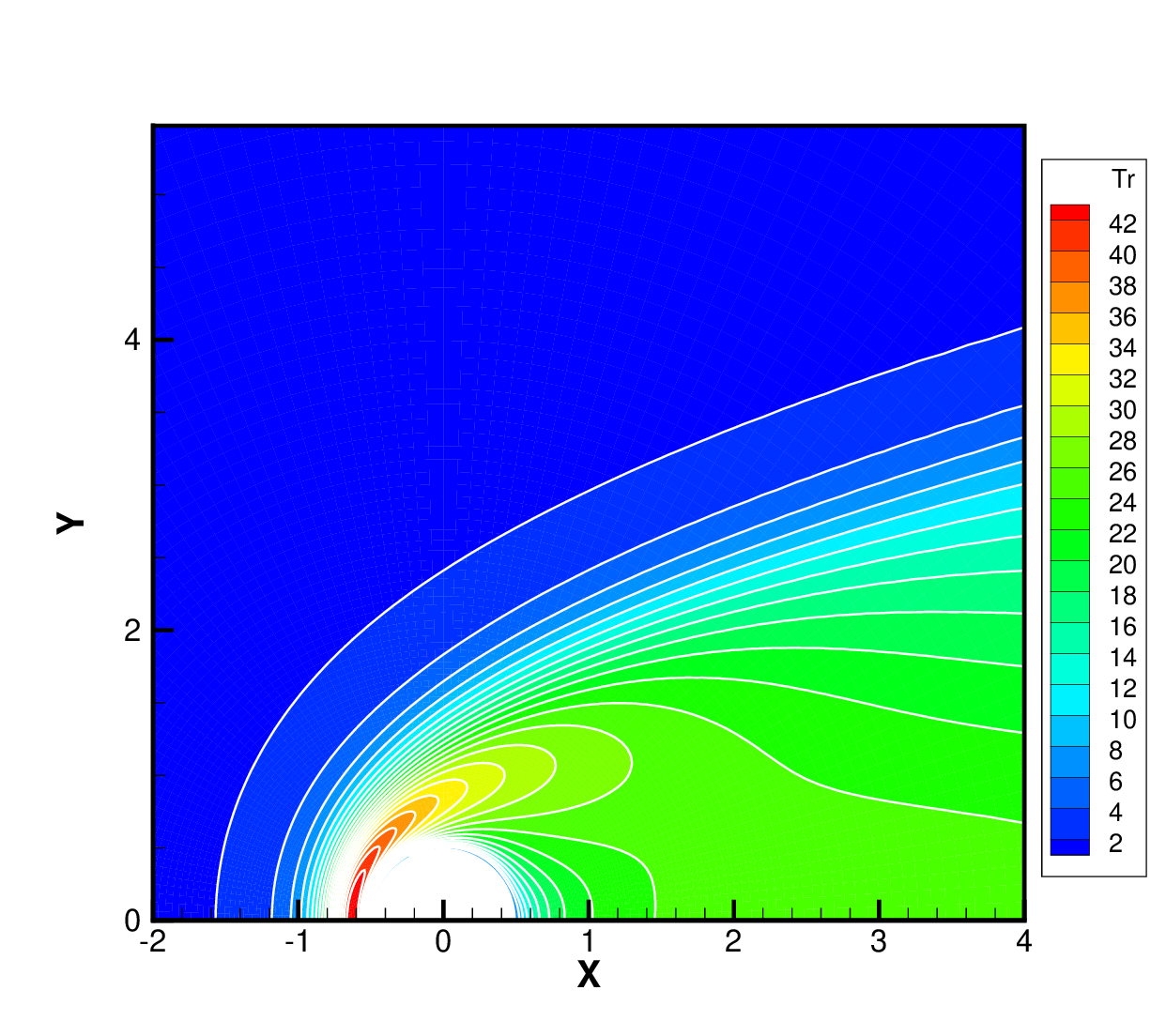}    
	\caption{
 First row: physical and velocity mesh in the supersonic cylinder flow at Ma = 20. (2nd, 3rd and 4th rows) Contours of velocity, translational temperature and rotational temperature when Kn = 0.001 (2nd row), 0.01 (3rd row) and 0.1 (4th row).
 }
	\label{Fig11}
\end{figure}

Finally, the hypersonic flow around a cylinder with the Mach number of 20 is simulated to investigate the efficiency and stability of GBT. Figure~\ref{Fig11} illustrates the meshes in physical and velocity spaces. To minimize the number of discrete meshes in the vast velocity space of $\left[-36, 36 \right]^{2}$, an technique of unstructured discrete velocity mesh~\cite{yuan_multi-prediction_2021} is introduced, with 5550 discrete velocity meshes and local refinement at the positions where the molecular velocities are equal to 0 and 20. In Fig.~\ref{Fig11}, the velocity and temperature contours obtained from the GSIS-GBT are presented at Kn=0.001, 0.01, and 0.1. In contrast to the hypersonic flow with Ma = 5, the case of Ma = 20 not only results in higher wall temperature and larger velocity variations, but also higher density variations, leading to significantly higher local Knudsen numbers in the wake region. Even at Kn = 0.01, strong rarefied flow occurs within the wake region, posing a considerable challenge for numerical algorithms due to the multiscale nature of the flow~\cite{liu_unified_2020}. 

Table~\ref{tb2} provides a comparison of the convergence steps in the CIS, GSIS, and GSIS-GBT. 
All methods employed the flow field obtained from 4000 steps of the first-order Euler equation for initialization. Prior to iteration, GSIS and GSIS-GBT adopt 20 steps of CIS iterations to initialize the VDFs. Due to the numerical dissipation and slow convergence, CIS costs over 12 hours to converge at Kn = 0.01, while GSIS takes  around 1 hour. Furthermore, benefiting from the independent evolution of macroscopic boundary conditions, GSIS-GBT finishes the computation by 51 iterations, and within 5 minutes. These results demonstrate the excellent stability and significant acceleration of GSIS-GBT in hypersonic flow.

\begin{table}[th]
	\centering
	\setlength{\abovecaptionskip}{10pt}
	\setlength{\belowcaptionskip}{10pt}
	\caption{Convergence step and computational time in CIS, GSIS and GSIS-GBT for the supersonic cylinder flow at Ma = 20.0 and various Kn. The simulations are conducted on a parallel computer of AMD Epyc 7742 with 20 cores.
 }
	\setlength{\tabcolsep}{3.6mm}
	\begin{threeparttable}
		\begin{tabular}{cccccccc}
			\hline \hline & \multicolumn{2}{c}{ CIS } & \multicolumn{2}{c}{ GSIS } & \multicolumn{2}{c}{ GSIS-GBT } & Speedup \\
			\cline { 2 - 7 } $\mathrm{Kn}$ & Steps & Times $(\mathrm{s})$ & Steps$^{\ast}$ & Times$^{\dagger}$ $(\mathrm{s})$ & Steps$^{\ast}$ & Times$^{\dagger}$ $(\mathrm{s})$ & Ratio$^{\ddagger}$ \\
			\hline 0.1 & 1778 & 5569 & 330 & 1647 & 83 & 490 & 11/3.3 \\
			0.01 & 15209 &  46615 & 655 & 3340 & 51 & 294 & 158/11 \\
			0.001 & - &  - & 820 & 3736 & 49 & 284 & -/13.2 \\
			\hline \hline
		\end{tabular}
		\begin{tablenotes}    
			\footnotesize     
			\item[$^{\ast}$]The steps contains the total steps for the 20-step CIS iterations and the subsequent GSIS iterations to convergence. Each iteration of GSIS comprises solving one time mesoscopic equation and 400 times of macroscopic inner iterations.
			\item[$^{\dagger}$] The computational time contains the total time for the 20-step CIS iterations and the subsequent GSIS iterations to convergence.
			\item[$^{\ddagger}$] The speed-up ratio comprises the acceleration ratio of GSIS-GBT relative to CIS and the acceleration ratio relative to the original  GSIS.
		\end{tablenotes}
		\label{tb2}
	\end{threeparttable}
\end{table}

\subsection{Pressure-driven flow in a variable-diameter circular pipe}
\label{sec4-3}

\begin{figure}[p]
	\centering
	\subfigure[global mesh]{\includegraphics[width=0.4\textwidth]{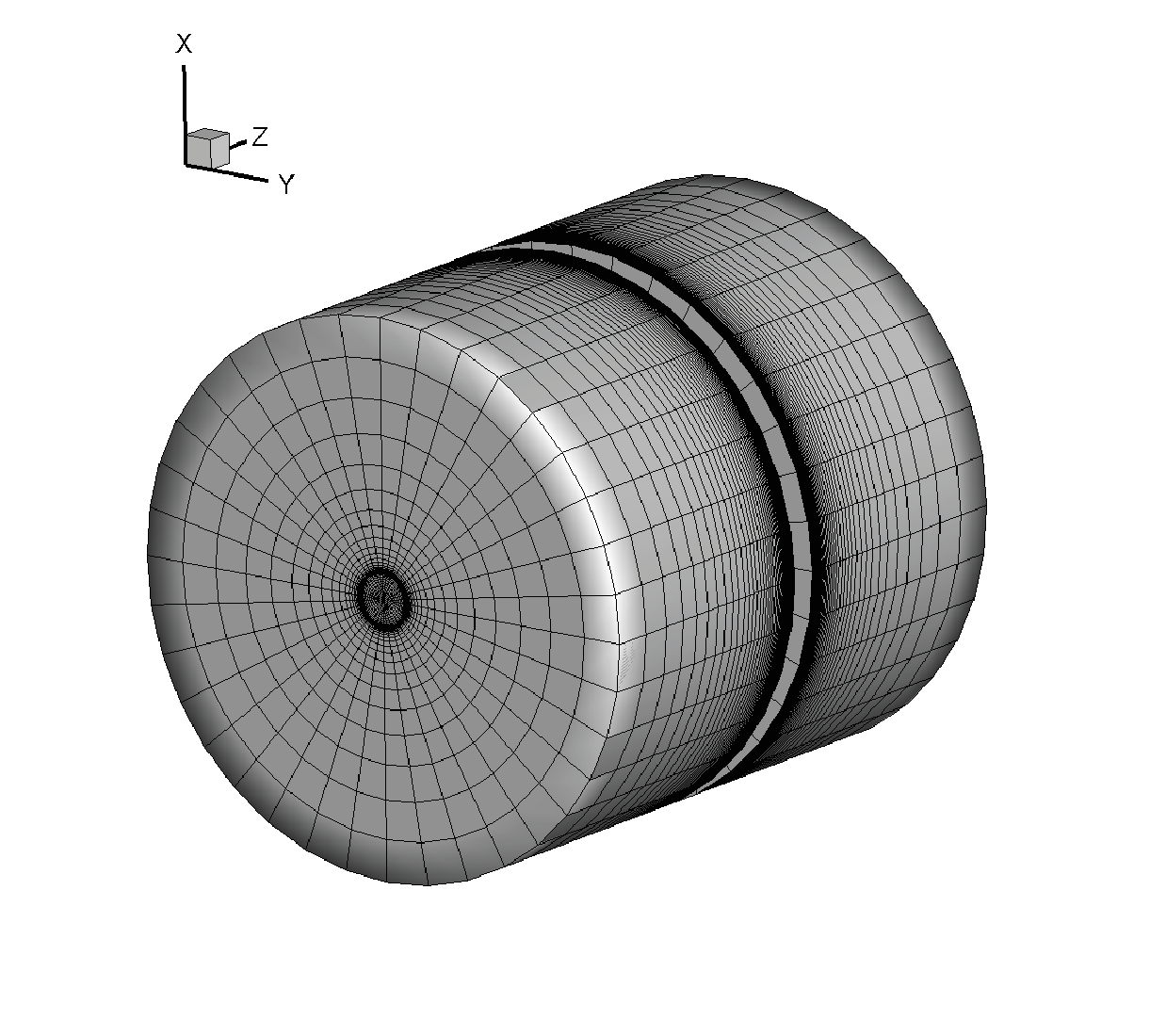}}
	\subfigure[cross-sectional mesh when $y=0$]{\includegraphics[width=0.4\textwidth]{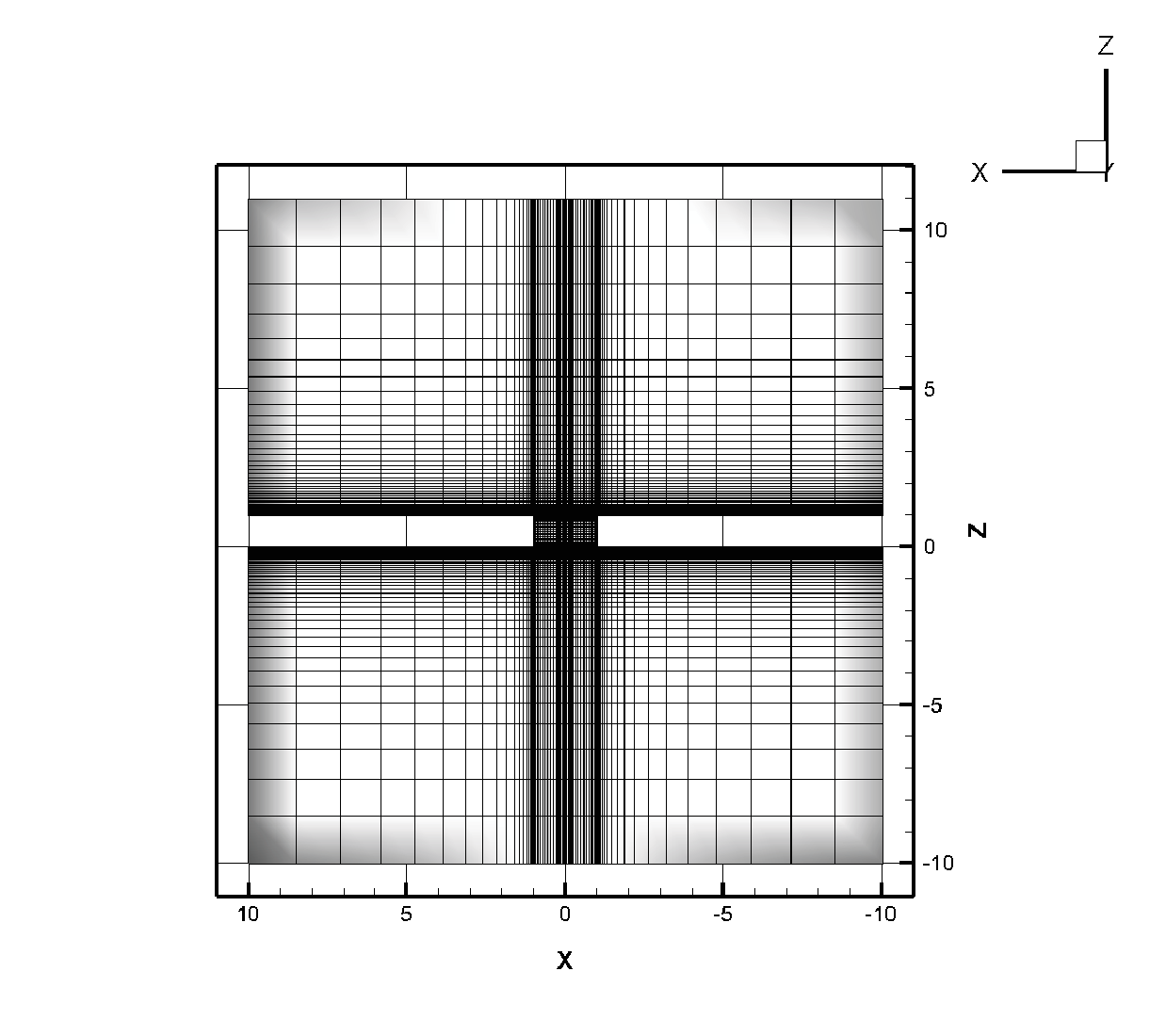}}
 \subfigure[$\delta_{rp} = 1000$]{
	\includegraphics[width=0.48\textwidth]{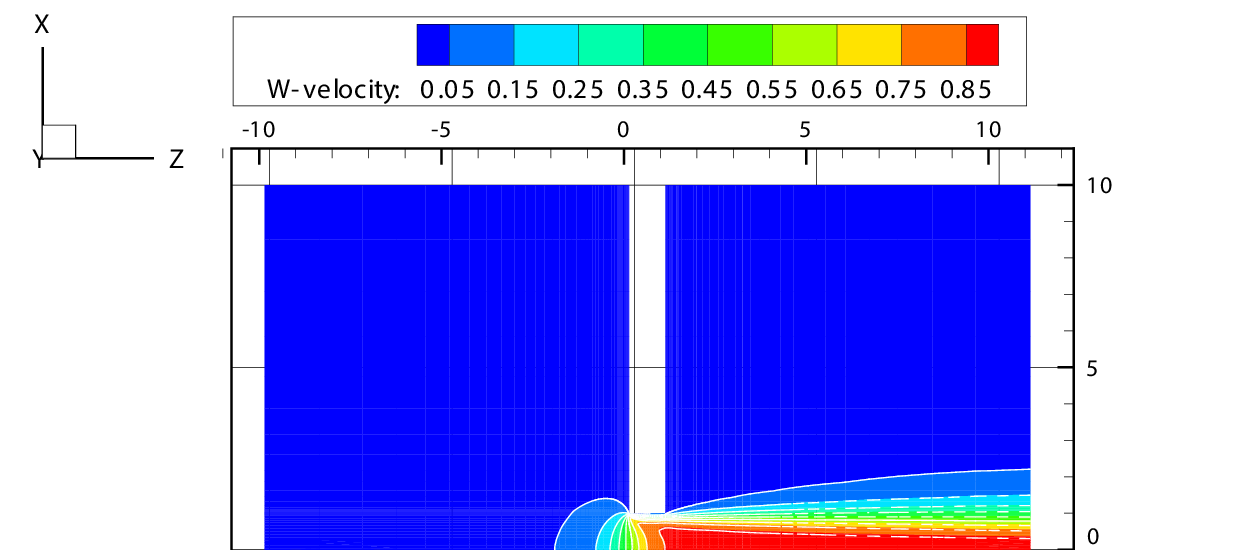}}
   \subfigure[$\delta_{rp} = 100$]
   { \includegraphics[width=0.48\textwidth]{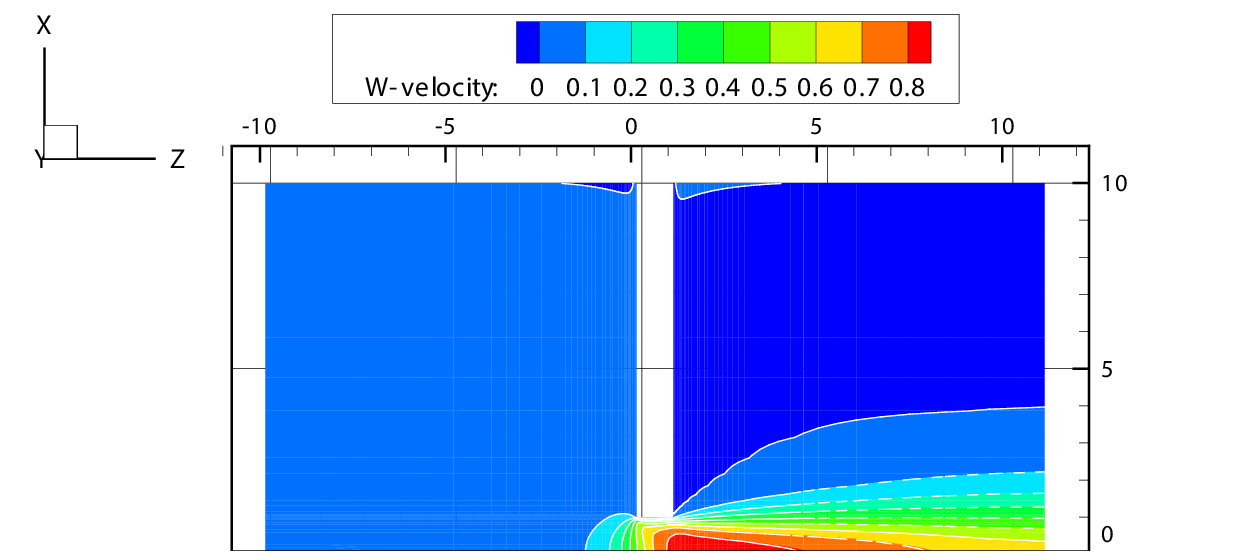}}\\
	\subfigure[$\delta_{rp} = 10$]
 {\includegraphics[width=0.48\textwidth]{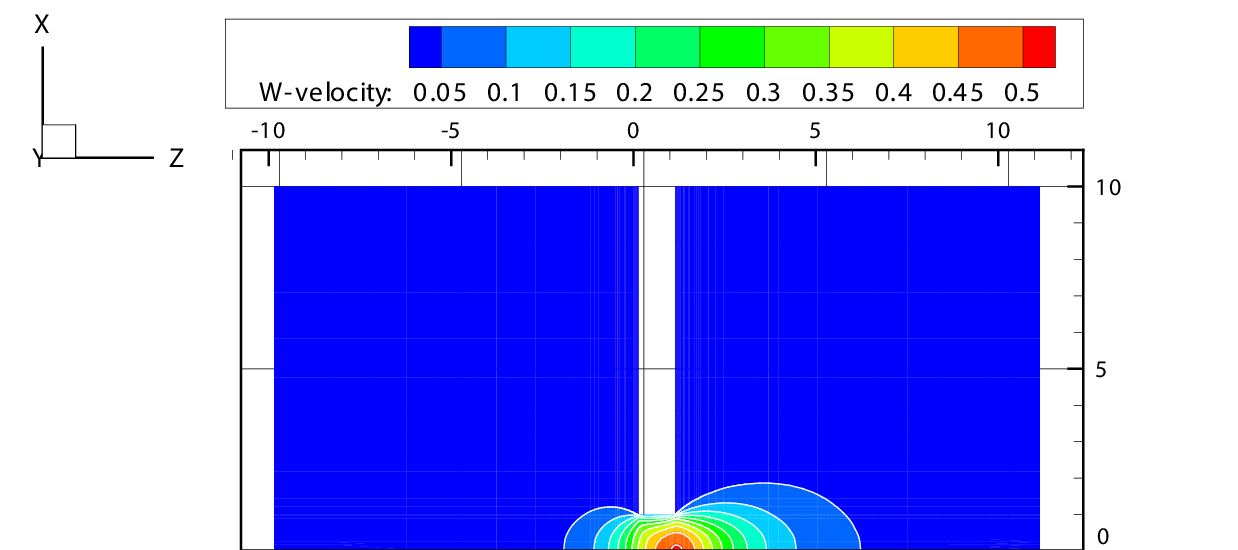}}
   \subfigure[$\delta_{rp} = 1$]
   { \includegraphics[width=0.48\textwidth]{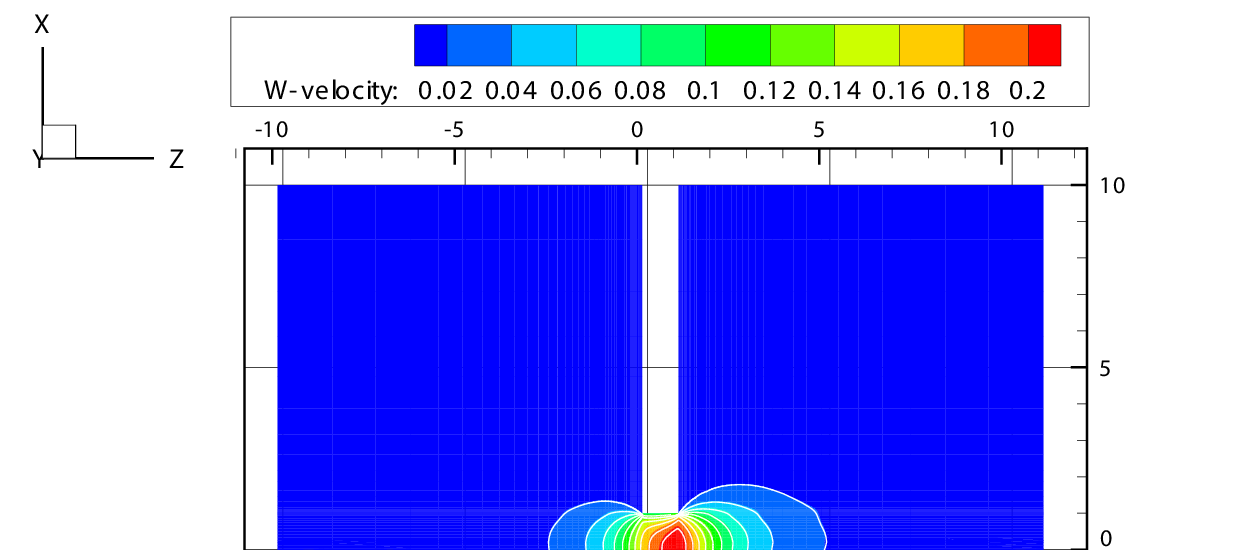}}
	\caption{(a,b) Spatial mesh for the pressure-driven pipe flow. (c-f) Contours of $W$-velocity in the pressure-driven pipe flow. Background contour: GSIS-GBT; White dashed lines: GSIS.
 }
	\label{Fig4.3.1}
\end{figure}

To assess the performance of GSIS-GBT in three-dimensional internal flow, the pressure-driven gas flow through a variable-diameter circular pipe is simulated. Figure~\ref{Fig4.3.1} illustrates the global geometry and the cross-sectional mesh. As shown in the figure, two cylindrical pipes with radii and lengths of 5 are connected to a small cylindrical pipe with a radius and length of 1. The diffuse gas-surface interaction is adopted. The reference temperature $T_0$ at all boundaries and the inner domain is set to 1.0. 
Inlet and outlet boundary conditions are applied at planes of $z = 0$ and $z = 11$, corresponding to the inlet and outlet pressures of $p_{in}=1.0$ and $p_{out} = 0.5$, respectively. 
Since it is difficult to give information about macroscopic quantities such as velocity and density in micro-flow problems, the pressure boundary conditions use characteristic relations to determine the variables at the boundary, assuming that the given boundary conditions are the inlet pressure $p_{in}$, the outlet pressure $p_{out}$ and the inlet temperature $T_{in}$. For the pressure inlet boundary conditions, the velocity is obtained by interpolation, the temperature is $T_L= p_{in}/\rho_L$, where the density is
\begin{equation}
    \rho_L = \rho_i + \frac{p_{in}-p_i}{\rho_i a_i},
\end{equation}
where the subscript $i$ is the cell index on the right side of the boundary and $a$ is local sound speed. A similar treatment is applied to the pressure outlet boundary:
\begin{equation}
    \rho_R = \rho_i + \frac{p_{out}-p_i}{\rho_i a_i}.
\end{equation}

\begin{table}[t]
	\centering
	\setlength{\abovecaptionskip}{10pt}
	\setlength{\belowcaptionskip}{10pt}
	\caption{Reduced flow rates computed by different methods for the pressure-driven pipe flow.}
	\setlength{\tabcolsep}{6.1mm}
	\begin{threeparttable}
            \begin{tabular}{ccccccc}
            \hline \hline
            $\delta_{rp}$ & TVD1D~\cite{titarev_construction_2014} & DSMC~\cite{varoutis_simulation_2009}  & GSIS-GBT \\
            \hline 
            1 & 0.419 & 0.405 & 0.397 \\
            10 & 0.871 & 0.866 & 0.858 \\
            100  & 1.292 & 1.29 & 1.292 \\
            1000 & - & 1.40 & 1.405 \\
            \hline \hline
            \end{tabular}
		\label{tb3}
	\end{threeparttable}
\end{table}

For cross-validation with other numerical methods, the hard-sphere model is adopted, where the viscosity index $\omega$ in Eq.~\eqref{eq2.1.3} is set to 0.5 and the rotational collision number set to 0.001 to recover the hard-sphere model. The reference length is chosen as the radius of the middle circular pipe $R = 1$. Gas flow is simulated under various rarefaction parameters 
\begin{equation}
    \delta_{rp}=\frac{\sqrt{\pi}}{2\text{Kn}}.
\end{equation}
When the rarefaction parameter is greater than or equal to 10, the molecular velocity space is discretized by a 16 $\times$ 16 $\times$ 16 points by Gauss-Hermite quadrature. In the case of a rarefaction parameter of 1, a uniform Newton-Cotes quadrature with 41 $\times$ 41 $\times$ 41 points is employed to discretize the molecular velocity space in the range of $\left[-6, 6\right]^{3}$. The computational domain is discretized by 143,370 hexahedral cells.

\begin{table}[th]
	\centering
	\setlength{\abovecaptionskip}{10pt}
	\setlength{\belowcaptionskip}{10pt}
	\caption{Convergence step and computational time in CIS, GSIS and GSIS-GBT for the pressure-driven pipe flow at various $\delta_{rp}$.
 The simulations are conducted on a parallel computer of AMD Epyc 7742 processor with 600 cores. 
 }
	\setlength{\tabcolsep}{3.6mm}
	\begin{threeparttable}
		\begin{tabular}{cccccccc}
			\hline \hline & \multicolumn{2}{c}{ CIS } & \multicolumn{2}{c}{ GSIS } & \multicolumn{2}{c}{ GSIS-GBT } & Speedup \\
			\cline { 2 - 7 } $\delta_{rp}$ & Steps & Times $(\mathrm{s})$ & Steps$^{\ast}$ & Times$^{\dagger}$ $(\mathrm{s})$ & Steps$^{\ast}$ & Times$^{\dagger}$ $(\mathrm{s})$ & Ratio$^{\ddagger}$ \\
			\hline 1 & 194 & 345.3 & 61 & 277.9 & 40 & 167.3 & 2.06/1.66 \\
			10 & 1395 & 2316.3 & 65 & 280.9 & 34 & 135.8 & 17.1/2.07 \\
			100 & 6793 & 10990.3 & 140 & 666.3 & 42 & 173.1 & 63.5/3.85 \\
               1000 & 36225 & 60154.3 & 298 & 1355.4 & 24 & 100.5 & 598.5/13.5 \\
			\hline \hline
		\end{tabular}
		\begin{tablenotes}    
			\footnotesize     
			\item[$^{\ast}$]The steps contains the total steps for the 10-step CIS iterations and the subsequent GSIS iterations to convergence. Each iteration of GSIS comprises solving one time mesoscopic equation and 600 times of macroscopic inner iterations.
			\item[$^{\dagger}$] The computational time contains the total time for the 4000-step iterations of first-order Euler solver, 10-step CIS iterations and the subsequent GSIS iterations to convergence.
			\item[$^{\ddagger}$] The speed-up ratio comprises the acceleration ratio of GSIS-GBT relative to CIS and the acceleration ratio relative to the original  GSIS.
		\end{tablenotes}
		\label{tb4_tubeflow}
	\end{threeparttable}
\end{table}

The contours of the $W$-velocity at the plane of $y = 0$ obtained from the GSIS-GBT are illustrated in Figure~\ref{Fig4.3.1}, which agree well with the results from the GSIS~\cite{zeng_general_2023}. Furthermore, the reduced mass flow rate $\hat{M}$ is calculated within the circular pipe:
\begin{equation}
\hat{M}=\sqrt{\frac{2}{\pi}} \int_{A(z)} \rho u_z d x d y,
	\label{eq_reduceMassrate}
\end{equation}
where ${A(z)}$ is the cross-section area at position $z$. Table \ref{tb3} presents a comparison of the reduced mass flow rates calculated by the locally structured reconstruction method~\cite{titarev_construction_2014}, DSMC~\cite{varoutis_simulation_2009}, and GSIS-GBT. The GSIS-GBT results exhibit good agreement with both numerical methods. Specifically, the errors of GSIS-GBT, compared to the results from DSMC, remain within 1\%.

Table~\ref{tb4_tubeflow} presents the convergence step and computational time in CIS, GSIS and GSIS-GBT. The GSIS-GBT achieves convergence within 40 iterations in this low-speed pressure-driven flows across different rarefaction parameters. Specifically, when $\delta_{rp}=1000$, the GSIS-GBT is faster than the CIS and GSIS by 598.5 and 13.5 times, respectively, demonstrating excellent acceleration performance.

\begin{figure}[p]
	\centering
	\includegraphics[width=0.4\textwidth]{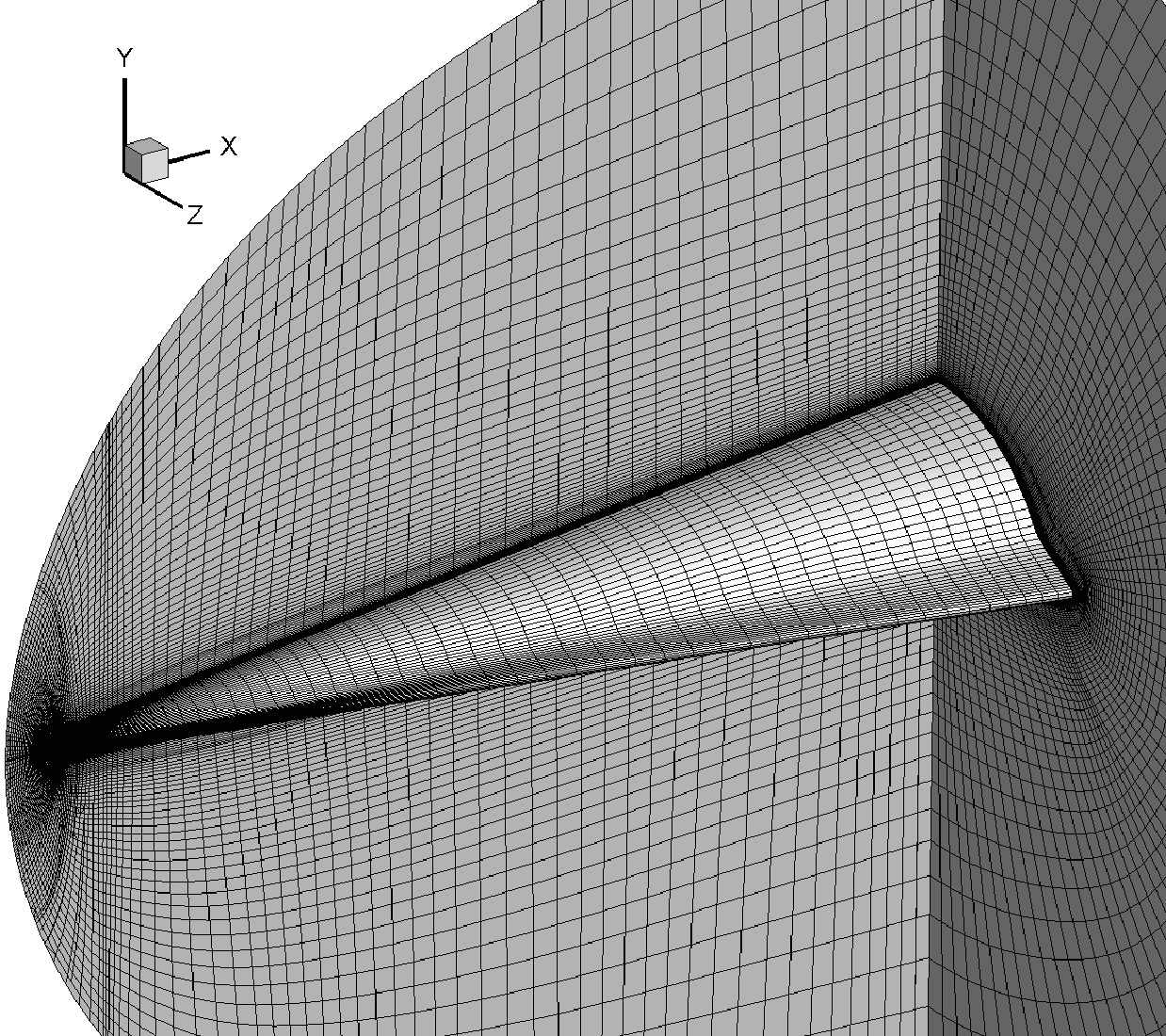}
	\includegraphics[width=0.4\textwidth]{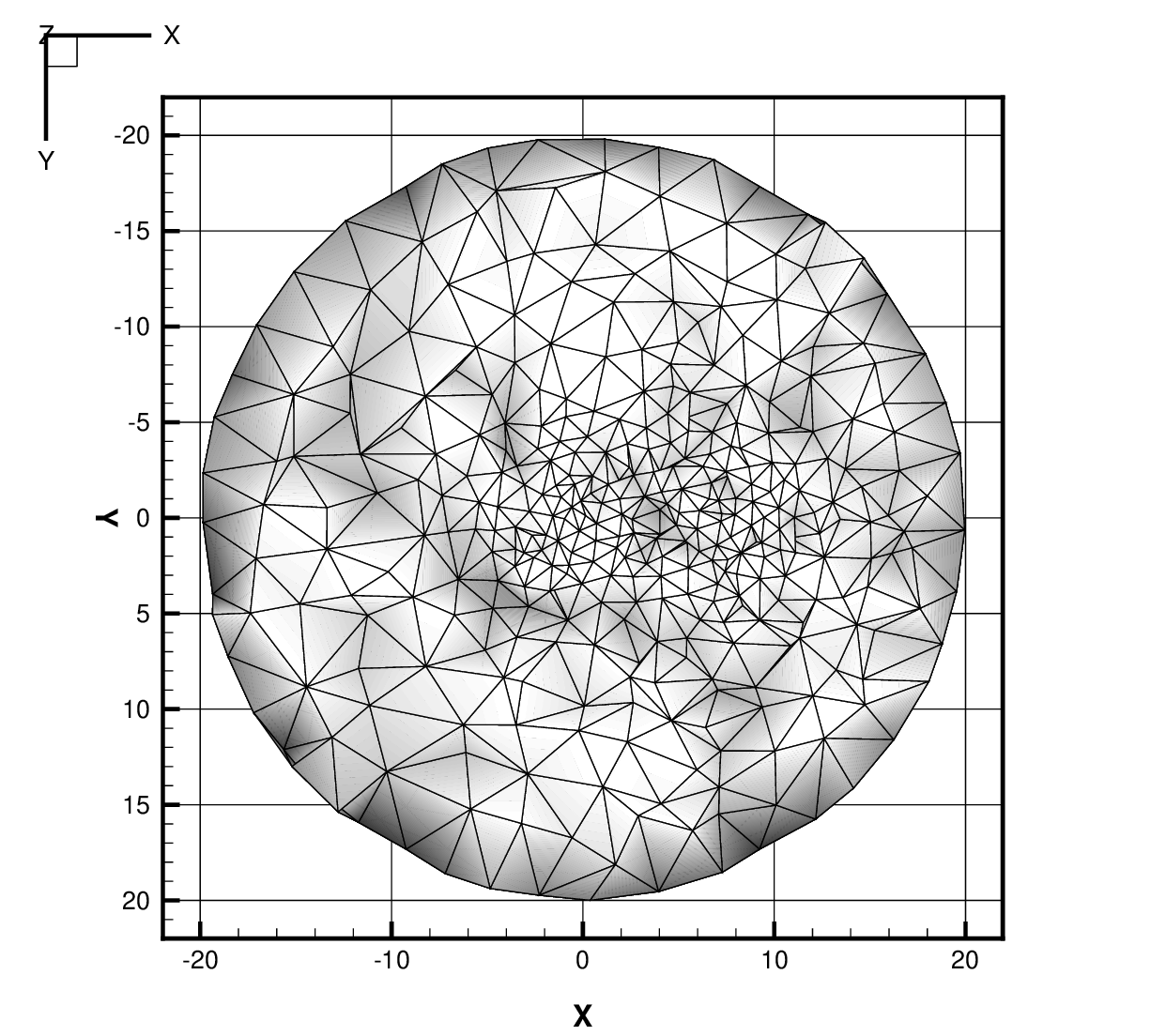}
	\caption{Meshes in the physical (left) and velocity (right) spaces for the flow around hypersonic technology vehicle at Ma = 6.0.}
	\label{Fig4.4.1}
\end{figure}

\begin{figure}[p]
	\centering
	\includegraphics[width=0.35\textwidth]{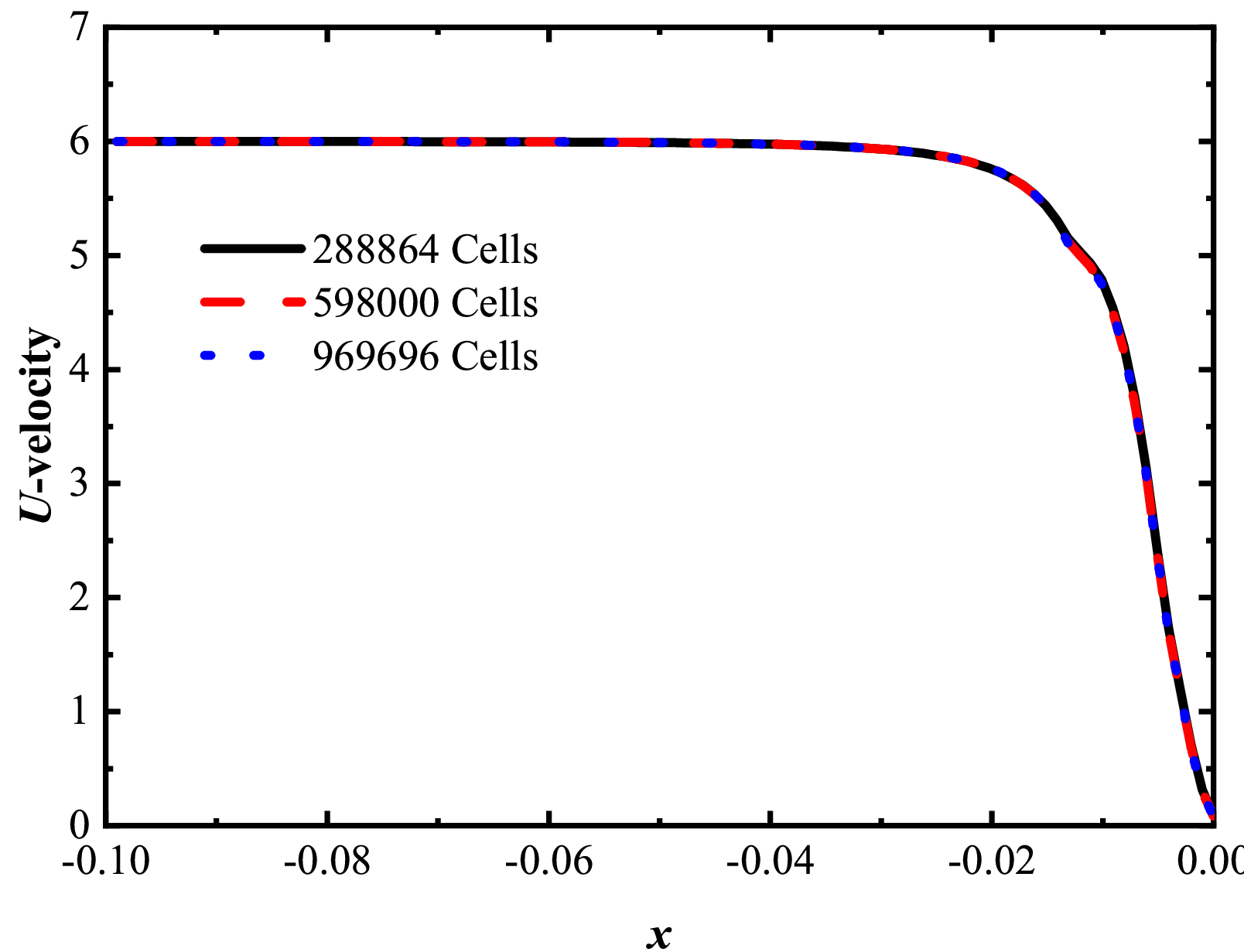}
 \quad
	\includegraphics[width=0.35\textwidth]{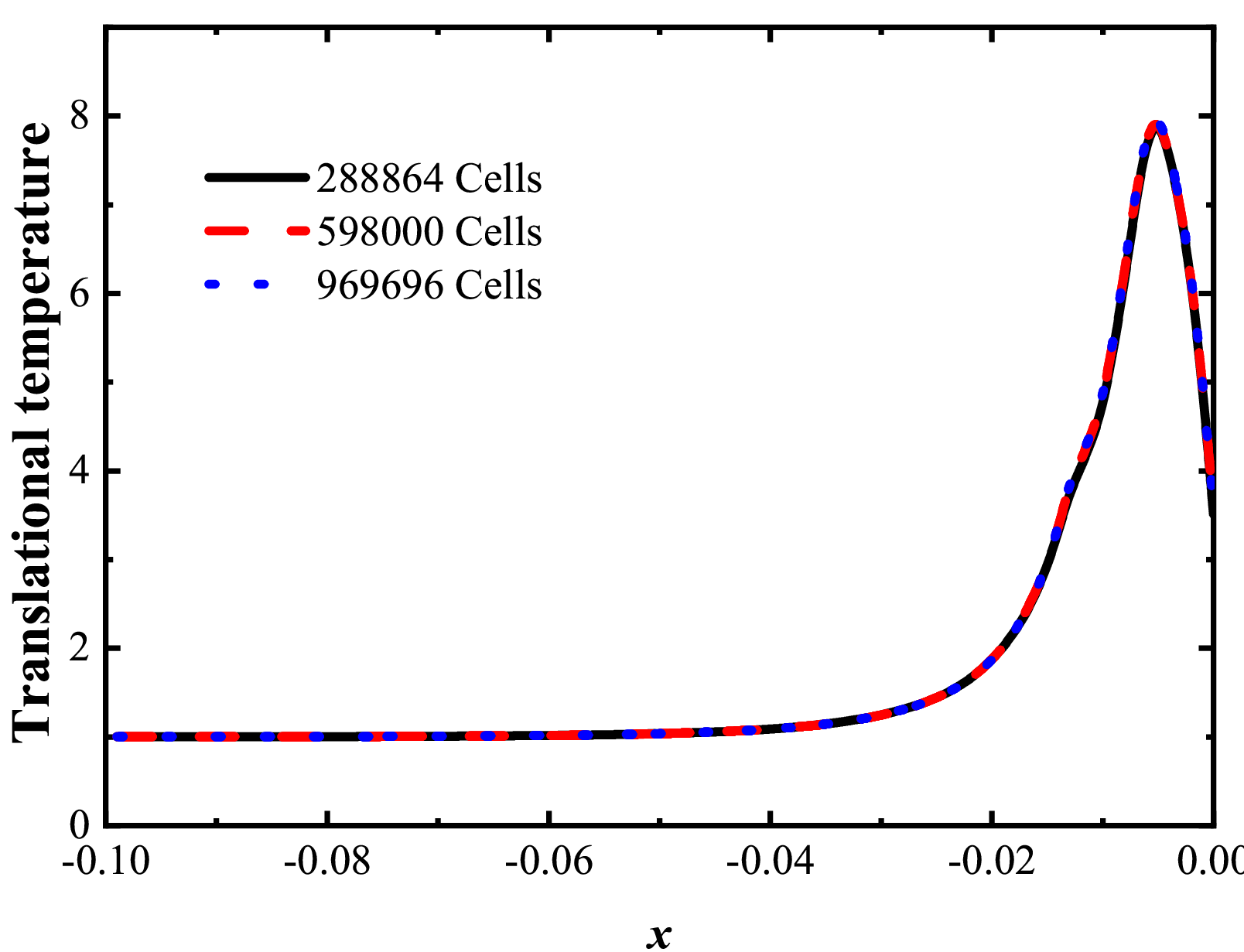}
	\caption{Comparison of (left) $U$-velocity normalized by the sound speed and (right) translational temperature along the stagnation line ($y = 0.02$) at Ma = 6.0 and Kn = 0.005 under different mesh numbers.} 
	\label{Fig4.4.2}
\end{figure}

\begin{figure}[p]
	\centering
	\includegraphics[width=0.4\textwidth]{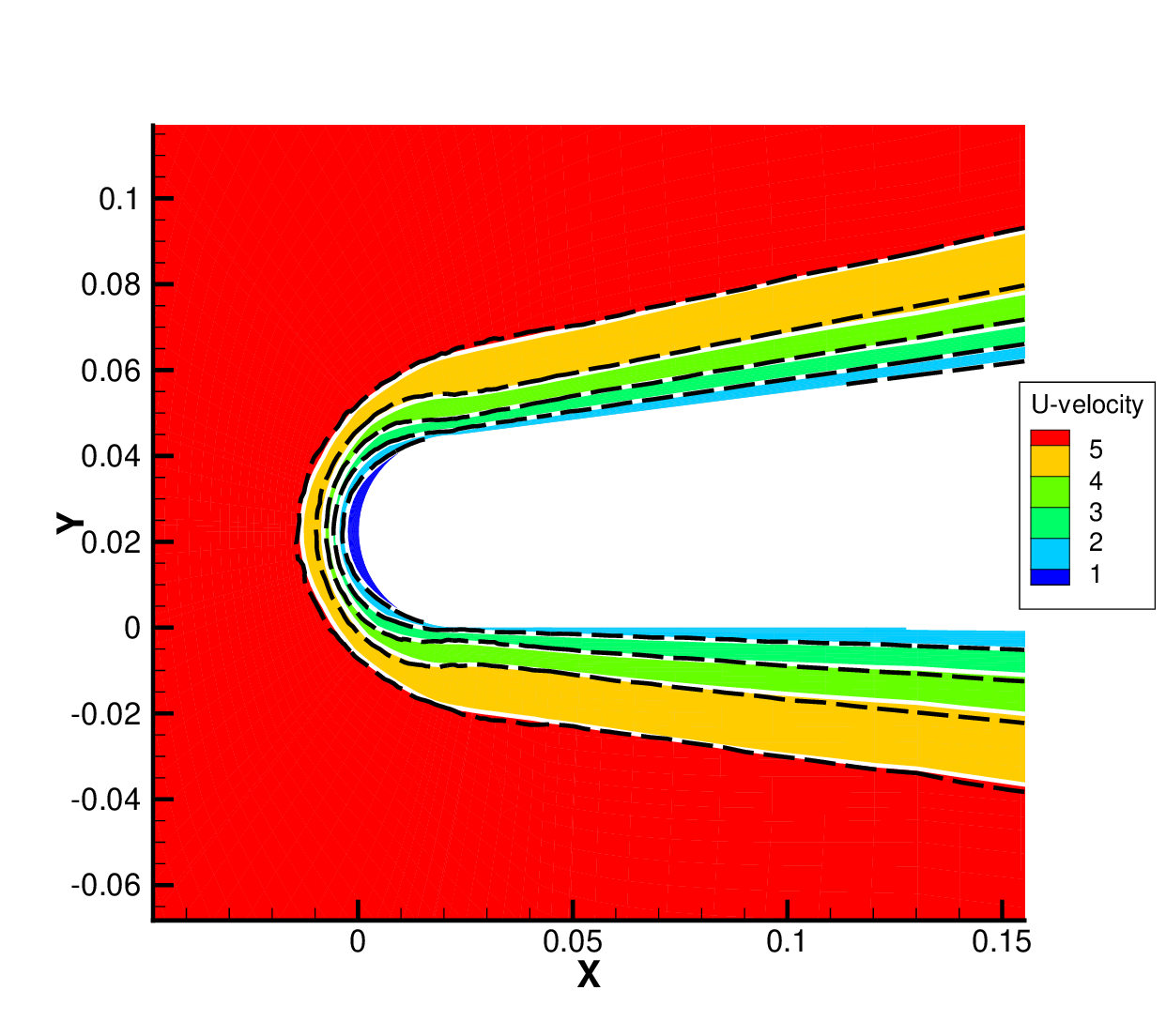}
  \quad
\includegraphics[width=0.4\textwidth]{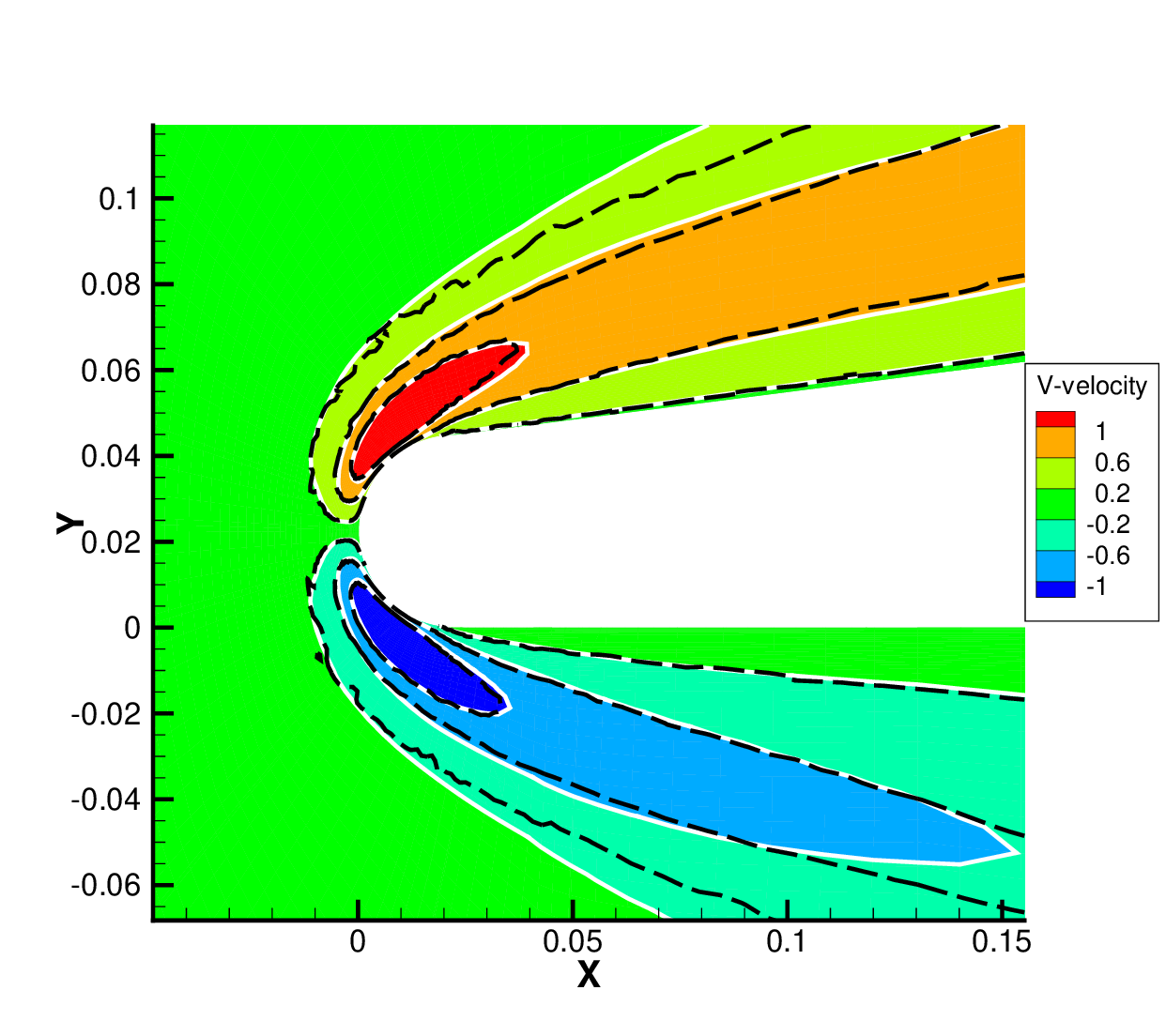}
	\caption{Comparison of (left) $U$-velocity  and (right) $V$-velocity, normalized by the sound speed, for the flow around the HTV at Ma = 6.0 and Kn = 0.005. Background contour with white solid lines: GSIS-GBT; Black dashed lines: UGKWP \cite{fan_implementation_2023}.} 
	\label{Fig4.4.3}
\end{figure}

\subsection{Multiscale flow around hypersonic technology vehicle}
\label{sec4-4}

The flow around a Hypersonic Technology Vehicle (HTV) is considered to assess the performance of GSIS-GBT in complex geometry, where the schematic representations of the physical and velocity meshes are illustrated in Fig.~\ref{Fig4.4.1}. The length and width of the HTV are 2~m and 0.3~m, respectively, and the diameter of the front edge is 0.05~m. The characteristic length is chosen as the length of the spacecraft. The HTV is surrounded by the air with a reference temperature of 188.8~K and a reference velocity of 1653~m/s, so the three-dimensional molecular velocity space is discretized using 14806 unstructured meshes, ranging from $\left[-36, 36 \right]^{3}$, with local refinement at positions where the molecular velocity is 0 and 6. The wall of HTV is characterized by a fully diffuse boundary condition with constant-temperature of 300~K. The incoming flow has a attack angle of zero-degree. Inlet and outlet boundaries are positioned at far-field regions. Again, the velocity results in this section are normalized with respect to the speed of sound.

To validate the mesh independence, 288864, 598000, and 969696 cells in physical space are simulated. The velocity and temperature profiles along the stagnation line ($y = 0.02$, this is the shock region where refined mesh is highly demanded) for the three mesh sets overlap with each other, see Fig.~\ref{Fig4.4.2}, demonstrating the mesh independence has been achieved, due to the asymptotic preserving property of the GSIS~\cite{su_fast_2020}.

\begin{figure}[t]
	\centering
	\includegraphics[width=0.35\textwidth]{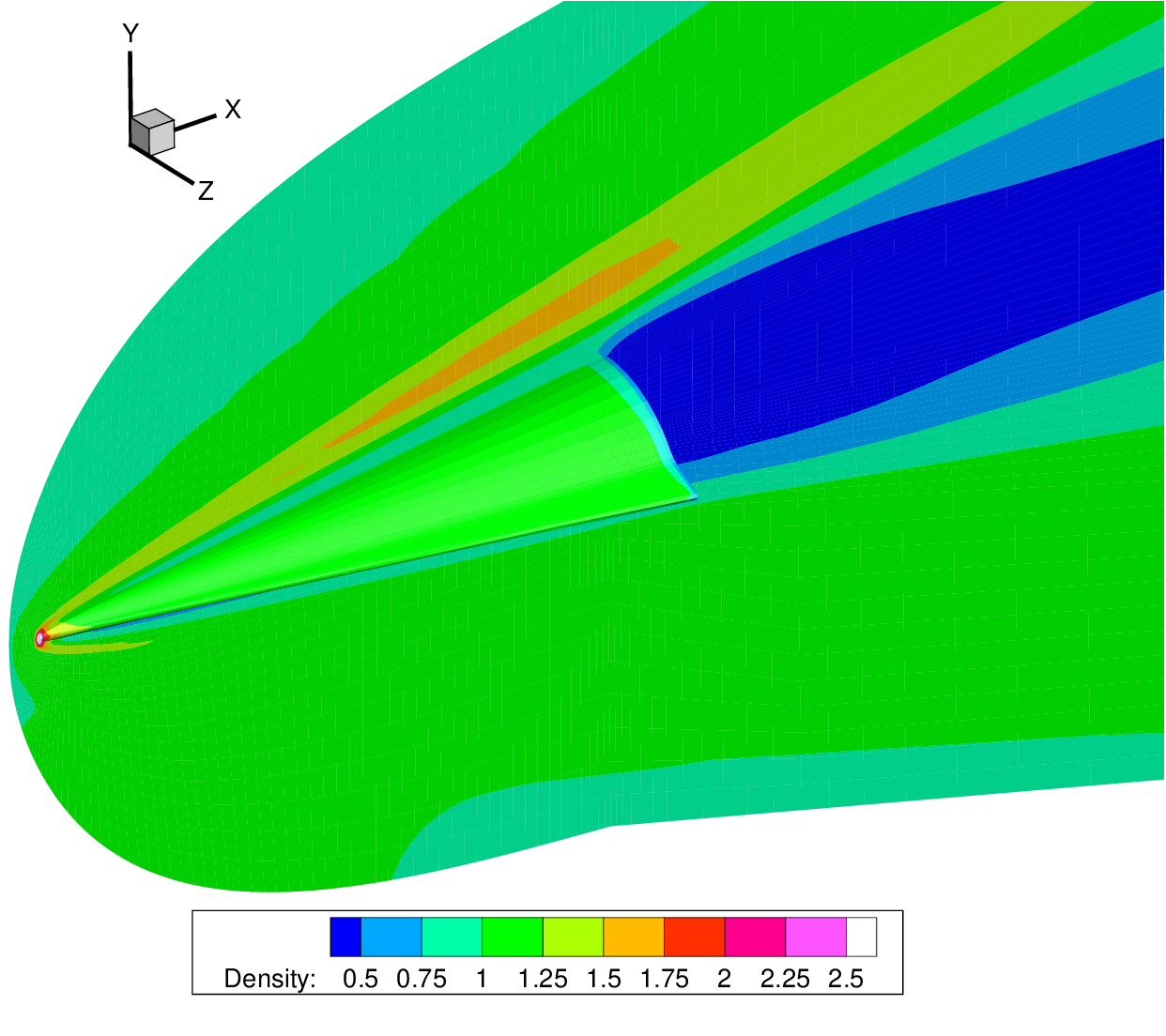}
 \quad
	\includegraphics[width=0.35\textwidth]{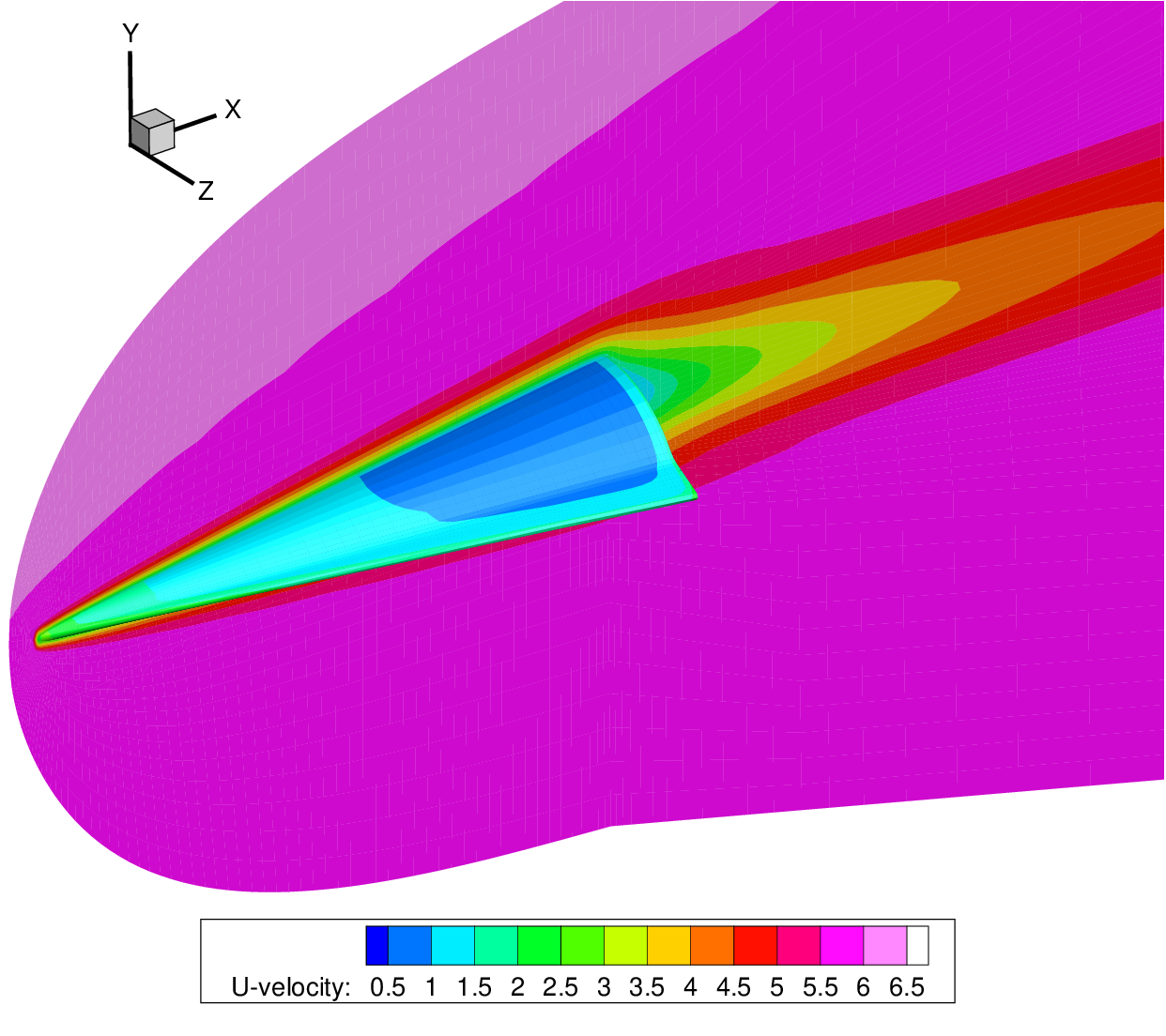}\\
 \vspace{0.3cm}
	\includegraphics[width=0.35\textwidth]{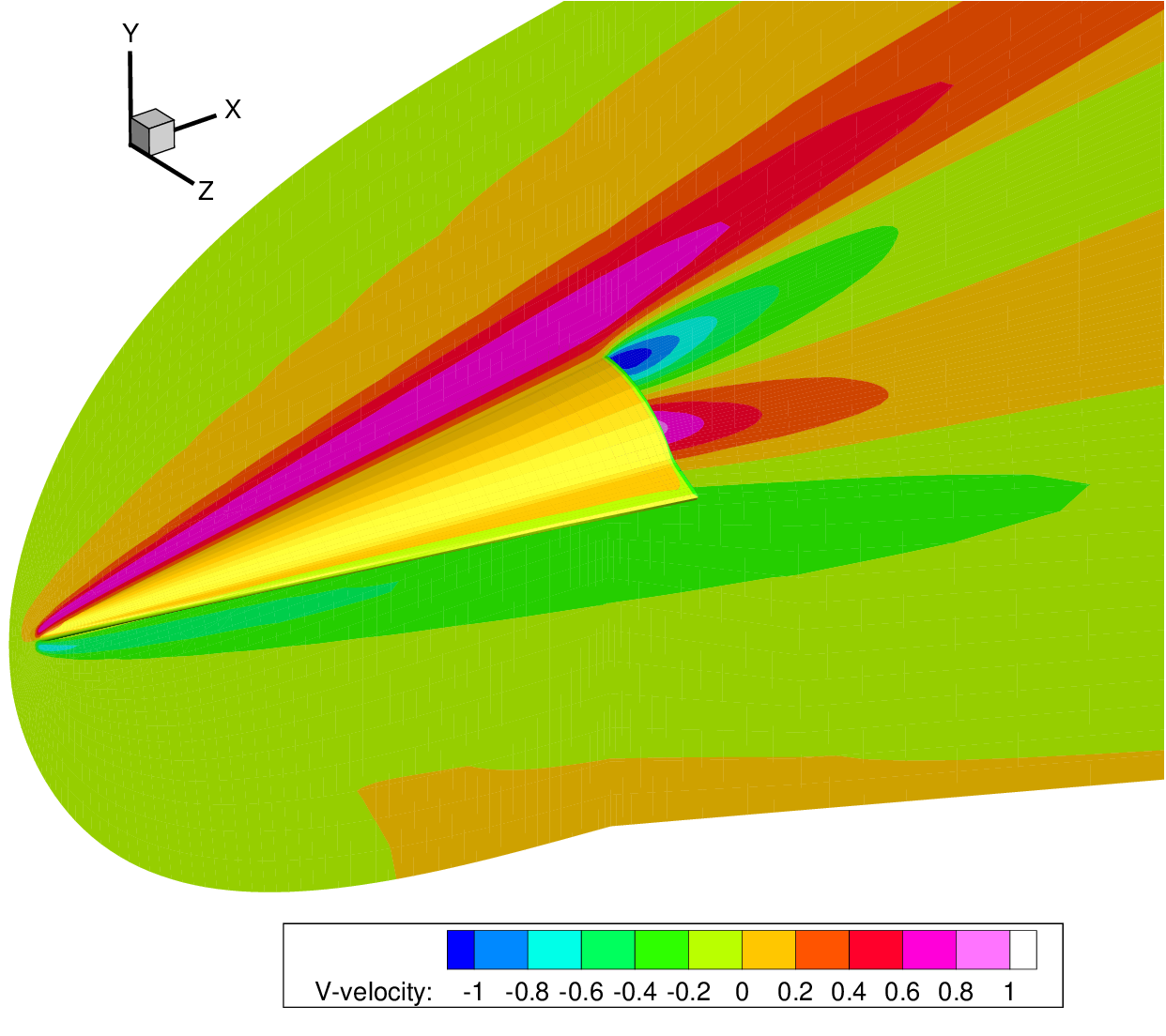}
  \quad
	\includegraphics[width=0.35\textwidth]{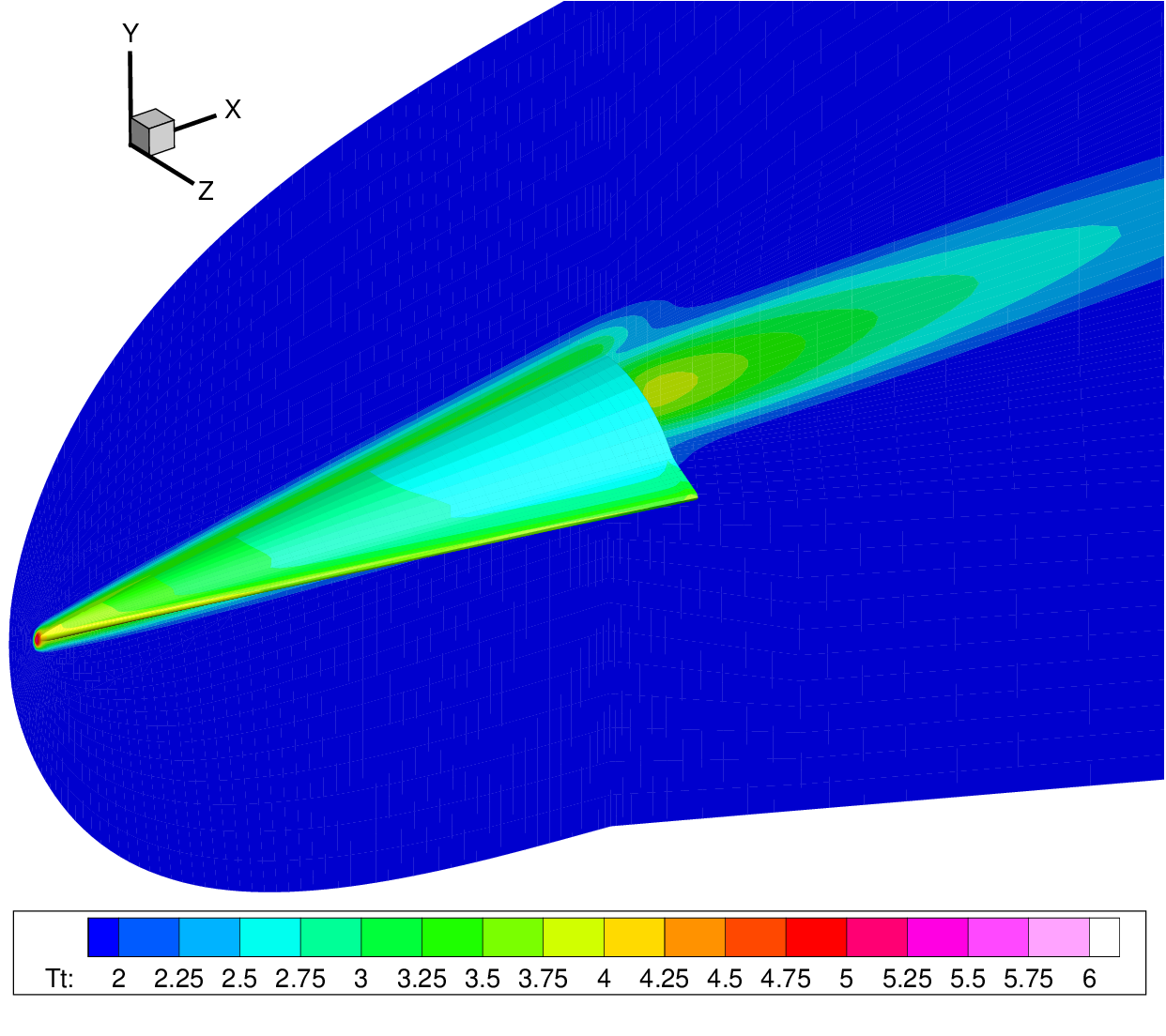}  
	\caption{Contours of the density,  $U$-velocity, $V$-velocity and  translational temperature for the supersonic cylinder flow around hypersonic technology vehicle Ma = 6.0 and Kn = 0.005.}
	\label{Fig4.4.4}
\end{figure}

Figure~\ref{Fig4.4.3} presents the velocity contours at the tip of the HTV. The black dashed lines are obtained from the unified gas-kinetic wave-particle (UGKWP) method~\cite{fan_implementation_2023}. By considering the collision and streaming of gas molecules simultaneously in the flux reconstruction, UGKWP has been demonstrated to possess the capability to overcome the limitations imposed by the mean free path while preserving the advantages of stochastic particle methods in hypersonic flow~\cite{liu_unified_2020, zhu_unified_2019, chen_three-dimensional_2020}. Due to the small diameter of the tip relative to the characteristic length of the aircraft, a pronounced rarefaction effect occurs at the tip. The study of Fan \textit{et al.} indicates that at a global Knudsen number of 0.005 and Mach number of 6, the local Knudsen number at the tip can reach 0.44~\cite{fan_implementation_2023}. Results obtained from the NS equations with first-order slip boundary conditions exhibit significant deviations under these conditions. In contrast, the GSIS-GBT results agree well with those from the UGKWP. Note that in UGKWP 2 million spatial cells are used, while due to the nice asymptotic preserving property of the GSIS~\cite{su_fast_2020}, 288,864 cells are adequate to get the converged solution.

\begin{figure}[t]
	\centering
	\includegraphics[width=0.35\textwidth]{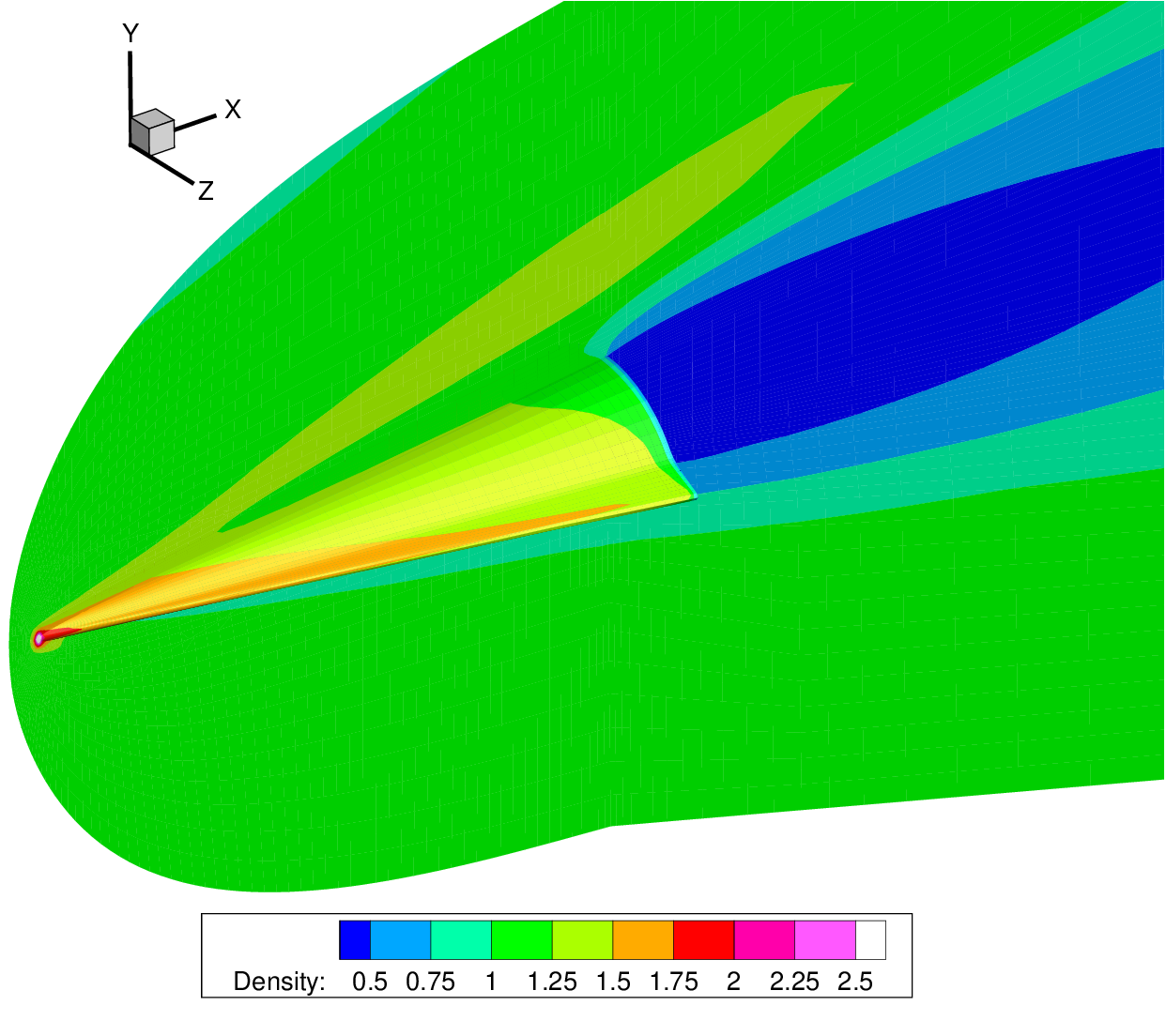}
  \quad
	\includegraphics[width=0.35\textwidth]{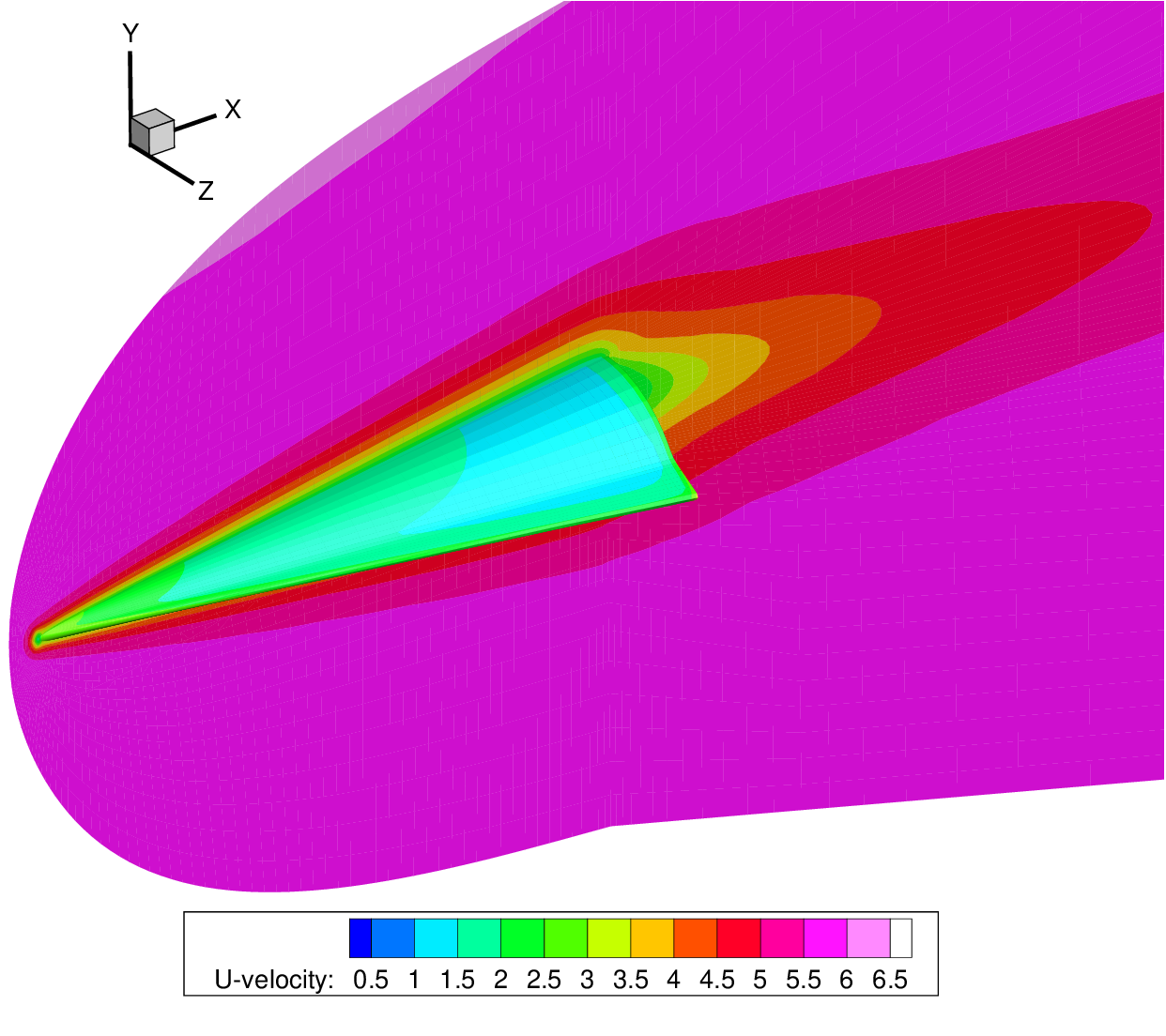}\\
  \vspace{0.3cm}
	\includegraphics[width=0.35\textwidth]{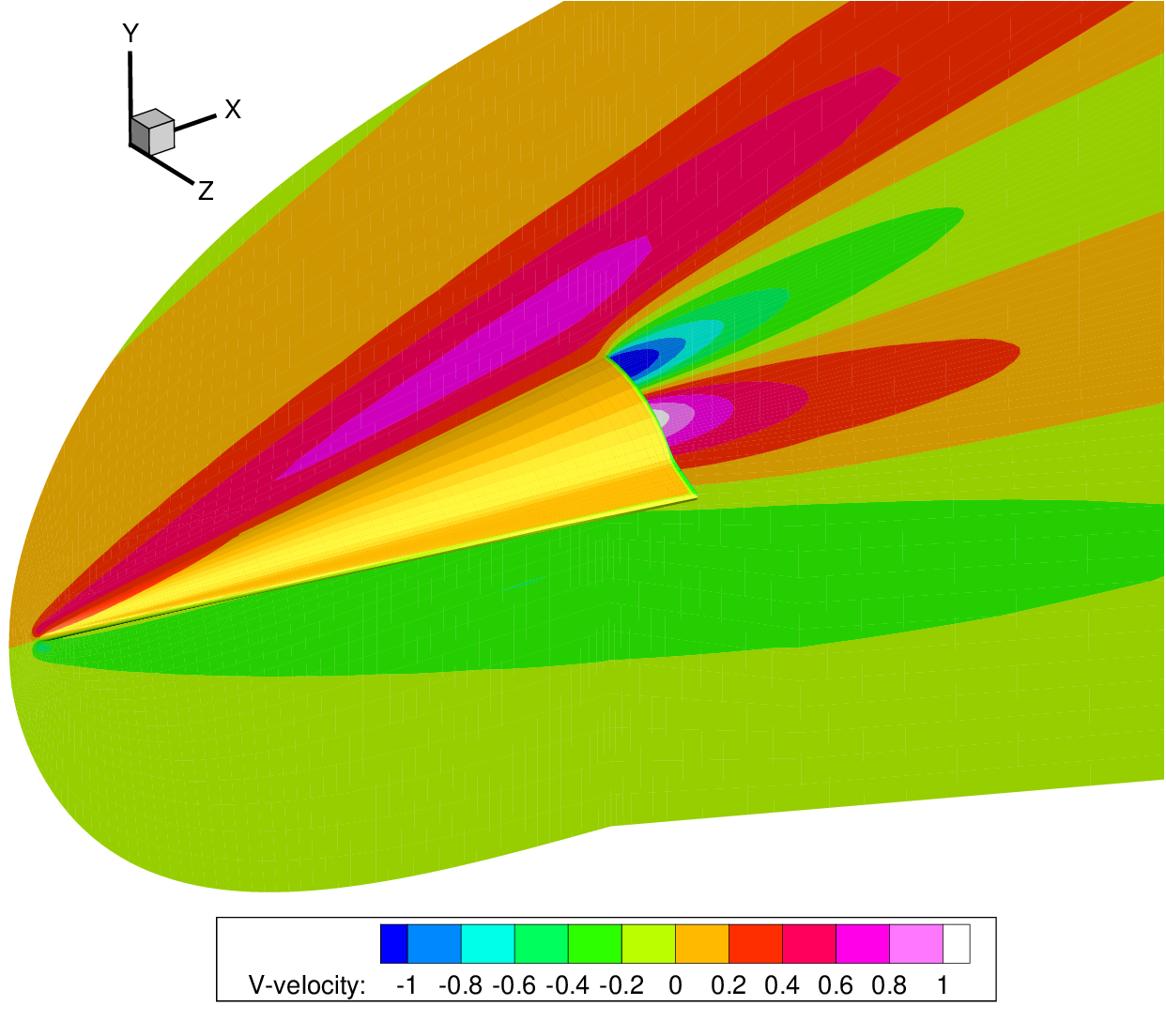}
  \quad
	\includegraphics[width=0.35\textwidth]{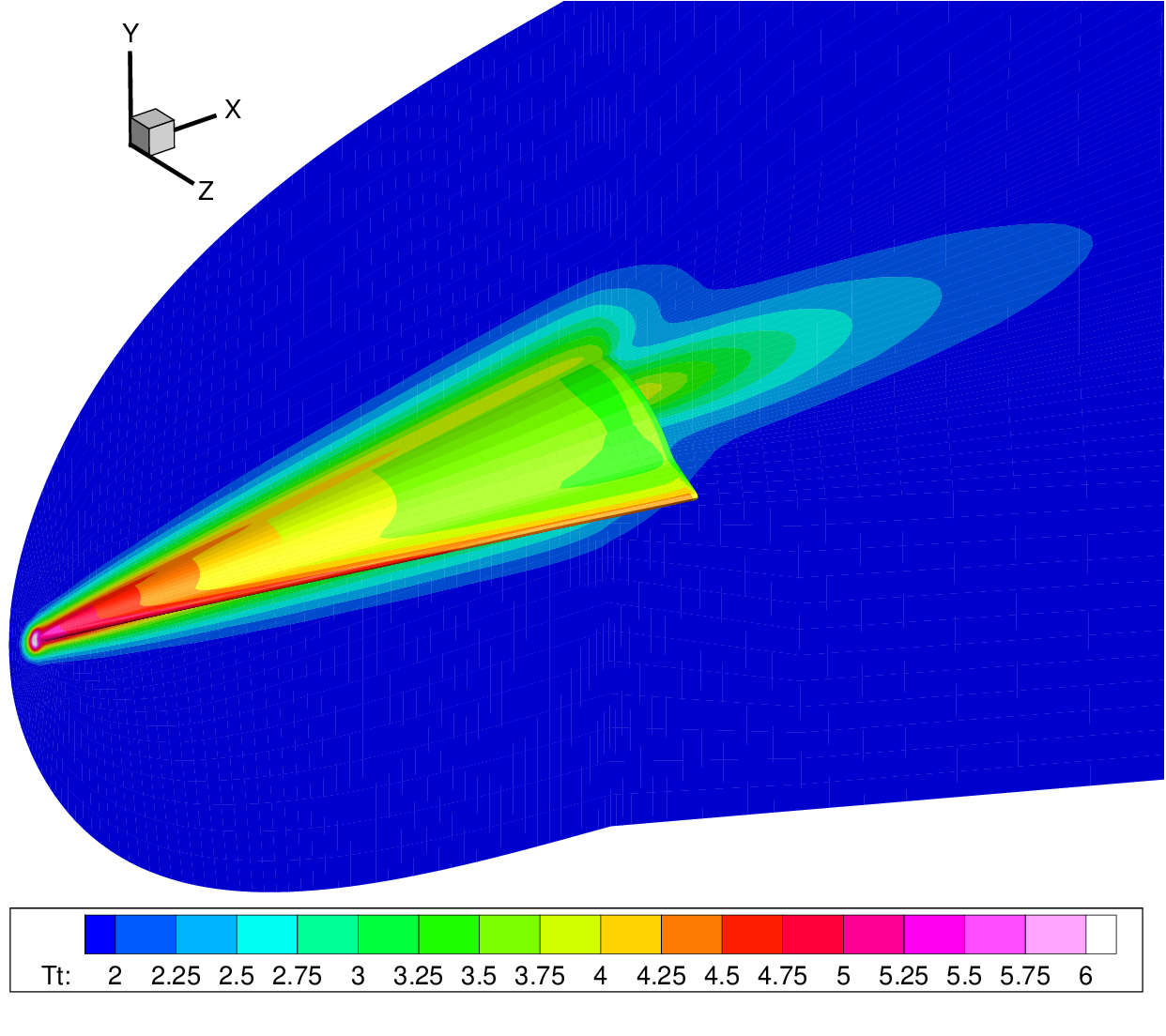}   
	\caption{Contours of the density,  $U$-velocity, $V$-velocity and  translational temperature for the supersonic cylinder flow around hypersonic technology vehicle Ma = 6.0 and Kn = 0.05.}
	\label{Fig4.4.5}
\end{figure}

\begin{figure}[th]
	\centering
	\includegraphics[width=0.35\textwidth]{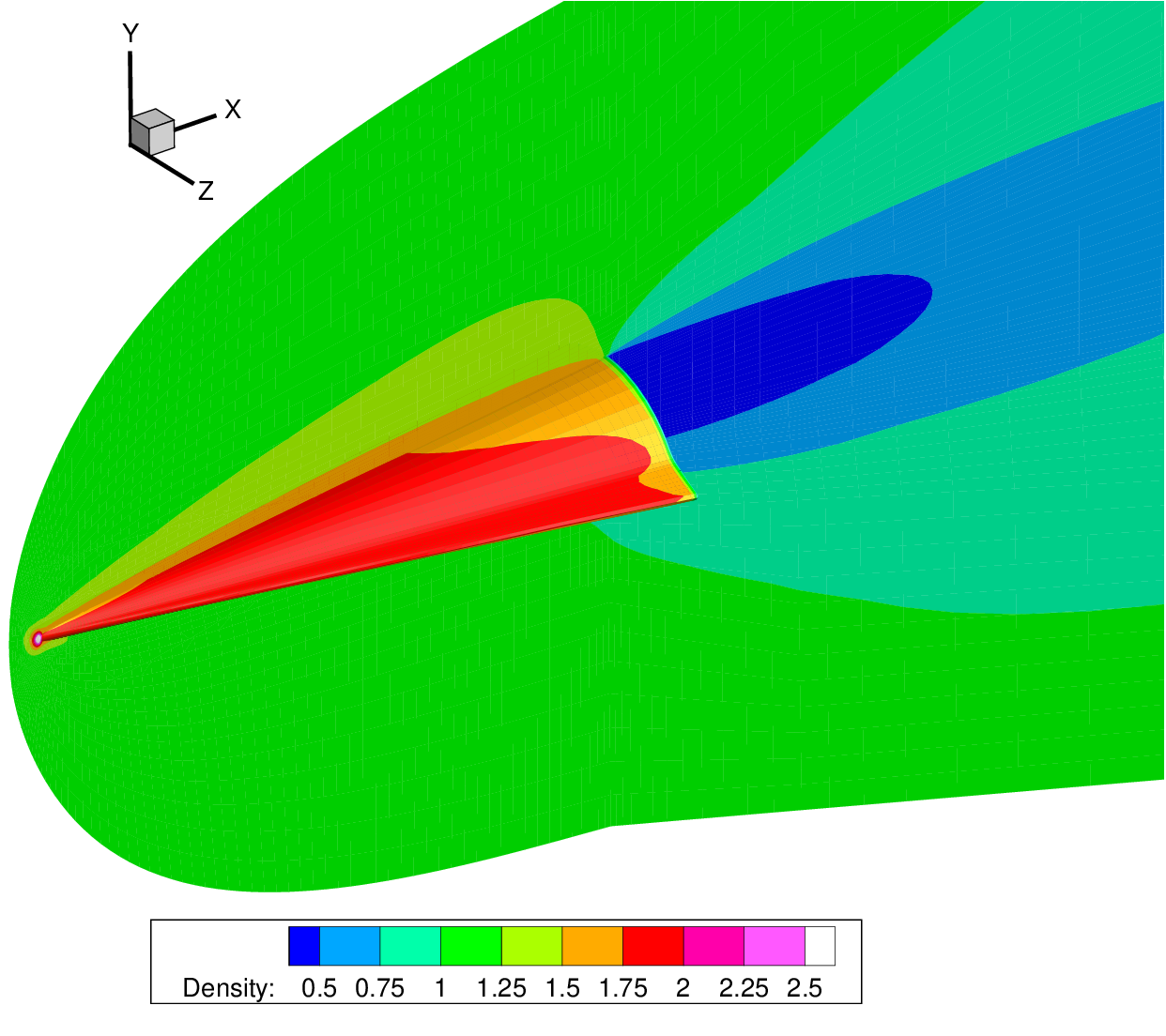}
  \quad
	\includegraphics[width=0.35\textwidth]{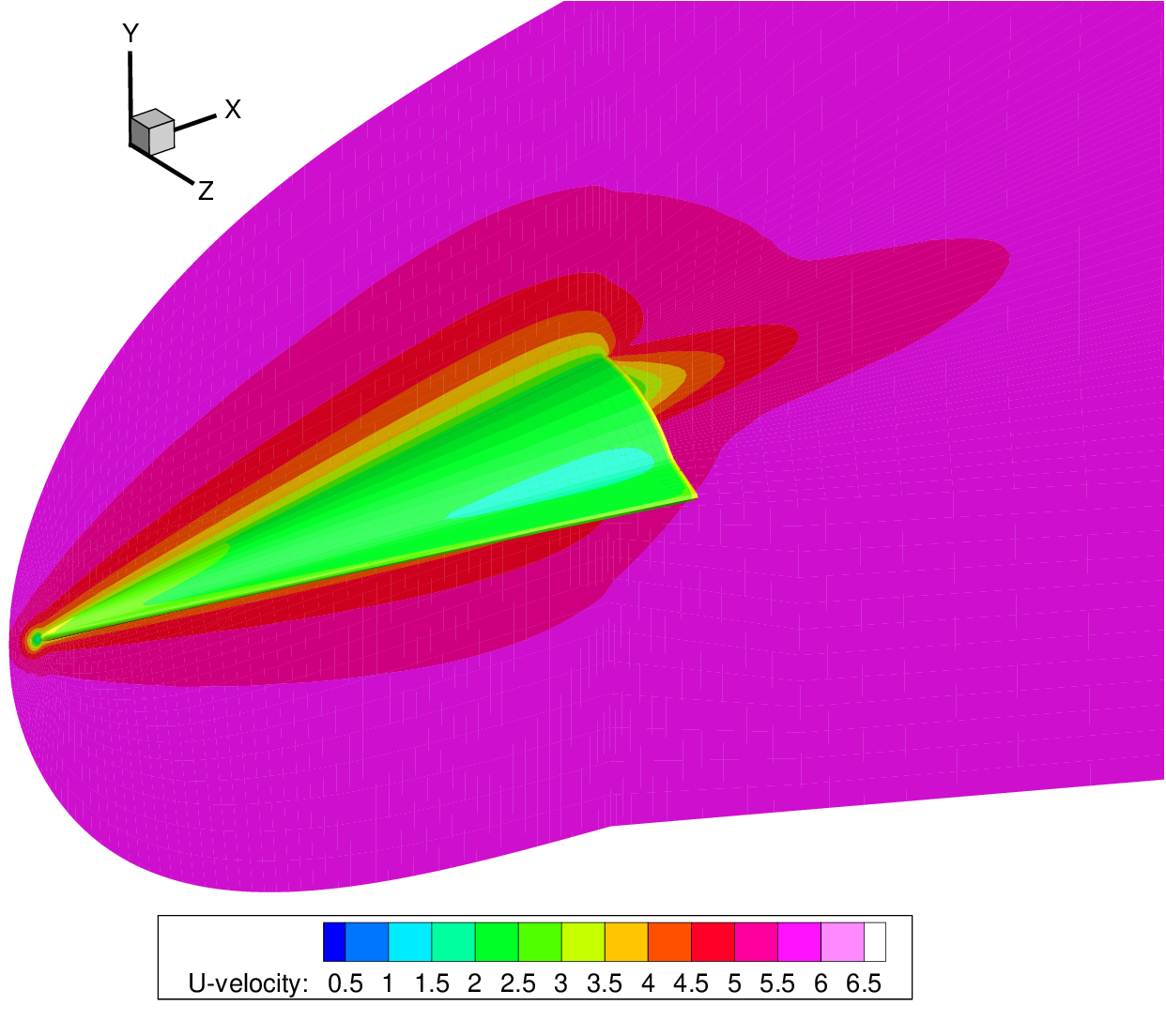}\\
  \vspace{0.3cm}
	\includegraphics[width=0.35\textwidth]{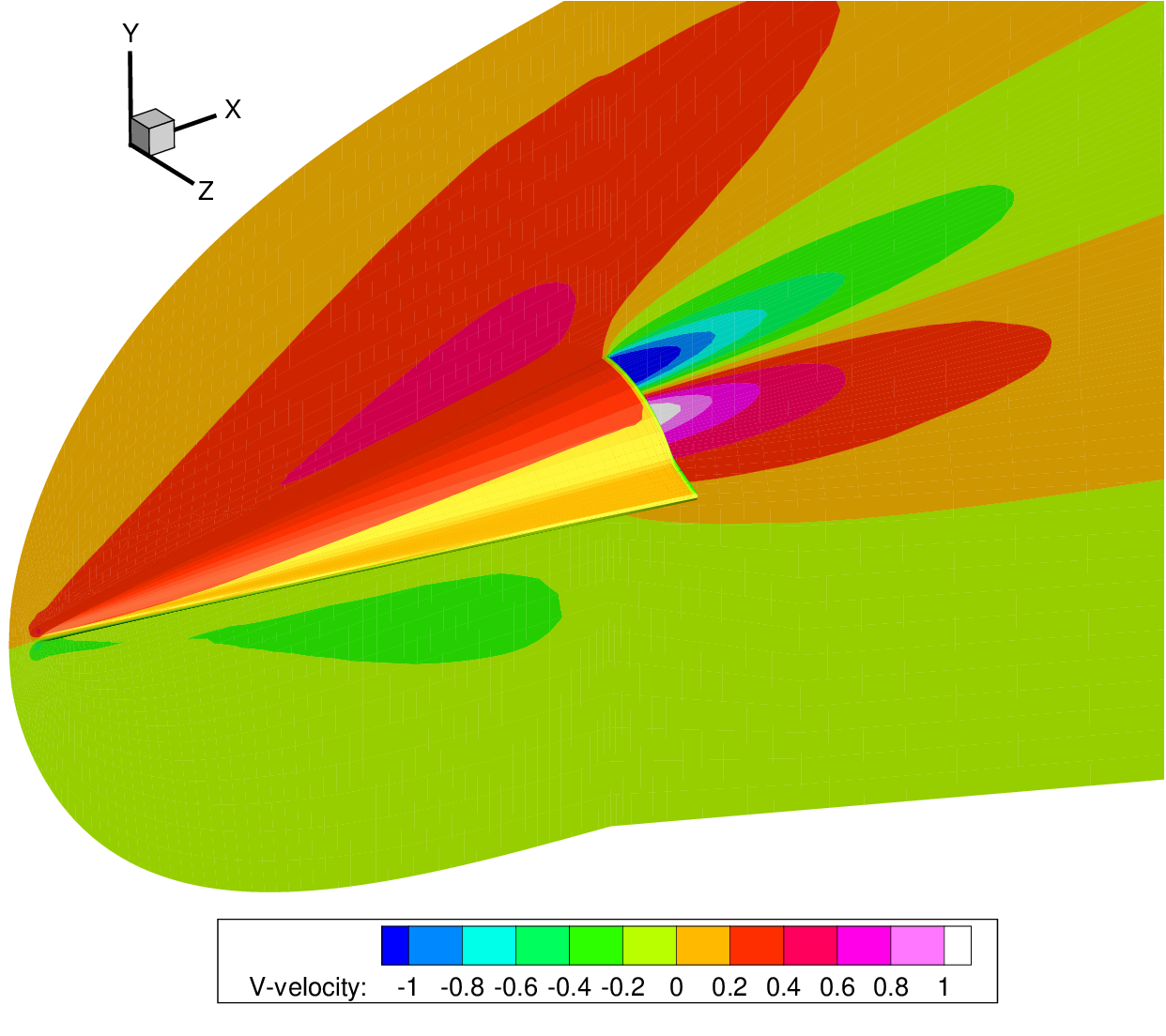}
  \quad
	\includegraphics[width=0.35\textwidth]{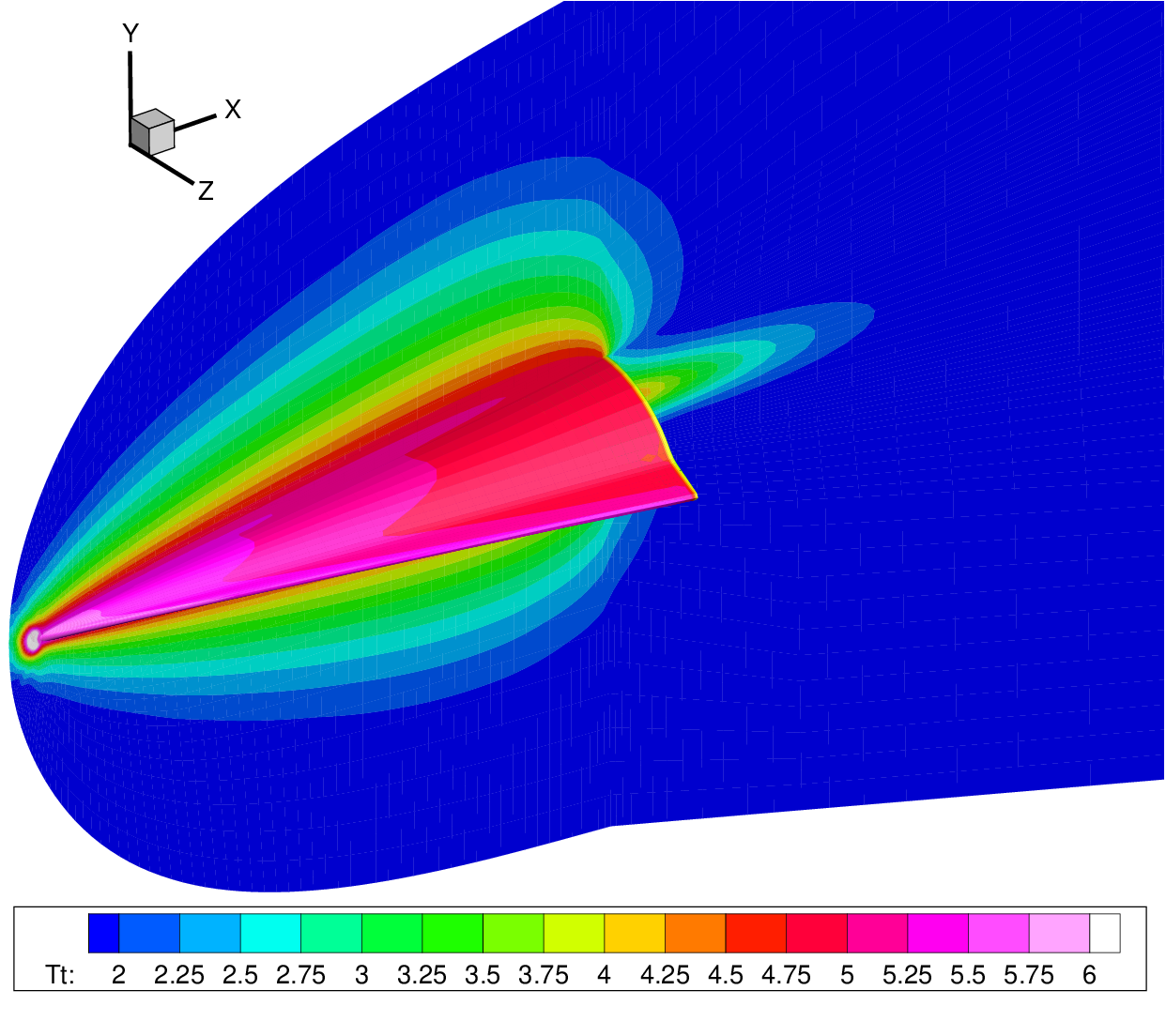}  
	\caption{Contours of the density,  $U$-velocity, $V$-velocity and  translational temperature for the supersonic cylinder flow around hypersonic technology vehicle Ma = 6.0 and Kn = 0.5.}
	\label{Fig4.4.6}
\end{figure}

Figures~\ref{Fig4.4.4}-\ref{Fig4.4.6} offer a comprehensive visualization of the hypersonic flow over the HTV at Kn = 0.005, 0.05 and 0.5, respectively. With the increasing influence of rarefaction effects, noticeable enhancements are observed in both velocity slip and temperature jump at the wall. Additionally, the thickness of shock waves continues to grow.

Table~\ref{tb3} summarizes the convergence steps of different numerical methods. The UGKWP, GSIS~\cite{zeng_general_2023}, and GSIS-GBT all utilized results from the Euler equations (solved by first-order scheme) as the initial field. Note that the UGKWP uses a total of 50,000 steps of evolution to reach the steady-state, followed by an additional 50,000 steps for statistical averaging. The convergence steps and times for GSIS and GSIS-GBT encompass 10 iterations of CIS to initialize the distribution function. Benefiting from the dual acceleration at the boundaries and in the internal field, GSIS-GBT achieves convergence within 50 steps across different Knudsen numbers.

The corresponding computational time of UGKWP based on MPI parallelization employing 92 cores and 2,208,000 cells in physical space is also given in Table~\ref{tb5}. Neglecting variations in CPU performance and considering the linear scalability, the computational time of the CPU-UGKWP, when normalized for the same number of parallel cores (600 cores) and cell number of 288,864 in physical space, is approximately 5.71 hours at Knudsen number 0.005; the computational time for GSIS-GBT is only 0.06 hours, which is 96 times faster than CPU-UGKWP and 6.33 times faster than GSIS. 

\begin{table}[t]
	\centering
	\setlength{\abovecaptionskip}{10pt}
	\setlength{\belowcaptionskip}{10pt}
	\caption{Convergence step and computational time in GSIS, GSIS-GBT and UGKWP for the supersonic HTV flow at Ma = 6.0 and various Kn.
The GSIS and GSIS-GBT simulations are conducted on a parallel computer of AMD Epyc 7742 processor with 600 cores.
 }
	\setlength{\tabcolsep}{3.1mm}
	\begin{threeparttable}
		\begin{tabular}{cccccccc}
			\hline \hline & \multicolumn{2}{c}{ CPU-UGKWP } & \multicolumn{2}{c}{ GSIS } & \multicolumn{2}{c}{ GSIS-GBT } & Speedup \\
			\cline { 2 - 7 } $\mathrm{Kn}$ & Steps & Times$^{\dotplus}$ $(\mathrm{h})$ & Steps$^{\ast}$ & Times$^{\dagger}$ $(\mathrm{h})$ & Steps$^{\ast}$ & Times$^{\dagger}$ $(\mathrm{h})$ & Ratio$^{\ddagger}$ \\
			\hline 0.5 & - & - & 49 & 0.14 & 46 & 0.13 & 1.07 \\
			0.05 & - & - & 77 & 0.23 & 37 & 0.07 & 3.28 \\
			0.005 & 100000 & 212.7 & 129 & 0.38 & 23 & 0.06 & 6.33 \\
			\hline \hline
		\end{tabular}
		\begin{tablenotes}    
			\footnotesize   
                \item[$^{\dotplus}$] The computational time contains the total time for the 50,000-step implicit NS solver and the subsequent UGKWP iterations to convergence. The simulations fo UGKWP are conducted on a parallel computer of AMD Epyc 7k83 with 92 cores. 
			\item[$^{\ast}$]The steps contains the total steps for the 10-step CIS iterations and the subsequent GSIS iterations to convergence. Each iteration of GSIS comprises solving one time mesoscopic equation and 400 times of macroscopic inner iterations.
			\item[$^{\dagger}$] The computational time contains the total time for the 2,000-step iterations of first-order Euler equations, 10-step CIS iterations and the subsequent GSIS iterations to convergence.
			\item[$^{\ddagger}$] The speed-up ratio comprises the acceleration ratio of GSIS-GBT relative to the GSIS.
		\end{tablenotes}
		\label{tb5}
	\end{threeparttable}
\end{table}

\section{Conclusions}
\label{S:5}

In summary, we have proposed a GBT to further accelerate the convergence to the steady state in GSIS. Inspired by the moment methods for handling boundary conditions, we have derived the truncated distribution function to approximate the incident distribution function at the wall, which contains the information of the conserved quantities and the Newton law of viscosity and the Fourier law of heat conductivity from the synthetic equation, as well as the higher-order stress and heat flux from the kinetic equation. Then, we have derived the macroscopic numerical flux at the boundary condition by explicitly integrating the truncated distribution function. Thus, the boundary conditions in the macroscopic synthetic equation is compatible with that in the mesoscopic kinetic equation. It has been found that, the explicit Newton and Fourier constitutive relations facilitate the macroscopic boundary in accelerating the evolution of flow, while higher-order corrections ensure accuracy of boundary treatment from the continuum to rarefied flow regimes. 

Representative and challenging numerical cases, ranging from one-dimensional to three-dimensional flows, and from low-speed to hypersonic flows, have been examined to asses the performance of GSIS-GBT. Results in the hypersonic cylinder flows have shown the accuracy and stability of GBT in strong non-equilibrium flows. The GSIS-GBT has also been applied to the Fourier heat transfer and pressure-driven pipe flow, demonstrating its superior efficiency in handling temperature jump phenomena and low-speed internal flows. 

More importantly, in the Fourier heat transfer and hypersonic cylinder flow, in addition to the previous versions of GSIS in accelerating the flow evolution in interior computational domain, it has been revealed that GSIS-GBT further accelerates the evolution of the flow field at the boundaries. Specifically, GSIS-GBT can yield speedup ratios ranging from 98 to 159 compared to the conventional iterative scheme in hypersonic cylinder flows with Mach numbers of 5 and 20. Compared to the previous versions of GSIS, a further acceleration of 5 to 13 times can be achieved by GSIS-GBT. This further-acceleration has also been seen in pressure-driven pipe flow, supporting the effectiveness of GBT on low-speed internal flows. Furthermore, GSIS-GBT has been applied to the flow around a hypersonic technology with three-dimensional complex geometries. The GBT enables the GSIS to converge within 50 steps across various flow regimes, thus giving a positive answer to the title of our initial work ``Can we find steady-state solutions to multiscale rarefied gas flows within dozens of iterations''~\cite{su_can_2020}, even for general nonlinear flows. 

Although only steady-state problems have been investigated to achieve accelerated convergence, the extension of GBT to unsteady GSIS is straightforward~\cite{zeng_general_2023-1}. Besides, the derivation of GBT is based on a polyatomic gas model with only the rotational degrees of freedom. Within the current framework, further extension to kinetic models with vibration and radiation is not overly challenging. We believe that this work presents significant potential for simulating practical engineering problems involving multiscale rarefied gas flows.


\section*{Acknowledgments}
\label{S:6}

This work is supported by the National Natural Science Foundation of China (12172162). Special thanks are given to the Center for Computational Science and Engineering at the Southern University of Science and Technology. The authors thank G. C. Fan in the Zhejiang University for sharing the mesh of hypersonic technology vehicle.

\appendix

\section{Nondimensionalization for the kinetic equation}
\label{A:1}

We provide the dimensionless process for the kinetic theory equations. Variables with a colon superscript denote dimensional quantities. The specific gas constant $R$, reference length $\hat{L}$, reference density $\hat{\rho}_0$, and reference temperature $\hat{T}_0$ are utilized in the dimensionless calculations:
\begin{equation}
	\begin{gathered}
		f_0=\frac{\hat{f}_0}{\hat{\rho}_0 /\left(R \hat{T}_0\right)^{3 / 2}}, \quad f_1=\frac{\hat{f}_1}{\hat{\rho}_0 R \hat{T}_0 /\left(R \hat{T}_0\right)^{3 / 2}}, \quad x=\frac{\hat{x}}{\hat{L}}, \quad \rho=\frac{\hat{\rho}}{\hat{\rho}_0}, \quad t=\frac{\hat{t}}{\hat{L} /\left(R \hat{T}_0\right)^{1 / 2}}, \\
		(\boldsymbol{\xi}, \mathbf{c}, \mathbf{u})=\frac{(\hat{\boldsymbol{\xi}}, \hat{\mathbf{c}}, \hat{\mathbf{u}})}{\left(R \hat{T}_0\right)^{1 / 2}}, \quad\left(T, T_t, T_r\right)=\frac{\left(\hat{T}, \hat{T}_t, \hat{T}_r\right)}{\hat{T}_0},\\
		\left(\boldsymbol{\sigma}, p, p_t\right)=\frac{\left(\hat{\boldsymbol{\sigma}}, \hat{p}, \hat{p}_t\right)}{\hat{\rho}_0 R \hat{T}_0}, \quad\left(q_t, q_r\right)=\frac{\left(\hat{q}_t, \hat{q}_r\right)}{\hat{\rho}_0 R \hat{T}_0\left(R \hat{T}_0\right)^{1 / 2}}.
	\end{gathered}
	\label{eqA1}
\end{equation}
The equilibrium distribution functions in Eqs.~\eqref{eq2.1.4} and \eqref{eq2.1.5} are normalized in the same way as $\hat{f}_0$ and $\hat{f}_1$.

The Knudsen number is defined as the ratio of the molecular mean free path $\lambda$ to the reference length $\hat{L}$:
\begin{equation}
    \text{Kn}=\frac{\lambda}{\hat{L}}, 
    \quad 
    \lambda=\frac{\mu(\hat{T}_0)}{\hat{\rho}_0 R \hat{T}_0 }\sqrt{\frac{\pi R \hat{T}_0}{2}},
\end{equation}
where $\mu(\hat{T}_0)$ is the gas shear viscosity at the reference temperature $\hat{T}_0$.

\section{Truncated distribution function for the non-vibrating polyatomic gas with rotational energy}
\label{B:1}

The truncation of distribution functions $f^{T}(t, \boldsymbol{x}, \boldsymbol{\xi}, I_r) $ with rotational energy $I_r$ is expressed as a linear combination of different orders of peculiar velocity as follows
\begin{equation}
	f^{T}(t, \boldsymbol{x}, \boldsymbol{\xi}, I_r)=f_0^{e q} f_{I_r}^{e q} \left(\begin{array}{l}
		a_0+\left(a_{11} c_x+a_{12} c_y+a_{13} c_z\right)\\
		+\left(a_{211} c_x^2+a_{222} c_y^2+a_{233} c_z^2\right) \\
		+\left(a_{212} c_x c_y+a_{213} c_x c_z+a_{223} c_y c_z\right)\\
		+\left(a_{31} c_x c^2+a_{32} c_y c^2+a_{33} c_z c^2\right) \\
		\textcolor{black}{+b_{01} I_r+\left(b_{11} c_x+b_{12} c_y+b_{13} c_z\right) I_r}
	\end{array}\right),
	\label{eqB1}
\end{equation}

\noindent The factors $a$ and $b$ are the coefficients need to be determined in the distribution function based on the moment relationship as follows
\begin{equation}
	\begin{aligned}
		& \rho=\iint f \mathrm{~d} \boldsymbol{\xi} \mathrm{d} I_r, 
   \quad
   0=\iint \mathbf{c} f \mathrm{~d} \boldsymbol{\xi} \mathrm{d} I_r, 
    \quad
    3 \rho T_t=\iint c^2 f \mathrm{~d} \boldsymbol{\xi} \mathrm{d} I_r, \\
		& \boldsymbol{\sigma}=\iint\left(\mathbf{c} \mathbf{c}-\frac{c^2}{3} \mathrm{I}\right) f \mathrm{~d} \boldsymbol{\xi} \mathrm{d} I_r, 
   \quad
   \mathbf{q}_t=\iint \mathbf{c} \frac{c^2}{2} f \mathrm{~d} \boldsymbol{\xi} \mathrm{d} I_r, \\
		& \frac{d_r}{2} \rho T_r=\iint I_r f \mathrm{~d} \boldsymbol{\xi} \mathrm{d} I_r, 
  \quad
  \mathbf{q}_r=\iint \mathbf{c} I_r f \mathrm{~d} \boldsymbol{\xi} \mathrm{d} I_r.
	\end{aligned}
	\label{eqB2}
\end{equation}

Then, we can get all coefficients $a$ and $b$ as 
\begin{equation}
	\begin{aligned}
		\begin{gathered}
			a_0=1, \quad b_{01}=0,\\
			a_{211}=\frac{\sigma_{xx}}{2 \rho\left(T_t\right)^2}, \quad a_{212}=\frac{\sigma_{xy}}{2 \rho\left(T_t\right)^2}, \quad a_{213}=\frac{\sigma_{xz}}{2 \rho\left(T_t\right)^2}, \\
			a_{212}=\frac{\sigma_{xy}}{\rho\left(T_t\right)^2}, 
   \quad
   a_{213}=\frac{\sigma_{xz}}{\rho\left( T_t\right)^2}, 
   \quad
   a_{223}=\frac{\sigma_{yz}}{\rho\left(T_t\right)^2},
		\end{gathered}
	\end{aligned}
	\label{eqB3}
\end{equation}

\noindent and
\begin{equation}
	\begin{aligned}
		& a_{11}=-\frac{q_{t, x}}{\rho\left( T_t\right)^2}-\frac{q_{r, x}}{\rho  T_t\left( T_r\right)}, \\
		& a_{12}=-\frac{q_{t, y}}{\rho\left( T_t\right)^2}-\frac{q_{r, y}}{\rho  T_t\left( T_r\right)}, \\
		& a_{13}=-\frac{q_{3, t}}{\rho\left( T_t\right)^2}-\frac{q_{3, r}}{\rho  T_t\left( T_r\right)},
	\end{aligned}
	\label{eqB4}
\end{equation}

\begin{equation}
	\begin{aligned}
		a_{31}=\frac{q_{t, x}}{5 \rho\left( T_t\right)^3}, 
  \quad
  a_{32}=\frac{q_{t, y}}{5 \rho\left( T_t\right)^3}, 
   \quad
   a_{33}=\frac{q_{3, t}}{5 \rho\left( T_t\right)^3},
	\end{aligned}
	\label{eqB5}
\end{equation}

\begin{equation}
	\begin{aligned}
		b_{11}=\frac{q_{r, x}}{\frac{d_r}{2}\left( T_r\right)^2 \rho  T_t}, 
   \quad
   b_{12}=\frac{q_{r, y}}{\frac{d_r}{2}\left( T_r\right)^2 \rho  T_t}, 
    \quad
    b_{13}=\frac{q_{3, r}}{\frac{d_r}{2}\left( T_r\right)^2 \rho  T_t}.
	\end{aligned}
	\label{eqB6}
\end{equation}

Finally, the expression of distribution functions $f^{T}(t, \boldsymbol{x}, \boldsymbol{\xi}, I_r) $ is given as
\begin{equation}
	f^T\left(t, \mathbf{x}, \xi, I_r\right)=f_0^{e q} f_{I_r}^{e q}\left[1+\frac{\boldsymbol{\sigma} \cdot \mathbf{c c}}{2 \rho\left(T_t\right)^2}+\frac{\mathbf{q}_t \cdot \mathbf{c}}{\rho\left(T_t\right)^2}\left(\frac{c^2}{5 T_t}-1\right)+\frac{\mathbf{q}_r \cdot \mathbf{c}}{\frac{d_r}{2} \rho T_t T_r}\left(\frac{I_r}{T_r}-\frac{d_r}{2}\right)\right].
	\label{eqB7}
\end{equation}

\section{Integration coefficients related to the equilibrium state}
\label{C:1}

It should be noted that the coefficients of $\left\langle\xi_{n}^{k}\right\rangle_{>0}^{eq}$ and $\left\langle\xi_{n}^{k}\right\rangle_{<0}^{eq}$ have different expressions. The parameters of $\left\langle\xi_{n}^{k}\right\rangle_{>0}^{eq}$ and $\left\langle\xi_{n}^{k}\right\rangle_{<0}^{eq}$ could be given as
\begin{equation}
	\left\langle\xi_{n}^{0}\right\rangle_{>0}^{e q}=\frac{1}{2}\left[1+\operatorname{erf}\left(\sqrt{\lambda_t^{L}} u_{n}^{L}\right)\right],
	\label{eqC1}
\end{equation}

\begin{equation}
	\left\langle\xi_{n}^{1}\right\rangle_{>0}^{e q}=u_{n}^{L}\left\langle\xi_{n}^{0}\right\rangle_{>0}^{e q}+\frac{1}{2} \frac{e^{-\lambda_t^{L}\left(u_{n}^{L}\right)^{2}}}{\sqrt{\lambda_t^{L} \pi}},
	\label{eqC2}
\end{equation}

\begin{equation}
	\left\langle\xi_{n}^{k+2}\right\rangle_{>0}^{e q}=u_{n}^{L}\left\langle\xi_{n}^{k+1}\right\rangle_{>0}^{e}+\frac{k+1}{2 \lambda_t^{L}}\left\langle\xi_{n}^{k}\right\rangle_{>0}^{e q}, \quad k=0,1,2, \cdots,
	\label{eqC3}
\end{equation}

\noindent and
\begin{equation}
	\left\langle\xi_{n}^{0}\right\rangle_{<0}^{e q}=\frac{1}{2} \operatorname{erfc}\left(\sqrt{\lambda_t^{R}} u_{n}^{R}\right),
	\label{eqC4}
\end{equation}

\begin{equation}
	\left\langle\xi_{n}^{1}\right\rangle_{<0}^{e q}=u_{n}^{R}\left\langle\xi_{n}^{0}\right\rangle_{<0}^{e q}-\frac{1}{2} \frac{e^{-\lambda_t^{R}\left(u_{n}^{R}\right)^{2}}}{\sqrt{u_{n}^{R} \pi}},
	\label{eqC5}
\end{equation}

\begin{equation}
	\left\langle\xi_{n}^{k+2}\right\rangle_{<0}^{e q}=u_{n}^{R}\left\langle\xi_{n}^{k+1}\right\rangle_{<0}^{e q}+\frac{k+1}{2 \lambda_t^{R}}\left\langle\xi_{n}^{k}\right\rangle_{<0}^{e q}, \quad k=0,1,2, \cdots.
	\label{eqC6}
\end{equation}

Following the binomial theory, part of even order of integration parameters $\left\langle c_{\tau}^{k}\right\rangle^{eq}$ and $\left\langle c_{s}^{k}\right\rangle^{eq}$ could be computed as
\begin{equation}
	\left\langle c_{\tau}^{0}\right\rangle^{e q}=\left\langle c_{s}^{0}\right\rangle^{e q}=1,
	\label{eqC7}
\end{equation}

\begin{equation}
	\left\langle c_{\tau}^{2}\right\rangle^{e q}=\left\langle c_{s}^{2}\right\rangle^{e q}=\frac{1}{2 \lambda_t},
	\label{eqC8}
\end{equation}

\begin{equation}
	\left\langle c_{\tau}^{4}\right\rangle^{e q}=\left\langle c_{s}^{4}\right\rangle^{e q}=\frac{3}{4 \lambda_t^{2}},
	\label{eqC9}
\end{equation}

\begin{equation}
	\left\langle c_{\tau}^{6}\right\rangle^{e q}=\left\langle c_{s}^{6}\right\rangle^{e q}=\frac{15}{8 \lambda_t^{3}},
	\label{eqC10}
\end{equation}

\begin{equation}
	\left\langle c_{\tau}^{8}\right\rangle^{e q}=\left\langle c_{s}^{8}\right\rangle^{e q}=\frac{105}{16 \lambda_t^{4}}.
	\label{eqC11}
\end{equation}

\noindent When the order of $k$ is odd, the moment integrals of $\left\langle c_{\tau}^{k}\right\rangle^{eq}$ and $\left\langle c_{s}^{k}\right\rangle^{eq}$ are equal to zero.

\bibliographystyle{reference_style.bst}
\bibliography{Zotero.bib}
\end{document}